\documentclass[apjl,twocolumn]{openjournal}

\usepackage{natbib} 
\usepackage[dvipsnames]{xcolor}
\usepackage{aas_macros} 
\usepackage{amssymb}
\usepackage{amsmath}
\usepackage{threeparttable}
\usepackage{multirow}
\usepackage[normalem]{ulem}
\usepackage{comment}
\usepackage[title]{appendix}
\usepackage{hyperref}	
\usepackage{orcidlink}
\usepackage{placeins}
\hypersetup{colorlinks=true,linkcolor=blue,citecolor=blue,filecolor=blue,urlcolor=blue}
\hypersetup{
    colorlinks=true,       
    linkcolor=cyan,          
    citecolor=cyan,         
    urlcolor=cyan        
}
\usepackage[caption=false]{subfig}
\usepackage{soul}

\newcommand{\mum}{\mu{\rm m}}

\newcommand{\mycol}[1]{\begin{tabular}{c}#1\end{tabular}}

\graphicspath{ {./figs/} }

\usepackage{relsize}
\newcommand{\authsize}{\fontsize{10}{14}\selectfont}

\defcitealias{MRN}{MRN77}

\usepackage{etoolbox}

\definecolor{darkgreen}{RGB}{0,200,0}

\begin{document}

\title[Short Title for Header]{\textsc{{{Dust and Grain Size Evolution in Galaxy Simulations: \\ What Matters and What Does Not}}} \vspace{-15mm}}

\author{{\authsize Massimiliano Parente} \orcidlink{0000-0002-9729-3721}$^{1\,*}$}
\author{{\authsize Desika Narayanan} \orcidlink{0000-0002-7064-4309}$^{1,2}$}
\author{{\authsize Paul Torrey} \orcidlink{0000-0002-5653-0786}$^{3,4,5}$}

\thanks{$^*$e-mail: \href{mailto:parente.m@ufl.edu}{parente.m@ufl.edu}}

\affiliation{$^{1}$Department of Astronomy, University of Florida, 211 Bryant Space Sciences Center, Gainesville, FL 32611, USA}
\affiliation{$^{2}$Cosmic Dawn Center at the Niels Bohr Institute, University of Copenhagen and DTU-Space, \\ Technical University of Denmark, Denmark}
\affiliation{$^{3}$Department of Astronomy, University of Virginia, 530 McCormick Road, Charlottesville, VA 22903, USA}
\affiliation{$^{4}$Virginia Institute for Theoretical Astronomy, University of Virginia, Charlottesville, VA 22904, USA}
\affiliation{$^{5}$The NSF-Simons AI Institute for Cosmic Origins, USA}

\shorttitle{\textit{Dust grain sizes in cosmological simulations} \hfill}
\shortauthors{\hfill \textit{Parente, Narayanan \& Torrey}}

\begin{abstract}
We present the first implementation of an evolving dust grain size distribution (GSD) within a semi-analytic cosmological model (SAM) of galaxy evolution. This flexible model self-consistently accounts for stellar dust production, shattering, coagulation, accretion of gas-phase metals, and destruction in supernova-driven shocks and hot gas, successfully reproducing key observational constraints.  The purpose of this paper is to present the key physical elements of this novel dust implementation in a SAM and to explore controlled numerical experiments to identify the mechanisms shaping the GSD and extinction law in galaxies. Our results show that the GSD evolves from a large-grain–dominated regime at high redshift to a flatter, MRN-like shape at low redshift. This transition occurs earlier for massive galaxies, at a characteristic metallicity determined by the galaxy depletion time.
The resulting extinction curves show an increase of the UV/optical slope and a pronounced $2175\,\text{\AA}$ bump toward lower redshift, in good agreement with the extinction properties of the MW. Through numerical experiments, we find that once stars provide the initial reservoir of large grains, shattering and ISM accretion are the principal mechanisms driving the growth of small grains. When accretion is included, the model robustly reproduces the observed $z \approx 0$ dust masses, largely independent of the specific assumptions adopted for grain-size physics. The extinction properties of MW–like galaxies are also generally recovered, except in extreme cases, such as when grain velocities in turbulent media are assumed to be independent of grain size.
\end{abstract}


\maketitle

\section{Introduction}
\label{sec:intro}

Interstellar dust grains are small solid particles, with radii $\approx 1\,\mu{\rm m} - 10\,\text{\AA}$,  that play a fundamental role in shaping both the emission and physical properties of the interstellar medium (ISM) of galaxies. By absorbing optical and ultraviolet photons from stars and active galactic nuclei (AGN) and re-emitting this energy in the infrared, dust strongly affects the observed spectral energy distributions (SEDs) of galaxies. In addition, dust grains are key agents in several physical processes occurring in the ISM, including the ejection of material through radiation pressure \citep[e.g.,][]{Thompson05}, the formation of molecules on grain surfaces \citep[e.g.,][]{Wakelam17}, and the cooling of the medium itself \citep[e.g.,][]{Burke74}.

These processes depend not only on the total dust abundance, but also on the physical properties of the grains, and in particular on their sizes, commonly described in terms of the grain size distribution (GSD). Grain size determines, for example, the surface area available for chemical reactions \citep{Yamasawa11, Harada17}, as well as the interaction between grains and radiation \citep[e.g.,][]{Draine03}. The latter gives rise to wavelength-dependent extinction, defined as the loss of radiation along a given line of sight due to both absorption and scattering out of the line of sight \citep{SalimNara20}. Extinction can be studied by comparing intrinsically known spectral energy distributions (e.g., stars) with those affected by dust, and its properties can be used to infer constraints on both the GSD and grain composition. In this context, a seminal example is the work by \cite*[][hereafter \citetalias{MRN}]{MRN}, who showed that the extinction properties of the Milky Way (MW) can be reproduced by a mixture of graphite and silicate grains whose size distribution follows a power-law $\partial n / \partial a \propto a^{-3.3}$. More recently, \cite{Hensley_astrodust} introduced a physical dust model featuring a composite “astrodust” component and small polycyclic aromatic hydrocarbons (PAHs) able to simultaneously reproduce the wavelength dependence of galactic extinction, as well as dust polarization and emission, yielding a bimodal grain size distribution for \textit{astrodust} grains together with a PAH population peaking at $\lesssim 1 \, {\rm nm}$.

The GSD, together with the overall dust abundance, is shaped by a variety of processes operating in galaxies. Following the production of typically large grains by stellar sources, dust grains evolve in both mass and size within the ISM (e.g., Figure 1 of \citealt{Parente25rev} and \citealt{Calura25_rev} for a recent review). In dense environments like giant molecular clouds (GMCs), low-velocity collisions can lead to coagulation (the sticking together of grains to form larger particles), as well as the accretion of free metals from the ISM.   In more diffuse regions of the ISM, where relative velocities may be higher, grain-grain collisions result in fragmentation, commonly referred to as shattering. Finally, grains are eroded and destroyed in highly energetic environments, such as supernova (SN) shocks and the hottest phases of the ISM and CGM, through the process of sputtering, returning metals to the gas phase.

While early theoretical studies focused primarily on the evolution of dust abundance and composition \citep[e.g.,][]{Dwek1980}, more recent analytical work demonstrated the importance of simultaneously modeling both the size and mass evolution of dust grains in galaxies \citep[e.g.,][]{Hirashita09, Hirashita11, Hirashita12, Asano13}. A key result of these studies is that, since many dust evolution processes depend on the grain surface area and are therefore size-dependent, modeling the size evolution of grains has a direct impact on their mass evolution as well \citep[e.g.,][]{Kuo12,Narayanan25}.

In the last decade, hydrodynamic simulations of galaxy evolution have begun to include treatments of dust mass evolution. However, only a limited number of simulations are currently able to simultaneously track the evolution of grain size \citep[][for a comprehensive summary]{Parente25rev}. Among these, most adopt the computationally efficient two-size approximation \citep{Hirashita15}, which follows two representative grain populations rather than the full GSD, substantially reducing the computational cost \citep[e.g.,][]{Aoyama18, Gjergo18, Granato21, Parente2022, Dubois24, Ragone24, Trayford25, CALIMA}. When the full grain size evolution is included, it enables more accurate predictions of observables such as extinction and attenuation curves \citep{Aoyama20, Li2021, Matsumoto26, Caleb_prep}, and provides a more detailed framework for implementing additional physics, for example dust diffusion \citep{Romano22} and PAHs (\citealt{Narayanan23}). The simultaneous modeling of dust abundance and full size evolution in hydrodynamic simulations is, however, computationally expensive, and existing implementations have therefore largely focused on idealized setups, zoom-in simulations of lower-mass galaxies, or galaxies limited to high-redshift \citep[e.g.,][]{Narayanan25b}.

On the other hand, observational studies aimed at constraining the grain-size–dependent attenuation curves of galaxies \citep[e.g.,][]{Salim18, Reddy18, Battisti2020, Shivaei20b, Shivaei20a} have seen substantial progress in recent years \citep[e.g.,][]{{Markov23, Markov25, Witstok23,  Fisher25, Ormerod25, Shivaei25}}, thanks to the capabilities of the \textit{James Webb Space Telescope} (JWST). Its sensitivity and spectral coverage has enabled the characterization of attenuation curves out to high redshift ($z \approx 9$), with some results suggesting a rapid production and/or processing of small carbonaceous dust grains in the early Universe, posing new challenges for current models of dust formation and evolution within galaxies.

In light of this, what is missing is the incorporation of the full size distribution into a computationally efficient cosmological framework of galaxy evolution, enabling the study of the GSD across wide galaxy populations and over cosmic time. Developing such a framework is the main goal of this work. We build on the semi-analytic modeling (SAM) approach to galaxy formation \citep[e.g.,][]{WhiteFrenk91, Somerville99}, which combines high computational efficiency with physically motivated prescriptions for the processes regulating galaxy evolution.\\
In this work, we exploit the \textsc{L-Galaxies} 2020 SAM \citep{Henriques2020}, and in particular the version introduced in \cite{2023MNRAS.521.6105P}, to implement a treatment of both dust abundance and grain size evolution, largely following the prescriptions of \cite{Hirashita19}, previously adopted mainly in hydrodynamic simulations. While SAMs have been widely and successfully employed to model dust abundances in cosmological contexts \citep[e.g.,][]{Popping17, Vijayan19, Triani20, Yates24}, this work represents the first application of SAMs to the modeling of a full dust GSD.

Our model successfully reproduces key observational constraints, including the Dust-to-Gas (DTG) vs. metallicity relation and the extinction curves of MW-mass galaxies, which are characterized by \citetalias{MRN}-like GSDs. The relative abundance of small and large grains is also consistent with observations in the local Universe, and its evolution depends on the star formation and metal enrichment history of galaxies.
Exploiting the high efficiency of the SAM approach, we investigate the impact of the main parameters and assumptions commonly adopted to model GSD in galaxy evolution studies. We find that this modeling is generally robust, with uncertainties in the treatment of the GSD having only a modest effect on the predicted total dust abundance. This framework provides a solid basis for future studies aimed at improving the predictive power of galaxy formation models, both in terms of observable properties and physical interpretation. Indeed, while the dust model is largely stand-alone -- with the exception of the removal of gas-phase metals locked into grains -- this treatment will enable future implementations of size-dependent dust–gas coupling, including grain-surface molecule formation and size-dependent radiation pressure feedback.

This paper is organized as follows. Sections \ref{sec:SAM} and \ref{sec:dustmodel} describe the SAM and the dust physics implemented within it. The main results on dust abundance are briefly discussed in Section \ref{sec:res:dustab}. In Section \ref{sec:res:GSD}, we analyze the GSD predicted by the model, its evolution and dependence on galaxy properties, while Section \ref{sec:res:ext} presents the resulting extinction curves. Section \ref{sec:res:dustproc} presents an extensive set of numerical experiments exploring the impact of assumptions and parameters related to GSD modeling. Finally, Section \ref{sec:summary} summarizes our results.

\section{The Semi-Analytic Model}
\label{sec:SAM}
We adopt the public\footnote{\url{https://lgalaxiespublicrelease.github.io/index.html}.} SAM of galaxy evolution \textsc{L-Galaxies 2020} (\citealt{Henriques2020}), alongside the modification introduced in \cite{2023MNRAS.521.6105P} in terms of disc instabilities and SMBH growth.
The SAM is designed to run on the Dark Matter (DM) merger trees of the \textsc{Millennium} and \textsc{Millennium-II} simulations (\citealt{Millennium}; \citealt{Boylan-Kolchin2009}). The model incorporates a number of astrophysical processes, including gas cooling, molecular-based star formation, feedback and chemical enrichment from evolved stellar populations, growth and feedback from supermassive black holes, merger-driven starbursts, disc instabilities and bulges formation. We refer the reader to the supplementary material available online\footnote{\url{https://lgalaxiespublicrelease.github.io/Hen20_doc.pdf}} for a complete and technical description of all processes, though summarize the elements of the model that are especially important to this work.

First, the inclusion of the galactic chemical enrichment (GCE) model of \cite{Yates2013} in \textsc{L-Galaxies 2020}  allows for the tracking of $11$ elements released into the gas phase by AGB stars, SNIa, and SNII. This metal tracking is the essential building block of the subsequent dust model, which re-processes and re-distributes these metals into both the gas phase and the solid (grain) phase.
Second, the cold ISM disk of model galaxies is discretized into $12$ concentric rings with radii $r_i = 0.01 \cdot 2^i h^{-1} \, {\rm kpc}$ ($i=0,...,11$) (\citealt{Fu2013}).  This discretization allows for some spatial resolution for the various ISM processes, including those associated with dust formation and evolution. 
Finally, star formation is linked to the $\mathrm{H}_2$ amount of each ring, which is modelled according to a metallicity-dependent description (\citealt{Krumholz2009}; \citealt{McKee2009}).
The L-Galaxies SAM has demonstrated successes in studies of dust within galaxy evolution across cosmic time, as evidenced by the range of models developed in recent years \citep[][the latter based on an even more detailed GCE model that includes the effects of binary stellar evolution]{Vijayan19, 2023MNRAS.521.6105P, Parente24, Parente25GV, Yates24}.

We run the model on the \textsc{Millennium} merger trees (original box size $500/h$, $2160^3$ particles), adopting a \textit{Planck} cosmology\footnote{The original \textsc{Millennium} cosmology has been scaled according to \cite{Angulo2010} and \cite{Angulo2015}.} (\citealt{Planck14}) with the following parameters: $h=0.673$, $\Omega_{\rm m}= 0.315$, $\Omega_{\rm b}= 0.0487$, and $\sigma_8=0.829$. We adopt a \cite{Chabrier2003} initial mass function (IMF).

\section{The Dust Model}
\label{sec:dustmodel}

In our model, we track the mass of dust grains with radii in the range $a_{\rm min}=10^{-4}\,\mu{\rm m}$ to $a_{\rm max}=10\,\mu{\rm m}$ using $N=32$ logarithmically spaced size bins. The adopted size range encompasses the grain radii required to reproduce observed extinction curves \citep{Weingartner01}, and is consistent with the range adopted in previous grain size evolution models \citep[e.g.][]{Hirashita19}. Our results are not strongly impacted by the choice of the number of bins. The convergence performances of the model with respect to this parameter are briefly discussed in Appendix \ref{app:convergence}.
Grains are assumed to be spherical and compact, so that each grain radius corresponds to a mass given by $m(a)=4\pi a^{3}s/3$, where $a$ denotes the midpoint of each size bin and $s$ is the material density. We separately follow carbonaceous and silicate dust components, with silicates assumed to be in the form of olivine (MgFeSiO$_4$), whose material densities are $s=2.2$ and $3.3\,{\rm g\,cm^{-3}}$, respectively.

We account for the main physical processes affecting both dust abundance and grain size, including grain production by asymptotic giant branch (AGB) stars and core-collapse SNe, shattering in the diffuse ISM, accretion of gas-phase metals and coagulation in dense GMCs, and grain sputtering in SN-driven shocks and hot gas. The rates of these processes are computed in each radial ring of the model gas discs or in the hot gas halo, and the GSD is updated accordingly at each simulation time step. We also account for the destruction of grains incorporated into newly formed stars, commonly referred to as astration. This process reduces the dust content of the cold ISM, while leaving the dust-to-gas ratio unchanged, since both components are transferred into stars during star formation events. Although star formation is proportional to the molecular gas fraction -- which we identify with the dense ISM phase (Sect.~\ref{sec:shacoa:vel}) -- astration removes dust from the entire cold gas ring, leaving the dust-to-gas ratio unchanged in both phases. The dust abundance in the dense and diffuse phases is not tracked explicitly, but is instead derived at each time-step from the local molecular gas fraction.

The parameters adopted for these processes (e.g., grain condensation in stellar ejecta, grain growth and SN destruction efficiency, grain collisions leading to shattering and coagulation) in the fiducial model are not fine-tuned, but instead are taken from previous studies in the literature \citep[][among the others]{Hirashita19, Hirashita11}. However, a broad and qualitative exploration of the impact of these parameters and modeling assumptions is presented in Section \ref{sec:res:dustproc}.

\subsection{Stellar Production}
\label{sec:stellarprod}
Stellar populations enrich their surrounding medium with gas metals and dust grains. Stellar ejecta enrich both the cold and hot gas phases following the SAM prescriptions\footnote{We note that the default chemical model of \citet{Yates2021} is known to produce overly metal-enriched galaxies at high redshift. We verified that this issue persists in our implementation; the inclusion of dust reduces $Z_{\rm gas}$ by only $\approx 0.1\,{\rm dex}$, which is insufficient to reconcile the model with 
observations. A modified prescription with enhanced metal ejection into the CGM 
\citep{Yates2021} would partially alleviate the over-enrichment, and we plan to 
explore this in a future work.} of \citet{Yates2021}. Disk SNII inject a fraction of metals and dust $f_{\rm SNII,hot}=0.3$ into the hot gas, while AGB stars inject all metals and dust into the cold phase  (i.e. $f_{\rm AGB,hot}=0$). Bulge and ICL stars enrich only the hot phase (i.e. $f_{\rm Bulge,hot}=1.0$ and $f_{\rm ICL,hot}=1.0$).

Dust is assumed to form only in AGB stars and core-collapse SNe, with no contribution from SNe Ia (e.g., \citealt{Gioannini2017}; \citealt{Li2019}; \citealt{Parente2022}). Moreover, the latter are expected to make only a minimal contribution to iron dust \citep[e.g.,][]{Zhukovska2008}, which is not considered in our work. The amount of carbonaceous and silicate dust produced by these sources follows the implementation of \cite{2023MNRAS.521.6105P} and its briefly detailed in the following.\\
    
AGB stars form either carbonaceous or silicate grains depending on the C/O ratio in their ejecta. Since CO forms efficiently at the microscopic level, only the atoms not locked in CO are available for dust condensation \citep[e.g.,][]{Dwek1998}. For ${\rm C/O}>1$, only carbon remains and carbon dust forms with efficiency $\delta_{\rm AGB,C}=0.1$:
\begin{equation}
    M_{\rm C\,dust} = 
    \max\!\left[\delta_{\rm AGB,C}\left(M_{\rm C\,ej}-0.75\,M_{\rm O\,ej}\right),\,0\right],
\end{equation}
where $0.75$ is the O/C atomic‐weight ratio.

For ${\rm C/O}<1$, AGB stars produce silicates in our model. Assuming an olivine-like composition MgFeSiO$_4$, the number of units formed is limited by the \emph{key element} -- the element that provides the smallest number of available atoms \citep[e.g.,][]{Zhukovska2008}:
\begin{equation}
\label{eq:Nsil}
    N_{\rm sil} = \delta_{\rm AGB,sil}
    \min_{X\in[\mathrm{Mg,Fe,Si,O}]}
    \left(
        \frac{M_{X\,\rm ej}}{\mu_X\,N^{X}_{\rm ato}}
    \right),
\end{equation}
with $\delta_{\rm AGB,sil}=0.1$, $\mu_{\rm X}$ is the atomic weight of the X element and $N^{\rm X}_{\rm ato}$ the number of X atoms in the compound. The mass of each element condensed into silicate dust then follows from:
\begin{equation}
    M_{X\,\rm dust} = N_{\rm sil}\,\mu_X\,N^{X}_{\rm ato}.
\end{equation}

Unlike AGB stars, SNII ejecta are not mixed at the microscopic level, so carbonaceous and silicate grains may form simultaneously. The mass of carbon dust is therefore:
\begin{equation}
    M_{\rm C\,dust} = \delta_{\rm SNII,C}\, M_{\rm C\,ej},
\end{equation}
while the silicate dust mass follows from:
\begin{equation}
    M_{X\,\rm dust} = N_{\rm sil}\,\mu_X\,N_{\rm ato}^X,
\end{equation}
with ${X \in \mathrm{[Mg,Fe,Si,O]}}$. Here, $M_{X\,\rm ej}$ is the mass of element $X$ ejected by SNII, and $N_{\rm sil}$ is computed as in Eq.~\ref{eq:Nsil}. We adopt condensation efficiencies $\delta_{\rm SNII,C}=\delta_{\rm SNII,sil}=0.1$.\\
We note that our adopted condensation efficiencies fall within the wide range of values reported in the literature \citep[ $\approx 0.00035-0.8$, among the most extreme\footnote{We note that the lower end of this range (e.g., \citealt{Zhukovska2008}) is intended to account for dust destruction by the SN reverse shock. Since our adopted value is significantly larger, we are implicitly neglecting this destruction channel.}; e.g.,][]{Zhukovska2008, Dwek1998}, reflecting the fundamental uncertainties in stellar dust yields. However, since ISM reprocessing dominates the dust budget at low redshift, variations in $\delta$ have only a modest impact on predicted dust masses and grain sizes, though they can still influence the resulting chemical composition.\\

Finally, the size of grains produced at this stage is the same for both sources, and it is assumed to follow a log-normal distribution with a peak at $a_{\rm peak}=0.1 \, \mum$ and width $\sigma = 0.5$ \citep{Asano13}, motivated by results suggesting the preferential survival of large dust grains within SN shock \citep{Winters97, Nozawa07}.

\subsection{Shattering and Coagulation}


\begin{figure*}[]

    \centering
    \includegraphics[width=1.9\columnwidth]{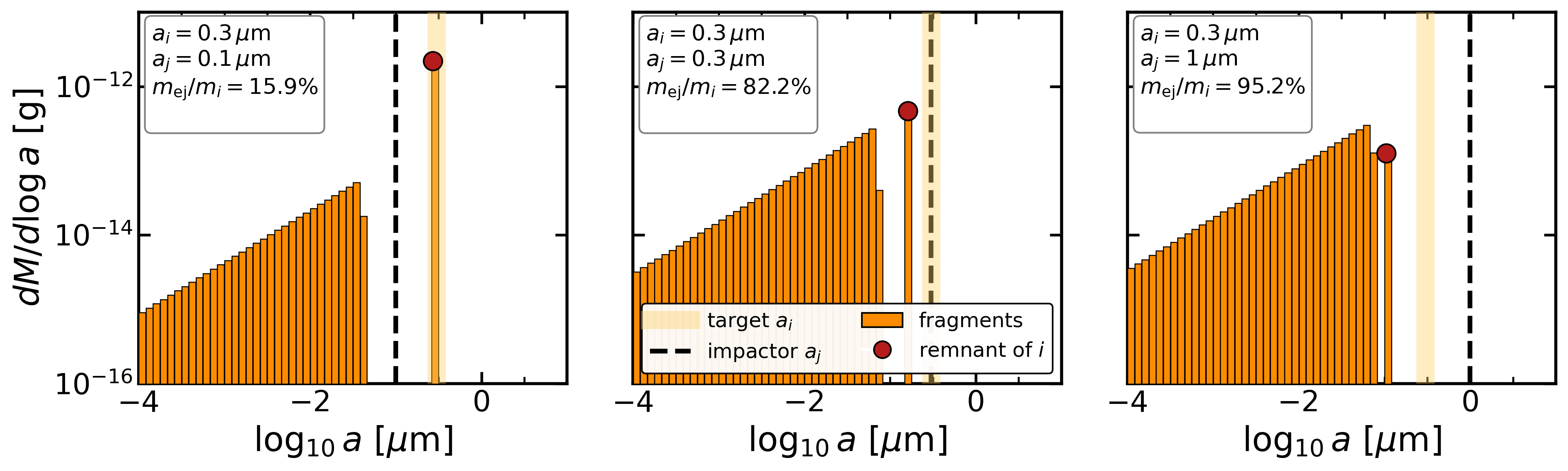}

    \caption{\textbf{Examples of shattered fragment distributions.} Three representative collisions between a target grain $i$ and an impactor $j$, for carbonaceous grains in diffuse-medium conditions ($n_{\rm gas}=0.1\,{\rm cm}^{-3}$, $T_{\rm gas}=10^{4}\,{\rm K}$). Orange histograms show the binned fragment distribution from Eq. \ref{eq:fragments}, while the red circle marks the remnant of grain $i$. As the impactor size increases, the collision becomes progressively more energetic (Eq. \ref{eq:mej}): the ejected mass grows toward complete disruption of the target, and the remnant mass correspondingly shrinks.}
    \label{fig:mufrag}
\end{figure*}

\begin{figure}
    \centering
    \includegraphics[width=0.9\columnwidth]{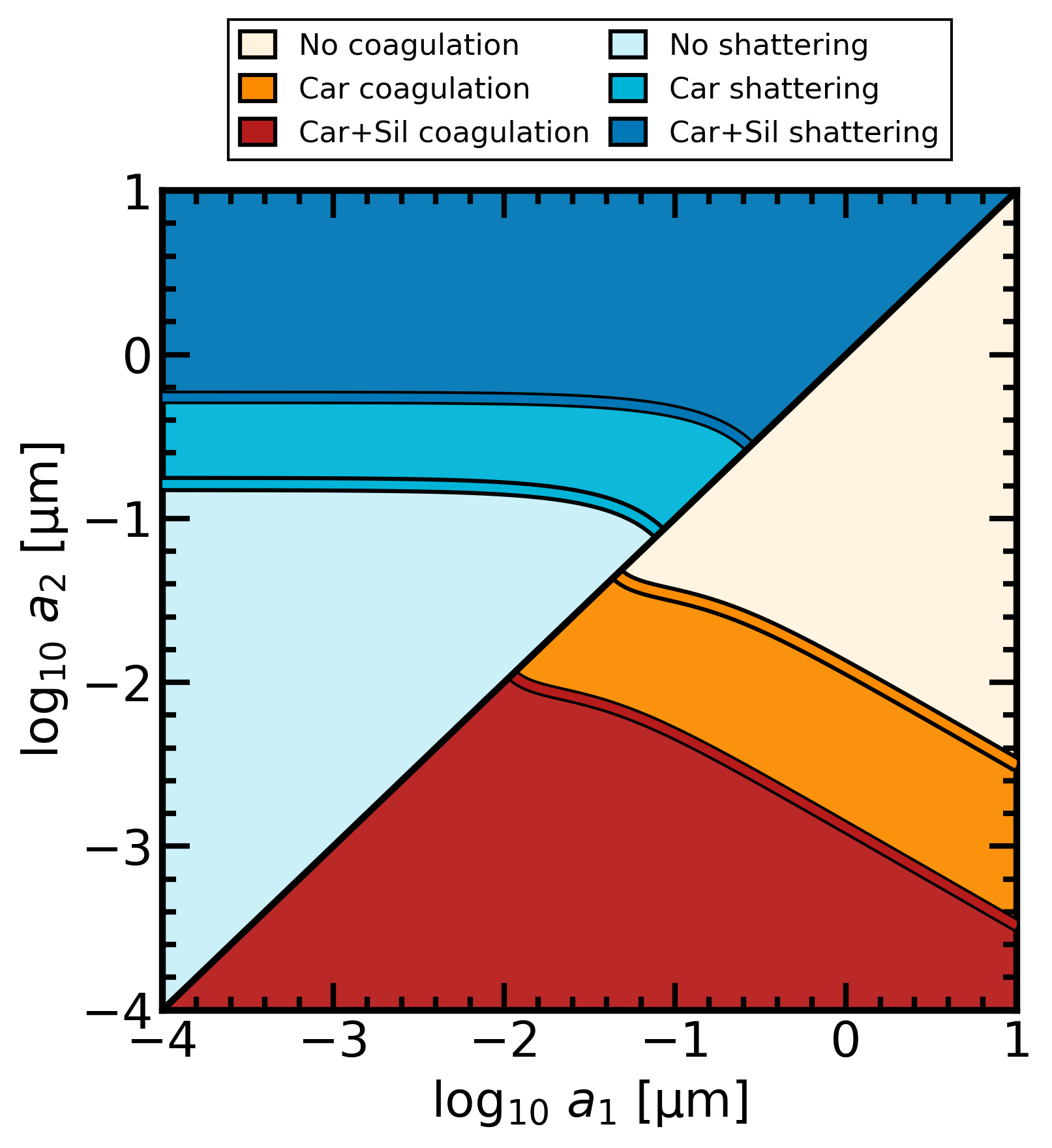}
    
    \caption{\textbf{Collision outcomes for pairs of dust grains in the diffuse and dense medium.} For each pair of colliding grains with radii $a_1$ and $a_2$, the colors indicate whether the relative velocity exceeds the shattering threshold (upper triangle; warm diffuse medium, $n_{\rm H}=0.1\,{\rm cm^{-3}}$, $T=10^3\,{\rm K}$) or falls below the coagulation threshold (lower triangle; cold dense medium, $n_{\rm H}=10^3\,{\rm cm^{-3}}$, $T=10\,{\rm K}$). Shattering affects the largest grains, which reach the highest turbulent velocities, whereas coagulation operates among the smallest grains. In both processes, carbonaceous grains are active over a broader region of diagram than silicates.}
    \label{fig:2phasedia}
\end{figure}

The evolution of the GSD due to shattering and coagulation -- both mass-conserving processes -- is implemented following \cite{Hirashita19}. 

Shattering modifies the dust mass density in each size bin $i$ according to:

\begin{equation}
    \frac{d \rho_{\rm dust, \, i}}{dt} = -m_i \rho_{ i} \sum_l \alpha_{li} \rho_{l} + \sum_{l, j} \alpha_{lj} \rho_{l} \rho_{j} m^{\rm shat}_{lj}(i),
    \label{eq:shacoa}
\end{equation}
where $m_i$ and $\rho_i$ are the grain mass and mass density associated with grains of radius $a_i$, respectively. The term $m^{\rm shat}_{lj}(i)$ represents the mass of fragments deposited into bin $i$ as a result of collisions between grains in bins $l$ and $j$. The collision frequency, normalized by grain mass, is

\begin{equation}
    \alpha_{lj} = \frac{\sigma_{lj} v_{lj}}{m_j m_{l}}
\end{equation}
where $\sigma_{lj}$ and $v_{lj}$ are the collisional cross-section and relative velocity of the colliding grains. \\

By construction, shattering conserves mass, such that

\begin{equation}
    \sum_i \frac{d \rho_{\rm dust, \, i}}{dt} = 0.
\end{equation}
This follows directly from 
\begin{equation}
    m^{\rm shat}_{lj}(i) = \int_i \mu_{\rm frag}(m; m_l, m_j)dm,
\end{equation}
where $\mu_{\rm frag}$ is the fragment mass distribution resulting from the collision of grains of masses $m_l$ and $m_j$. The integral of $\mu_{\rm frag}$ over all bins equals the mass of the disrupted grain.

The fragment mass distribution depends on the physical properties of the collision, in particular the relative velocity. Following \citet{HirashitaKoba13}, shattering exhibits a threshold behavior associated with catastrophic fragmentation, defined as the regime in which half of the grain mass is disrupted. This threshold is characterized by the material-dependent parameter $Q_D^*$, proportional to the critical pressure \citep{Kobayashi10}. The fraction of mass disrupted in a collision is determined by
\begin{equation}
    \phi = \frac{E_{\rm imp}}{m_1 Q_D^*},
\end{equation}
where the impact energy is $E_{\rm imp} = \frac{1}{2} \frac{m_1 m_2}{m_1+m_2} \cdot v_{\rm rel}^2$ and the ejected mass is then given by
\begin{equation}
    m_{\rm ej} = \frac{\phi}{1+\phi} m_1.
    \label{eq:mej}
\end{equation}
Once $m_{\rm ej}$ is determined, the fragment mass distribution is assumed to follow a power law between $m_{\rm min} = 10^{-6} m_{\rm max}$ and $m_{\rm max} = 0.02 m_{\rm ej}$ \citep{Guillet11}, such that
\begin{equation}
\label{eq:fragments}
\begin{split}
\mu_{\rm frag} (m; m_1, m_2) = {} & 
\frac{(4-\alpha_f) m_{\rm ej} m^{(1-\alpha_f)/3}}
     {3 \left[ m_{\rm max}^{\frac{4-\alpha_f}{3}} - m_{\rm min}^{\frac{4-\alpha_f}{3}} \right]} \\
& + (m_1 - m_{\rm ej}) \, \delta(m - m_1 + m_{\rm ej}),
\end{split}
\end{equation}
where $\alpha_f = 3.3$ is the fragment size-distribution exponent \citep{Jones96}, and the second term accounts for the surviving remnant of mass $m_1 - m_{\rm ej}$. An explicit example of a shattered fragment distribution is shown in Fig.~\ref{fig:mufrag} for a target grain of size $a_i = 0.03\,{\rm \mu m}$ colliding with three impactors of different sizes.\\

Coagulation is implemented in an analogous manner, with the fragment distribution term replaced by
\begin{equation}
m_{lj}^{\rm coag} ( i) = 
\begin{cases}
m_l \quad {\rm if } \quad m_l + m_j \in [m_i \pm \Delta m_i] \\
0 \quad \, \, \, \, {\rm otherwise}, \\
\end{cases}
\end{equation}
where $[m_i \pm \Delta m_i]$ defines the mass range of the $i$-th size bin.

\subsubsection{Grains velocities and diffuse/dense phase treatment}
\label{sec:shacoa:vel}

A key ingredient in determining the efficiency of both shattering and coagulation is the relative velocity between colliding grains. We adopt the prescription of \citet{Hirashita19}, according to which the grain velocity is given by\footnote{Unlike in their original formulation, we fix the Mach number to unity. Since we do not explicitly resolve the medium, this would just be another (degenerate) free parameter.}
\begin{equation}
\begin{split}
    v_{\rm gr} = 1.1 \left( \frac{a}{0.1 \, {\mu \rm  m}}\right)^{0.5}  \left( \frac{T_{\rm gas}}{10^4 \, {\rm K}}\right)^{0.25} \left( \frac{n_{\rm H}}{1 \, {\rm cm^{-3}}}\right)^{-0.25} \\ 
    \times \left( \frac{s}{3 \, {\rm g cm^{-3}}}\right)^{0.5} \, {\rm km \cdot s^{-1}},
\end{split}
\label{eq:vgrain}
\end{equation}
which relates grain velocity to grain size and material density, as well as to the gas density and temperature.
Although approximate, this formula qualitatively captures the expected trends of grain dynamics: (i) larger grains attain higher velocities as they couple to larger-scale gas motions; (ii) grain velocities increase in hotter environments; (iii) grains move faster in lower-density media, where gas drag is weaker; and (iv) grains composed of denser materials have greater inertia and are therefore less efficiently decelerated\footnote{We note that the higher equilibrium velocity of denser grains arises because, in the Epstein drag regime, a larger stopping time $\tau_{\rm stop} \propto s \cdot a / (\rho_{\rm gas} c_{\rm s})$ corresponds to preferential coupling to larger, more energetic turbulent eddies \citep[see][for a more detailed treatment]{Ormel07, Ormel09}.} by gas drag \citep[see][for a more comprehensive description]{Hirashita19}.

Unlike hydrodynamic simulations, which directly resolve gas densities and temperatures at high spatial resolution \citep[e.g.,][]{McKinnon18, Aoyama20}, our SAM requires adopting representative values for the dense and diffuse ISM phases. For the dense phase, identified with GMCs, we assume fixed values\footnote{Although the physical conditions within GMCs are known to vary, we adopt fixed values here that lie within the typical ranges reported in the literature \citep[e.g.,][]{Hirashita00, Tielens05}. We note, however, that their impact on our results is minimal (Sect. \ref{sec:shacoa:exp}).} of $n_{\rm H,\,dense}=10^{3}\,{\rm cm^{-3}}$ and $T_{\rm dense}=10\,{\rm K}$. For the diffuse phase, we adopt a constant temperature $T_{\rm diffuse}=10^{4}\,{\rm K}$ and estimate the gas density from the properties of each ring as $n_{\rm H,\,diff} = \frac{\rho_{\rm diff}}{\mu m_{\rm H}}$, where the diffuse gas mass density is computed as
\begin{equation}
    \rho_{\rm diff} = \frac{ (1-f_{\rm H_2})M_{\rm gas}}{\pi \Delta R^2 \times (0.1 \cdot R_{\rm cold, gas})}.
    \label{eq:rhodiff}
\end{equation}
Here, $(1-f_{\rm H_2})M_{\rm gas}$ represents the diffuse (i.e. non-molecular) gas mass in each ring, while the denominator is the volume of the annulus, with surface area $\pi\Delta R^2$ and a scale height equal to one tenth of the cold gas disc radius. We stress that the dense and diffuse ISM phases are described by single representative values of
($n_{\rm H}, \,T$), shared by all grain sizes and chemical compositions. This effective description is adopted within our SAM framework, which does not explicitly resolve the internal substructure of the ISM, but instead tracks only the mass fractions associated with the two phases. Hydrodynamic studies have nevertheless shown that the GSD can differ between the diffuse and dense media \citep{Aoyama20}.

Once the grain velocities in the dense and diffuse phases are determined for each grain size, the relative velocity between two grains is computed as
\begin{equation}
    v_{\rm rel} = \sqrt{v^2_1 + v_2^2}.
\end{equation}
Shattering is allowed only when the relative velocity exceeds a material-dependent threshold, namely $v^{\rm C}_{\rm sh,\,th}=1.2\,{\rm km\,s^{-1}}$ for carbonaceous grains and $v^{\rm Sil}_{\rm sh,\,th}=2.7\,{\rm km\,s^{-1}}$ for silicate grains \citep{Jones96}.

Similarly, coagulation is assumed to occur only when the relative velocity is below a size- and material-dependent threshold \citep{Chokshi93, Dominik97, Yan04}, given by
\begin{equation}
    v^{ij}_{\rm th, \, coa} = 21.4 \left( \frac{a_i^3 + a_j^3}{(a_i + a_j)^3} \right)^{1/2} \frac{\gamma^{5/6}}{E^{1/3}R^{5/6}s^{1/2}},
\end{equation}
where $\gamma$ is the surface energy per unit area, $E$ depends on the Young’s modulus and Poisson’s ratio, and $R$ is the reduced grain radius. We adopt the material parameters listed in Table 3 of \citet{Chokshi93}, using quartz and graphite as proxies for silicate and carbonaceous grains, respectively.

The combined adoption of velocity thresholds and the size dependence of grain velocities naturally results in shattering being most efficient for larger grains, while coagulation preferentially affects smaller grains. This is explicitly shown in the phase diagram in Fig. \ref{fig:2phasedia} for both silicate and carbonaceous grains.

Finally, since Equation (\ref{eq:shacoa}) yields shattering and coagulation rates per unit volume, we assign volumes to the dense and diffuse phases by assuming that they coexist in pressure equilibrium within each annulus, such that
\begin{equation}
n_{\rm diff} T_{\rm diff} = n_{\rm dense} T_{\rm dense},
\end{equation}
with the temperatures taken to be the same as those used in the grain velocity calculations. Typical filling factors of the dense phase are $\lesssim 10^{-4}$, which justifies the assumption that the diffuse medium uniformly fills each ring (Eq. \ref{eq:rhodiff}).

\subsection{Dust-gas metals exchange}

The number-conserving processes that exchange mass between dust grains and gas-phase metals -- accretion in molecular clouds, destruction in SN-driven shocks, and sputtering in hot gas -- are modeled through the continuity equation
\begin{equation}
\label{eq:continuity}
    \frac{\partial n(a, t)}{\partial t} + \frac{\partial \left( \dot{a}n(a,t)\right) }{\partial a} = 0,
\end{equation}
where $n(a,t)$ is the number density of grains of size $a$ at time $t$. 

This equation is solved using a Lagrangian formulation in grain-size space, in which the edges of each size bin are advected along the accretion characteristics, $a_{\rm edge}(t) = a_{\rm edge}(t_0) + \dot{a}\,\Delta t$ with
$\Delta t$ being the adopted time-step. Timesteps are chosen by comparing the SAM galaxy evolution time-step with the timescales of grain accretion and SN destruction. Specifically, for each process (accretion or SN destruction), we set $\Delta t = {\rm min} \left( \Delta t_{\rm SAM}, \tau_{\rm acc/SN}\right)$, where $\tau_{\rm acc}$ and $\tau_{\rm SN}$ are evaluated for the smallest grain size, which corresponds to the fastest timescales. 
Since the evolved bins generally do not coincide with the fixed logarithmically spaced grid, the updated number distribution is obtained through a conservative remapping procedure. Specifically, the content of each advected bin is redistributed onto the fixed bins according to the fractional overlap in logarithmic size space, thereby ensuring conservation of the total number of grains. Grains that are moved beyond (below) the largest (smallest) bin are accumulated into the final (first) bin.

The key quantity in the continuity equation (Eq. \ref{eq:continuity}) is the grain growth or erosion rate, $\dot{a}=a/\tau(a)$, where $\tau(a)$ is the characteristic, size-dependent timescale of the relevant physical process. We detail its derivation for each process in the following.

\subsubsection{SN destruction}

Dust destruction by SN explosions is modeled by assuming that each SN event sweeps a mass $M_{\rm swept}$ of the ISM, out of the total gas mass $M_{\rm gas}$. Grains in the shocked gas are destroyed with an efficiency $\epsilon(a)$ which is larger for smaller grains \citep{Hirashita19}:
\begin{equation}
    \epsilon (a) = 1 - {\rm exp} \left[ -0.1  \left( \frac{a}{0.1 \mu{\rm m}}\right)^{-1} \right].
    \label{eq:SNeps}
\end{equation}
The resulting size-dependent destruction timescale is then
\begin{equation}
    \tau_{\rm des, \, SN} (a) = \frac{M_{\rm gas}}{\epsilon(a)M_{\rm swept} \mathcal{R}_{\rm SNII+SNIa}},
\end{equation}
where $M_{\rm gas}$ and $\mathcal{R}_{\rm SNII+SNIa}$ are the cold gas mass and the combined core-collapse and Type Ia SN rates, respectively, as computed by the SAM in each radial ring. We adopt $M_{\rm swept}=6800\,M_\odot$, following \citet{McKee89} and \citet{Nozawa06}. We note that the destruction efficiency $\varepsilon(a)$ is a size-dependent function, normalised so that $\varepsilon(0.1\,\mu{\rm m})=0.1$ and reaching values of $0.1$--$0.6$ over the $0.01$--$0.1\,\mu{\rm m}$ range that dominates the dust mass, consistent with the original suggestion of $\approx 0.3-0.4$ by \citet{McKee89}.

We neglect the coupled effects of sputtering and shattering in SN shocks. This combined process has been shown to be important by \cite{Kirchschlager22}, who used hydrodynamic simulations of SN blast waves and found that the joint action of sputtering and shattering can enhance the effective grain destruction by up to an order of magnitude. This effect has instead been recently incorporated into hydrodynamic galaxy evolution simulations by \cite{Caleb_prep}, where it was found to be crucial for preventing the survival of very large ($\gtrsim 0.3\,\mu\mathrm{m}$) grains.

\subsubsection{Accretion}

The grain accretion timescale is formulated to account for the enhanced efficiency of accretion in high-metallicity environments, its preferential action on small grains, and its dependence on the local molecular gas content. We follow the element-by-element prescription of \citet{Hirashita11} and \citet{Granato21}, adopting the size-dependent accretion timescale
\begin{equation}
\label{eq:acc}
    \tau_{{\rm acc}, {\rm X}}(a) = \frac{f_{\rm X} s \mu_{\rm X} }{3 n Z_{\rm X} \bar{\mu} S} \left(\frac{2 \pi}{m_{\rm X} k T}\right)^{0.5} \frac{a}{0.005 \mum} \frac{1}{f_{\rm H_2}}.
\end{equation}
Here, $S=0.3$ is the sticking coefficient, while $T=10\,{\rm K}$ and $n=10^3\,{\rm cm^{-3}}$ are the temperature and number density assumed for unresolved GMCs. The quantities $\mu_X$ and $m_X$ denote the atomic weight and atomic mass of element $X$, respectively, $\bar{\mu}$ is the mean molecular weight of the gas, $f_X$ is the mass fraction of element $X$ in the grain, and $Z_X$ is its gas-phase mass fraction. The molecular gas fraction in the annulus is given by $f_{\rm H_2}$, and $a$ is the grain radius.\\
For carbonaceous grains, we adopt the accretion timescale computed for $X={\rm C}$. For silicate grains, composed of O, Si, Mg, and Fe, we adopt the longest accretion timescale among these elements. This choice ensures the preservation of the olivine-like stoichiometry assumed for silicate dust in this work.\\
We note that the size dependence implies that small grains grow proportionally faster than large ones -- despite their smaller individual surface area -- and are therefore the main sites of metal accretion. In combination with shattering, which continuously replenishes the small-grain population, this builds up the small-grain tail of the GSD rather than simply shifting the large-grain peak to larger sizes.

\subsubsection{Thermal Sputtering}
\label{sec:dustspu}

The erosion of dust grains in hot gas due to thermal sputtering is modeled following \citet{Tsai95} and \citet{2023MNRAS.521.6105P}. The corresponding size-dependent sputtering timescale is:
\begin{equation}
\label{eq:tauspu}
\begin{split}
\tau_{\rm spu}(a) &= \tau_{\rm spu,0} \, 
\frac{a_{0.1 \, \rm \mu m}}{\rho^{\rm gas, hot}_{10^{-27} \, \rm g \, cm^{-3}}} \\
&\quad \times 
\Biggl[ 
\left(\frac{T_{\rm spu,0}}{{\rm min}(T_{\rm gas,hot}, 3 \cdot 10^7 \, \rm K)} \right)^\omega 
+ 1
\Biggr].
\end{split}
\end{equation}
where $\tau_{\rm spu,0}=1.7/3\,{\rm Gyr}$, $T_{\rm spu,0}=2\times10^{6}\,{\rm K}$, and $\omega=2.5$. The quantity $a_{0.1\,\mu{\rm m}}$ is the grain radius in units of $0.1\,\mu{\rm m}$, $\rho^{\rm gas,hot}{10^{-27}\,{\rm g\,cm^{-3}}}$ is the hot gas density , and $T_{\rm gas,hot}$ is the temperature of the hot gas.

The hot gas density and temperature are estimated from the halo properties as
\begin{equation}
    \rho^{\rm gas, hot} = \frac{M_{\rm gas, hot}}{4\pi R^3_{\rm vir}/3}
\quad {\rm and} \quad
    T_{\rm gas, hot} = 35.9 \cdot \left( \frac{V_{\rm vir}}{\rm km/s}\right)^2 \, [{\rm K}],  
\end{equation}
where $R_{\rm vir}$ and $V_{\rm vir}$ are the virial radius and virial velocity of the halo for central galaxies, or the corresponding values at the time of infall for satellite systems. The gas temperature comes from the virial thorem \citep[e.g.,][]{MVWbook}, with mean molecular weight $\mu = 0.59$. The same prescription is applied to dust residing in the ejected gas reservoir.\\
Finally, we note that dust residing in the hot gas is allowed to return to the ISM during cooling episodes, following the same prescription adopted for the cooling gas and gas-phase metals. Specifically, it is distributed among the ISM rings according to an exponential profile whose scale radius is proportional to the ratio of the DM halo spin to its circular velocity \citep{Henriques2020}. Both the dust-to-gas ratio and GSD are therefore preserved during the cooling episode.

\section{Modelled dust abundance}
\label{sec:res:dustab}

\begin{figure*}[!htb]

    \centering
    \includegraphics[width=0.65\columnwidth]{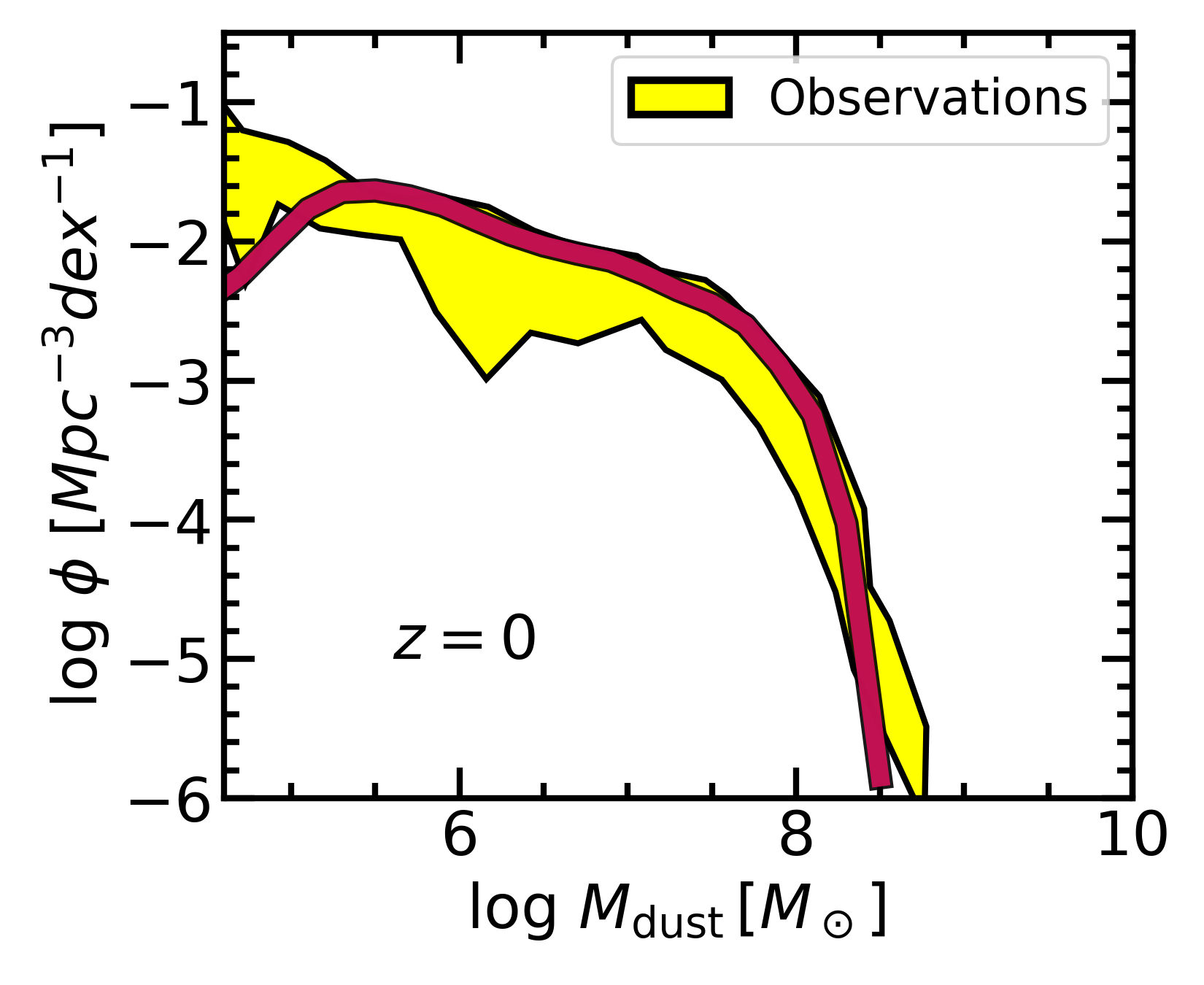}\quad
    \includegraphics[width=0.66\columnwidth]{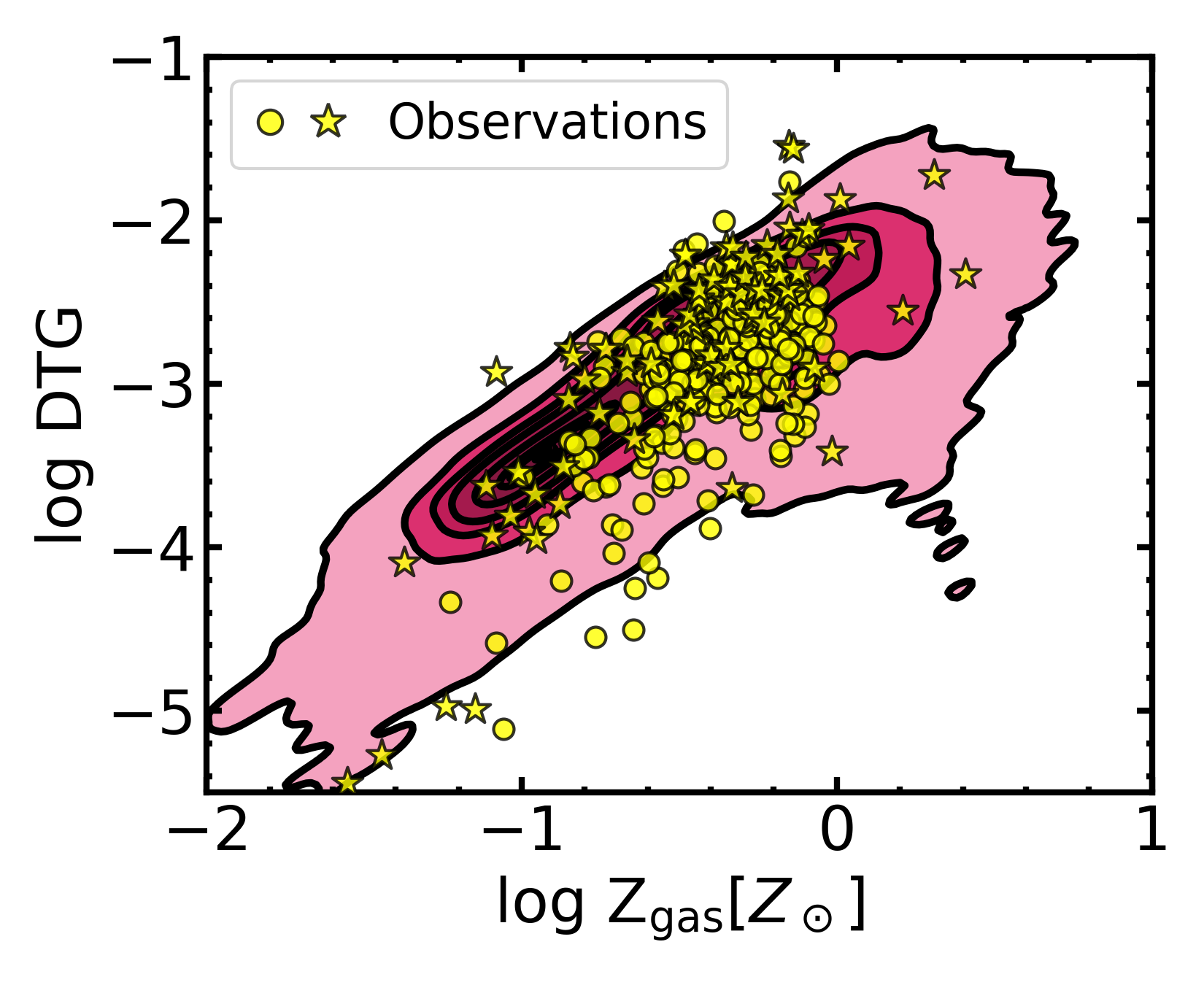}\quad
    \includegraphics[width=0.65\columnwidth]{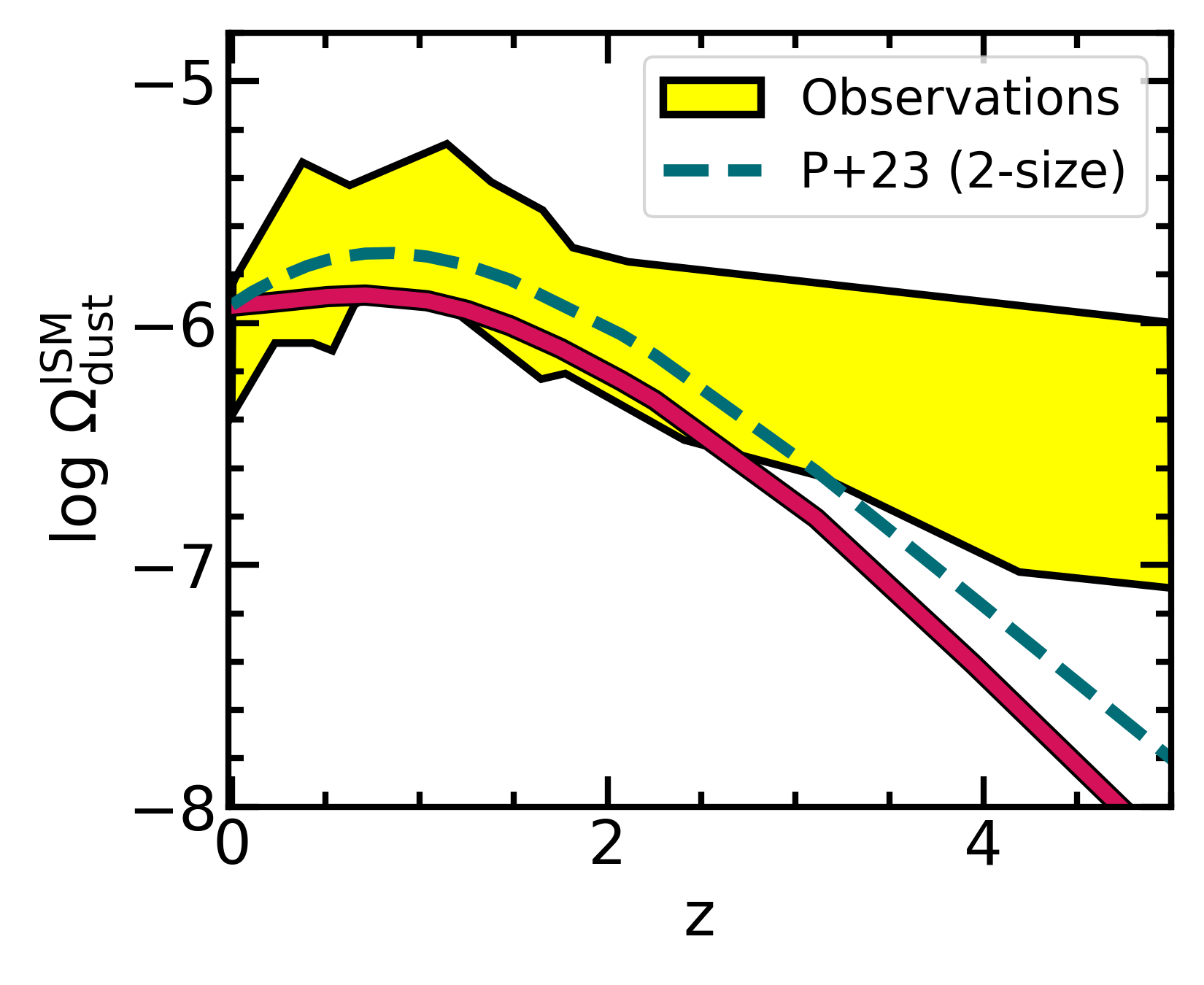}\quad

    \caption{\textbf{Model predictions of dust abundance are in good agreement with observations.} The local DMF (left panel), DTG-$Z_{\rm gas}$ relation at $z=0$ (middle panel) and cosmic dust denisty evolution with redshift (right panel) are shown. 
    Observational data are shown as yellow points or shaded regions and refer to \cite{Vlahakis2005, Dunne2011, Beeston2018a} for the DMF; \cite{Remy-Ruyer2014a, DeVis2019} for the DTG$-Z_{\rm gas}$ relation; and \cite{Vlahakis2005, Dunne2011, Beeston2018a, Dudzeviciute2020, Pozzi2020, Berta25} for $\Omega^{\rm ISM}_{\rm dust}$. In the latter we also show predictions from the 2-size dust model implemented in the same SAM by \citet[][dashed line]{2023MNRAS.521.6105P}.} 
    \label{fig:fid:main}
\end{figure*}

We begin by investigating the performance of our model against key observational benchmarks of dust abundance across cosmic time. These include the local dust mass function (DMF), the relation between dust-to-gas ratio (DTG) and ISM metallicity, and the cosmic evolution of the dust density parameter $\Omega_{\rm dust}$\footnote{The cosmic dust parameter is $\Omega_{\rm dust}(z)=\frac{\rho_{\rm dust}(z)}{\rho_0}$, with $\rho_{\rm dust}(z)$ the cosmic dust density and $\rho_{\rm c,0}=2.775 \,h^2 \times 10^{11} \, M_\odot/{\rm Mpc}^3$ the critical density of the Universe today.}. Unless specified otherwise, we refer to ISM dust, which is present in the cold gas disc of our model galaxies.

These comparisons for our fiducial model are shown in Figure \ref{fig:fid:main}, together with a compilation of observational data. The model reproduces $z = 0$ observations very well, particularly the DTG--$Z_{\rm gas}$ relation, which is a key indicator of the role of grain accretion in the ISM \citep[e.g.,][]{Asano13}. We note, however, a mild overestimate of the DTG at ${\rm log}\,Z_{\rm gas}/Z_\odot \lesssim -1$, where our model produces a flatter relation than suggested by the available observational data. This is not uncommon among cosmological dust models \citep{PP22, Parente25rev}, and may reflect an overestimation of stellar dust production, whose efficiency is admittedly uncertain (Sect. \ref{sec:stellarprod}), or a too efficient ISM accretion. On the other hand, we also note that the scarcity of observational constraints in this regime limits our ability to assess the quality of the modeling.\\
The agreement with observations is weaker when examining the redshift evolution of the total dust abundance, particularly at $z \gtrsim 2$, where our model systematically underpredicts the observed trend. Note however that this comparison should be interpreted with caution since observations at $z \gtrsim 0.5$ are limited to massive systems, introducing biases in the inferred DMF and, consequently, in $\Omega_{\rm dust}$ \citep{Traina24, Parente25rev}. In particular, at high$-z$ only the most dust-rich systems are detected, leaving the low-mass end of the DMF poorly constrained. Since $\Omega_{\rm dust}$ is obtained by integrating the DMF, its value relies heavily on the extrapolation of the latter, potentially leading to an overestimate of the total dust abundance.\\
A comparison with the two-size approximation of \citet{2023MNRAS.521.6105P}, implemented in the same SAM configuration, reveals only modest differences in dust abundance across cosmic time: that model lies slightly above the present one at most redshifts.
Also, the drop in $\Omega_{\rm dust}$ suggested by observations at $z\lesssim 1$, already weak in \citet{2023MNRAS.521.6105P}, is now almost absent. While that work attributed this drop mainly to supermassive black holes feedback, we find that the more detailed dust treatment adopted here also contributes to shaping the low-redshift dust abundance.
These differences are not surprising, given that the treatments of shattering, coagulation, accretion, and SN destruction differ substantially between the two implementations -- in particular, the multi-size scheme replaces the timescale-based mass transfer of the two-size approach, and both accretion and destruction are now explicitly size-dependent. Although isolating the specific dust process responsible for this difference is not trivial, the overall broad agreement between these two independent prescriptions (i.e., two-size and multi-size) is reassuring, and attests to the overall robustness of dust modeling within the \textsc{L-Galaxies} framework.

\section{Grain Size Distribution}

\label{sec:res:GSD}

\begin{figure*}[!htb]

    \centering
    \includegraphics[width=0.85\columnwidth]{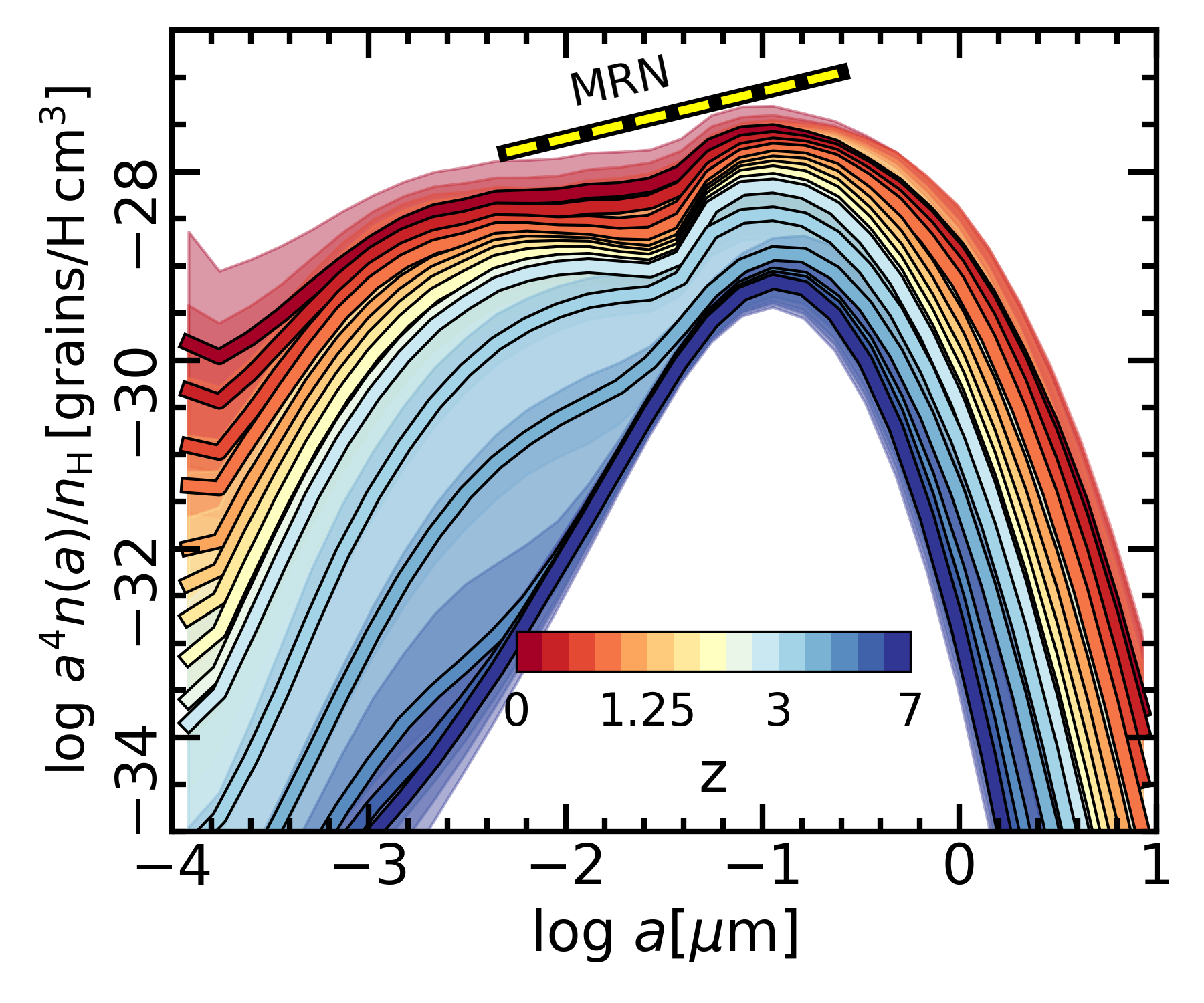}\quad
    \includegraphics[width=1\columnwidth]{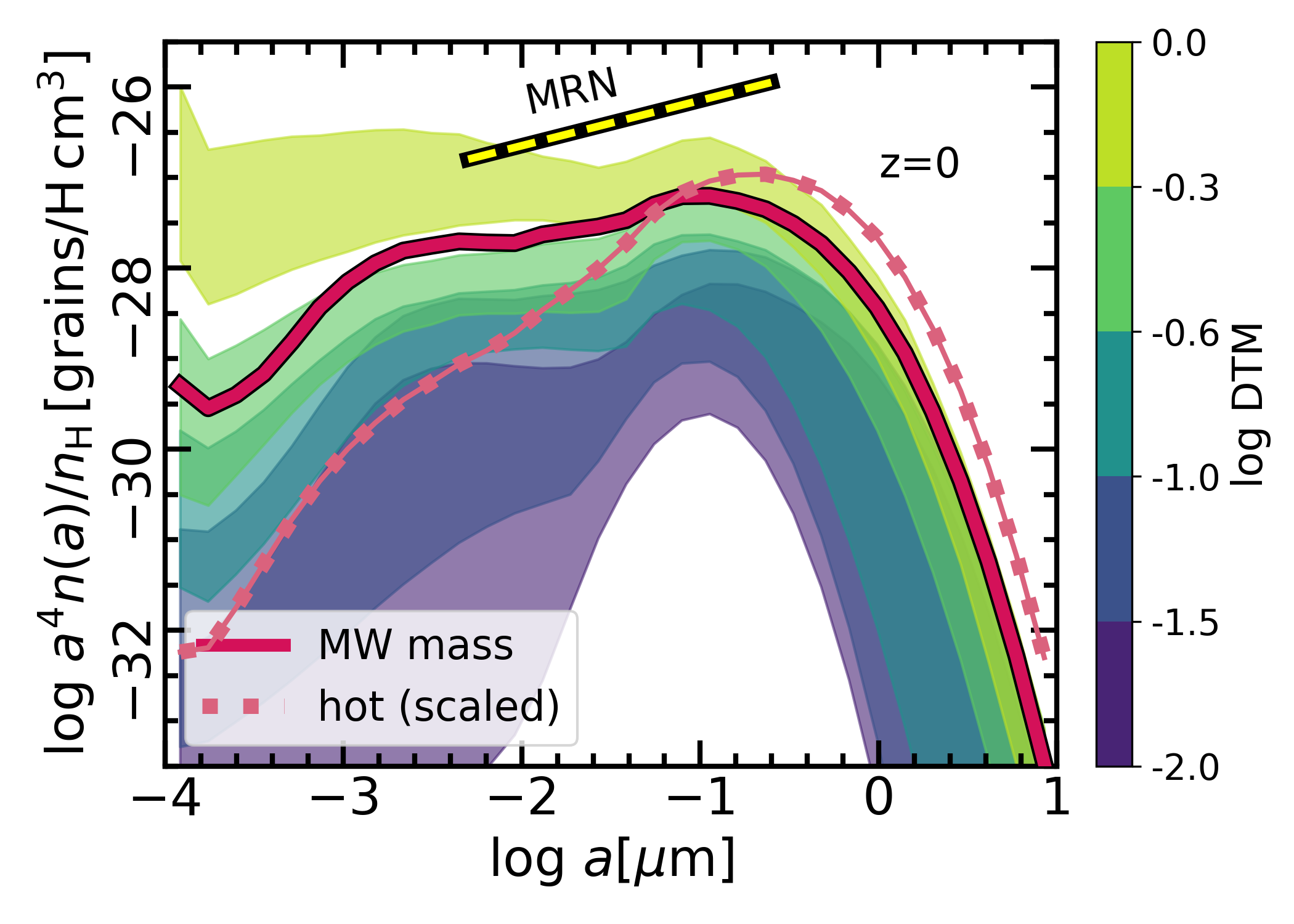}\quad

    \caption{\textbf{The GSD evolves from being dominated by large grains toward an \citetalias{MRN}-like shape as redshift decreases and the DTM increases.} GSD of the full simulated sample across redshift (left) and at $z=0$ (right). In the right panel, MW-mass galaxies are shown in red, the rescaled GSD of dust in the hot gas for the full sample is shown as a dotted line, and GSDs for different DTM ratios are color-coded. Solid lines and shaded regions indicate medians and 16th–84th percentiles. The \citetalias{MRN} slope is reported for reference.} 
    \label{fig:GSD:main}
\end{figure*}


\begin{figure}
\centering
\includegraphics[width=0.9\columnwidth]{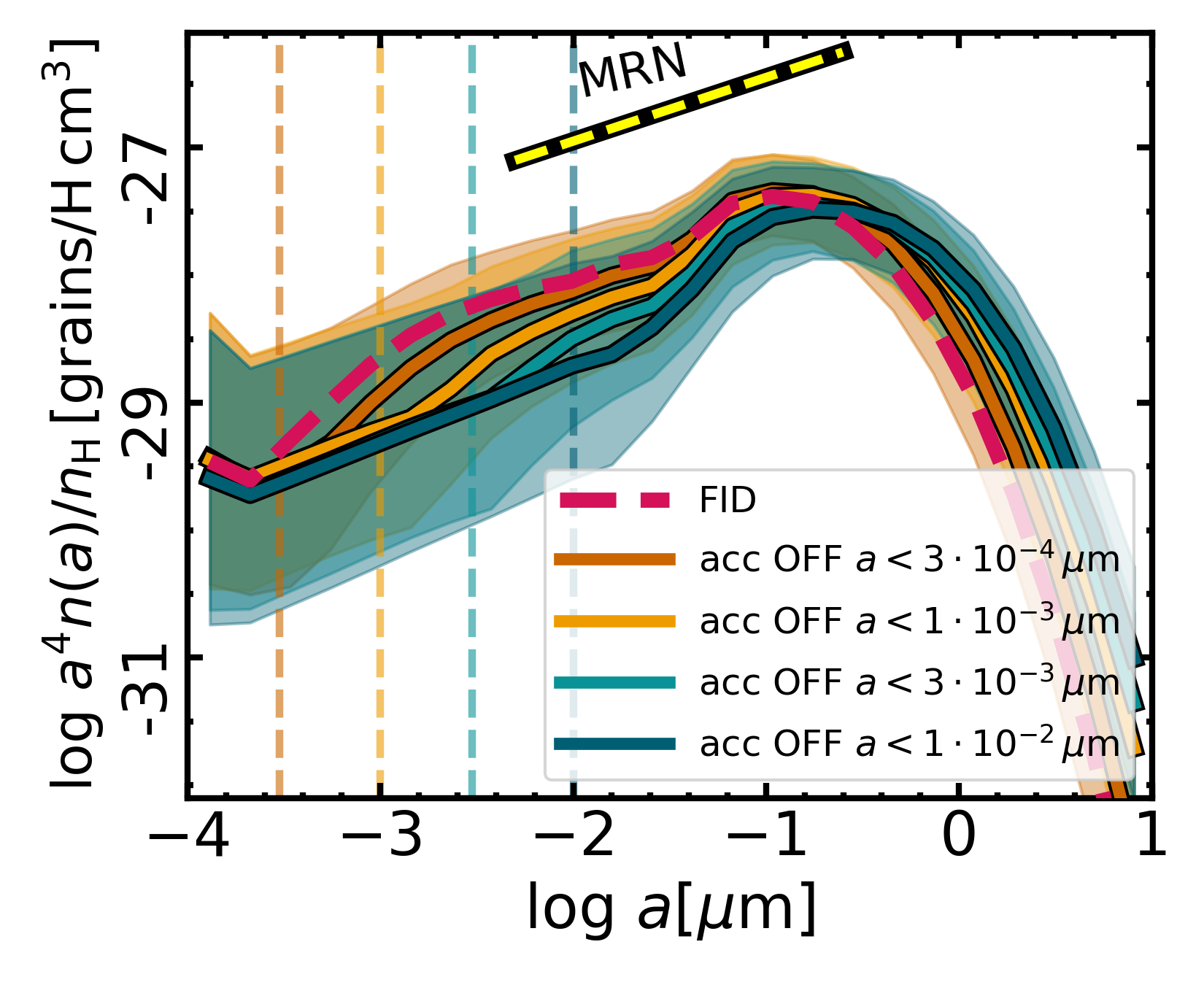}
\caption{\textbf{The small size tail of the GSD is due to the combined action of accretion and shattering} Effect of suppressing accretion below a size threshold on the $z=0$ GSD of the full simulated sample. The fiducial model (dashed red) is compared to runs in which accretion is switched off below the grain radii marked by the vertical dashed lines of matching color. As accretion onto small grains is suppressed, the small-grain tail is reduced and the large grain peak shifts to larger sizes. Small grains are the preferential sites of metal accretion, and the small grain tail of the GSD arises from the combined action of accretion and shattering.}
\label{fig:acc_threshold}
\end{figure}


\begin{figure*}[!htb]

    \centering
    \includegraphics[width=0.95\columnwidth]{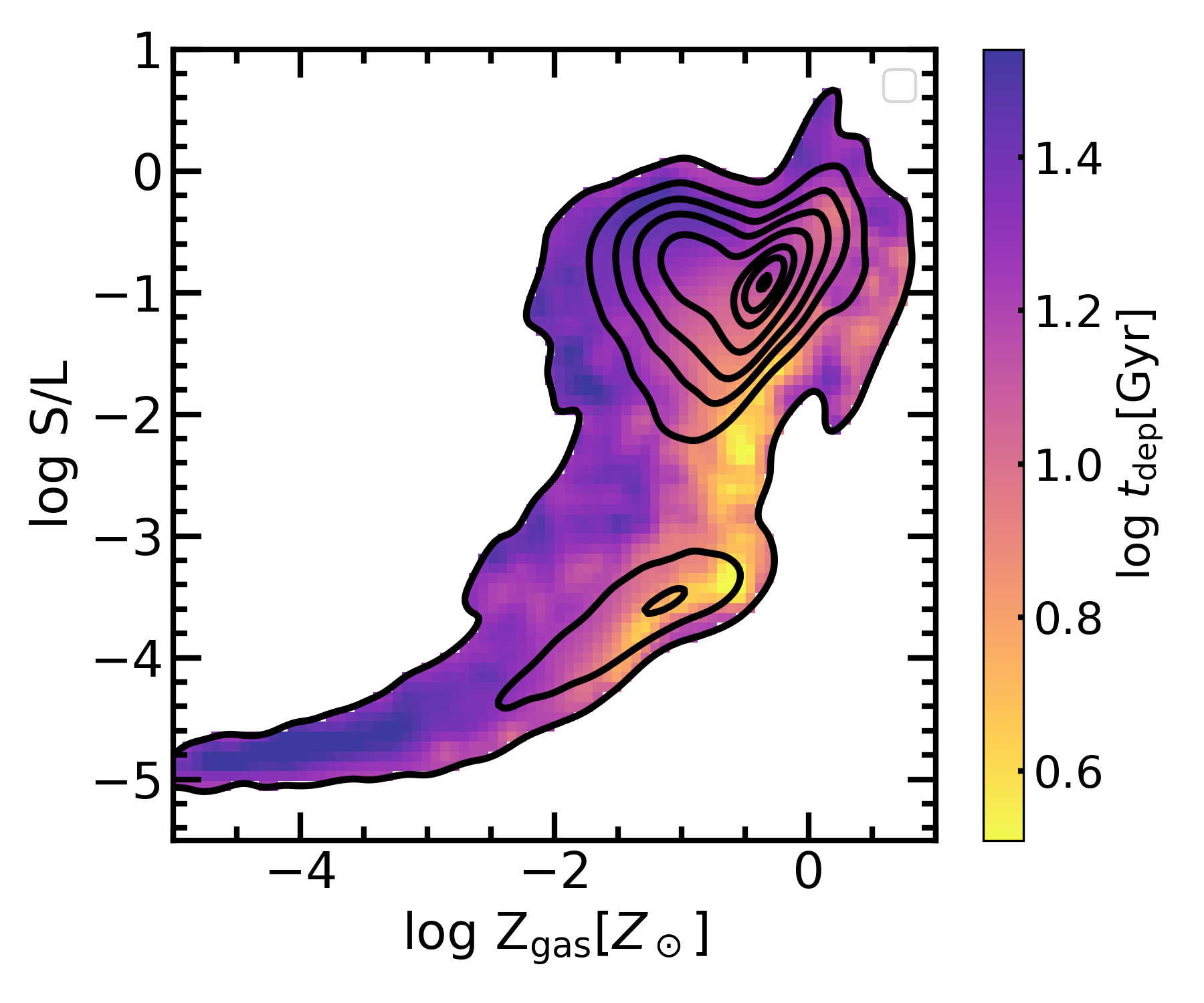}\quad
    \includegraphics[width=0.95\columnwidth]{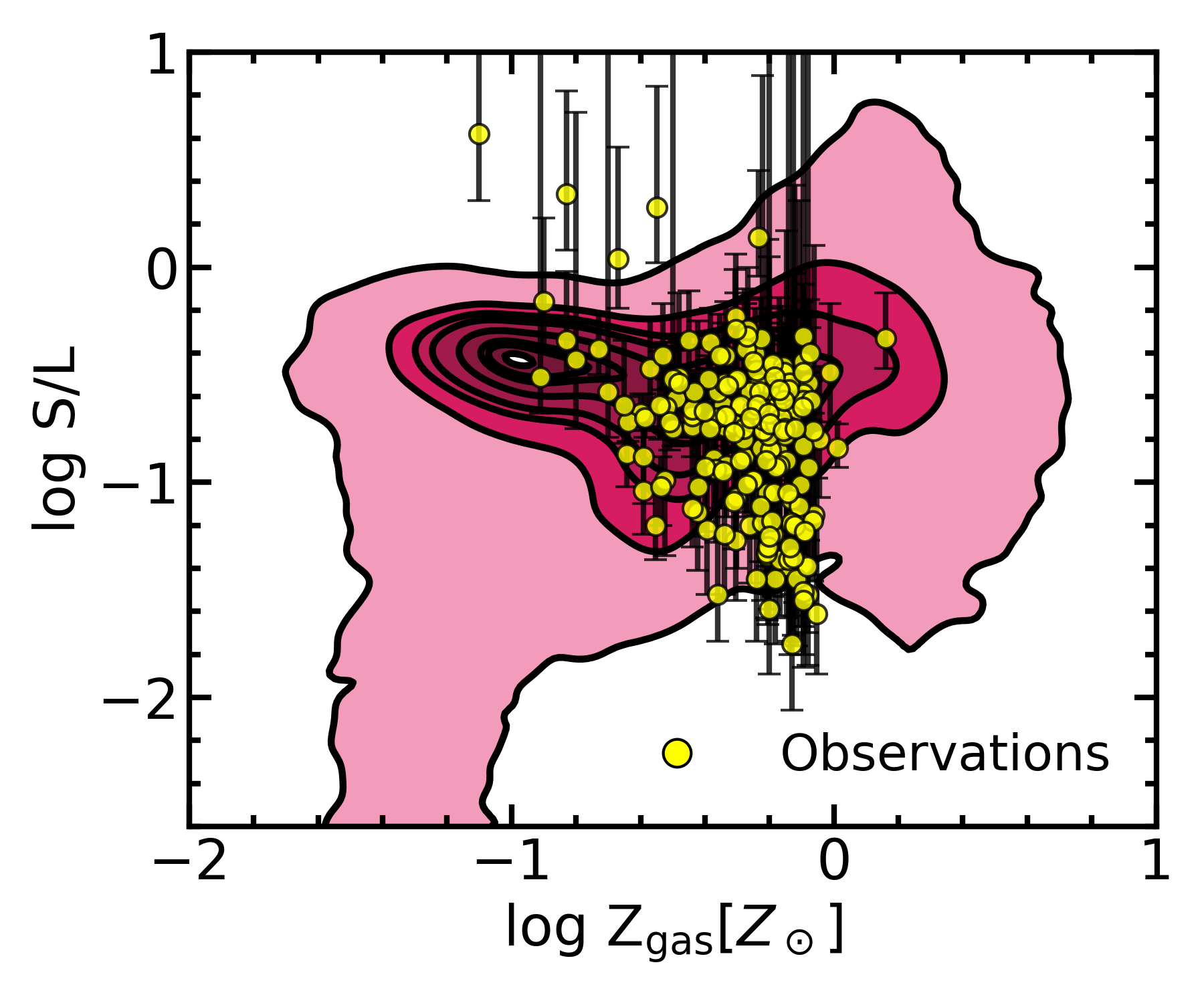}\quad

    \caption{\textbf{The small-to-large grain mass ratio grows with ISM metallicity.} We report the S/L mass ratio at all $z$ (left panel) and at $z=0$ (right panel) as a function of ISM gas metallicity. In the left panel, the color scale refers to the depletion time $t_{\rm dep}$ of galaxies and the black contours indicate the galaxies distribution. In the right panel, the distribution of model galaxies is shown as contours while observations refer to the results by \cite{Relano22}.} 
    \label{fig:SL:main}
\end{figure*}

In this section we discuss the GSD predicted by our model and the main properties of galaxies influencing it. We report the distribution $a^4 n(a) / n_{\rm H}$ -- where $n(a) = \partial n / \partial a$ -- in standard units of dust grains volume density per H atom\footnote{Although the GSD and dust mass are tracked for each ring into which the disk ISM is discretized, we present here ISM-integrated quantities, deferring the analysis of spatial properties to future work.}. This quantity is proportional to the mass of dust in each size bin. Figure \ref{fig:GSD:main} illustrates the GSD of our model galaxies -- both at various redshift (left panel) and $z=0$ (right panel), where also the MW mass galaxies\footnote{In this work we consider MW mass galaxies those with total stellar mass $10.4 \leq {\rm log} M_*/M_\odot \leq 10.6$ \citep[e.g.,][]{Liang25}. We have verified that imposing additional constraints to better select MW-like galaxies -- namely $0.5 \leq {\rm SFR}/(M_\odot\,{\rm yr}^{-1}) \leq 3$ and a stellar bulge-to-total mass ratio $\leq 0.3$ \citep{MWpaper} -- has a negligible impact on the GSD and extinction curves discussed in this work.} are reported.

The redshift evolution of the GSD shows a transition from a large-grain-dominated regime at high redshift, where stellar production is the primary dust source, to a flatter, \citetalias{MRN}-like distribution at low redshift. At $z \lesssim 2$, a nearly bimodal distribution emerges, with a secondary peak at $a \approx 0.005\,\mu\mathrm{m}$.
This bimodal behavior is similar to what is presented in the model developed by \citet{Hensley_astrodust}, which successfully reproduces the wavelength dependence of polarization extinction in galactic dust. A similar bimodality has also been reported in previous theoretical studies (e.g., \citealt{Aoyama20}), whose dust model, based on \citet{Hirashita19}, is conceptually similar to the one adopted here. \cite{Caleb_prep} also report a well-defined bimodal grain size distribution in MW–like galaxies, which they attribute to their simulations ability to resolve the multiphase ISM. This feature is commonly interpreted as the result of the combined action of shattering and accretion, with shattering producing small grains and accretion enhancing their abundance. This happens because the size dependence of the accretion timescale $\tau_{\rm acc} \propto a$ (Eq. \ref{eq:acc}) makes small grains the main sites of metal accretion, building up the small-grain tail of the GSD. Shattering, in turn, continuously replenishes this small-grain population from fragmentation of large grains, so that the two processes act together instead of simply shifting the GSD to larger sizes. This is explicitly demonstrated in Figure \ref{fig:acc_threshold}, where we show the effect of switching off accretion below a chosen grain size threshold. Preventing small grains to be accreted by gas phase metals reduces (but, owing to shattering, does not completely eliminate) the small-grain tail, and instead shifts the large-grain peak to larger sizes, as the available gas-phase metals are accreted onto the larger grains where growth can happen.

The GSD of the simulated MW–mass galaxies resembles that of the median $z=0$ full sample and is therefore also in good agreement with the \citetalias{MRN} slope. We note, however, that the overall spread in $z=0$ GSDs is wide and strongly correlates with the dust-to-metals (DTM) ratio of galaxies, with higher DTM values associated with flatter, small-grains–dominated distributions, as shown in Figure \ref{fig:GSD:main}. This clearly remark the dominant role of ISM accretion in enhancing the abundance of small grains and flattening the GSD, since it is the main process capable of converting gas-phase metals into dust, hence increasing the DTM. Finally, we also present the GSD of grains in the hot gas of galaxies, which in the semi-analytic framework broadly corresponds to the circumgalactic medium. This high temperature component is affected by sputtering (Sect. \ref{sec:dustspu}). Compared to the ISM, the resulting distribution is skewed toward larger grain sizes, as sputtering preferentially destroys small grains.

The strong link between the evolution of the GSD and the ISM accretion is further illustrated in Figure \ref{fig:SL:main}. The left panel shows the small-to-large (S/L) grain mass ratio as a function of ISM metallicity for a sample of simulated galaxies followed across their evolutionary histories, where the boundary between small and large grains is set at $a_{\rm S/L}=0.015\,\mu$m. The small-to-large ratio shows a clear evolution, moving from a stellar-dominated regime characterized by low and nearly constant S/L values to an ISM reprocessing–dominated regime with significantly higher S/L ratios -- of the order $\approx 0.1$ -- corresponding to flatter GSD curves. The transition takes place at a characteristic metallicity of $\approx 0.01-0.1\,Z_\odot$, corresponding to the point at which grain accretion in the ISM becomes the dominant process of dust mass growth. As the color code suggests, galaxies with lower depletion times $t_{\rm dep} = M_{\rm gas}/{\rm SFR}$ have larger transition metallicities. This is due to the fact that galaxies with shorter depletion times are able to convert their gas into stars faster, hence producing metals with a faster rate. 
In these cases, $Z_{\rm gas}$ grows faster than accretion can build up the small-grain population. As a result, the transition to an accretion-dominated regime is shifted toward higher metallicities.
This phenomenon is specular to what happens with the DTG$-Z_{\rm gas}$ relation, with shorter depletion times (or faster star formations) being associated with larger transition metallicites \citep[e.g.,][their Fig. 3]{Asano13acc}.

In the local Universe, the small-to-large grain ratio shows no clear trend with metallicity, taking values of $\sim 0.1$–$1$, as shown in the right panel of Figure \ref{fig:SL:main}. We compare our results with the observational estimates of \citet{Relano22} for a sample of nearby galaxies. In that study, the S/L grain mass ratio is inferred from SED fitting using the dust model of \citet{Desert90}, which assumes three grain populations -- PAHs, very small grains, and large silicate grains -- with a transition radius of $0.015\,\mu\mathrm{m}$ between very small and large grains.

For the purpose of this comparison, we adopt the same transition radius between small and large grains. Nevertheless, we caution that while our S/L ratio is a direct estimate of the two grain mass components, observational values derived from SED fitting are model-dependent and sensitive to the interaction of different grain populations with radiation, as well as to the star–dust geometry, which is not accounted for here. With this caveat in mind, we find reasonable agreement between simulations and observations, both in terms of normalization and scatter ($\approx 1\,\mathrm{dex}$). As discussed above, this scatter primarily reflects the dispersion in the DTM ratio, with the additional contribution of coagulation further broadening the S/L distribution at fixed DTM. A brief discussion of the S/L–DTM relation and its dependence on coagulation is provided in Appendix \ref{app:SL-DTM}.

\subsection{Chemical composition of dust}

\begin{figure}[]

    \centering
    \includegraphics[width=0.9\columnwidth]{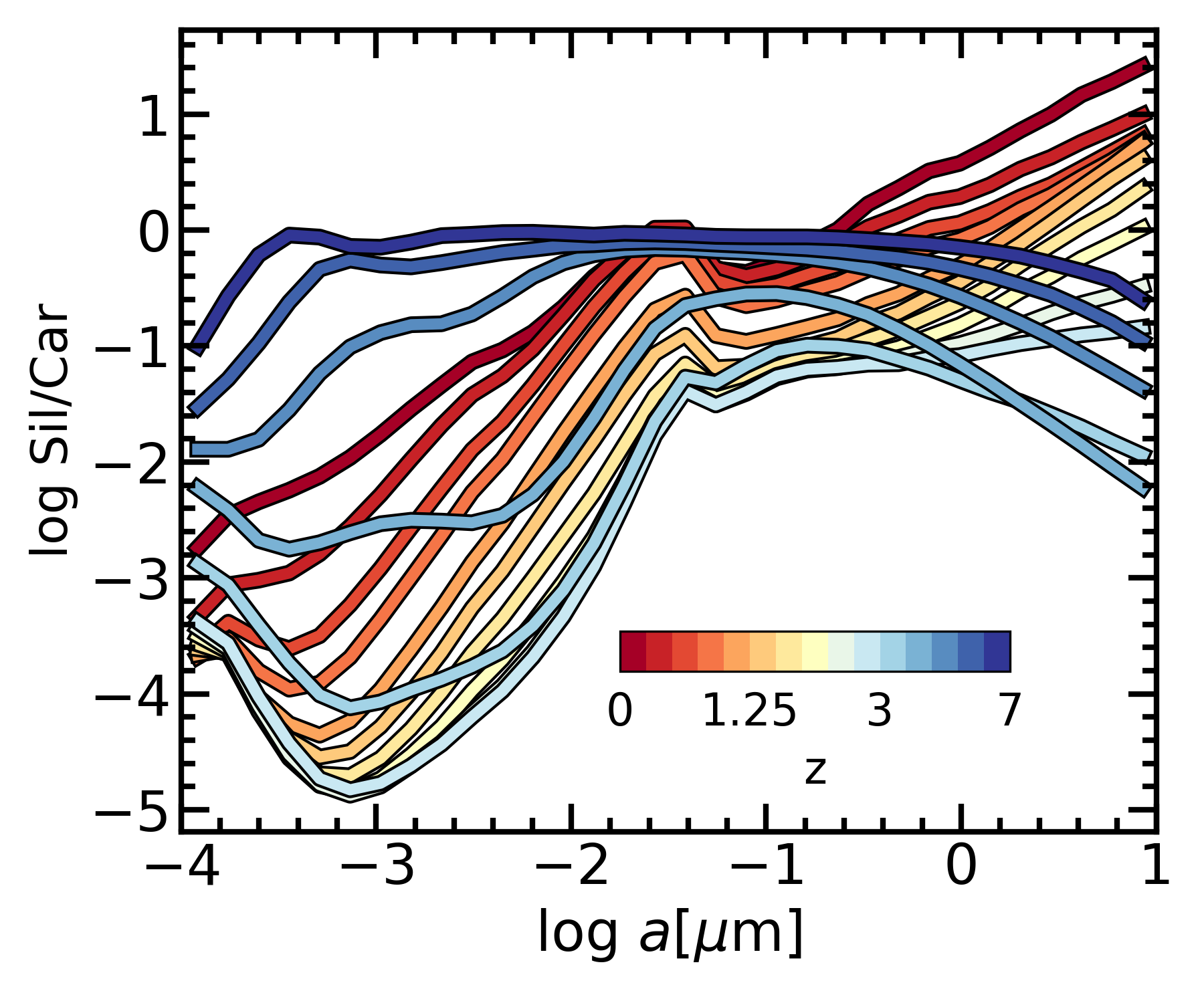}

    \caption{\textbf{Cosmic evolution of grain composition: carbonaceous grains at small sizes, silicates at large.} 
    Evolution of the silicate-to-carbonaceous mass ratio with redshift, as a function of grain size for our full sample of simulated galaxies. Lines indicate the ratio derived from the median size distributions of silicate-only and carbonaceous-only grains, with redshift color-coded.} 
    \label{fig:GSD:SilCar}
\end{figure}

Our model traces the evolution of the GSD for two grain species, carbonaceous (C) and silicate (MgFeSiO$_4$). The predicted redshift evolution of the silicate-to-carbonaceous (Sil/Car) mass ratio as a function of grain radius is shown in Fig. \ref{fig:GSD:SilCar}.
Initially, at $z \approx 7$, the distribution is nearly flat, with $\mathrm{Sil/Car} \approx 0.1-1$, as a result of the dominance of stellar dust production. As time evolves, carbonaceous grains come to dominate the mass budget of small grains $(\lesssim 0.01 \, \mum)$, due to shattering, which is more efficient for carbon grains (their lower relative velocities are more easily attained; see Sect. \ref{sec:shacoa:vel}). As the time goes on, dust production from AGB stars becomes increasingly important, further enhancing the carbonaceous mass. However, silicate grains accrete more mass in the ISM owing to the higher abundance of silicate elements, and they also grow to larger sizes. As a result, the $\mathrm{Sil/Car}$ ratio steepens toward $z \approx 0$, with carbonaceous grains dominating the small-size end and silicate grains prevailing at larger radii. This evolution is briefly discussed in terms of extinction curves in Sect. \ref{sec:res:ext}.

\subsection{Evolution of GSD in representative galaxies}


\begin{figure*}[!htb]

    \centering
    \includegraphics[width=0.65\columnwidth]{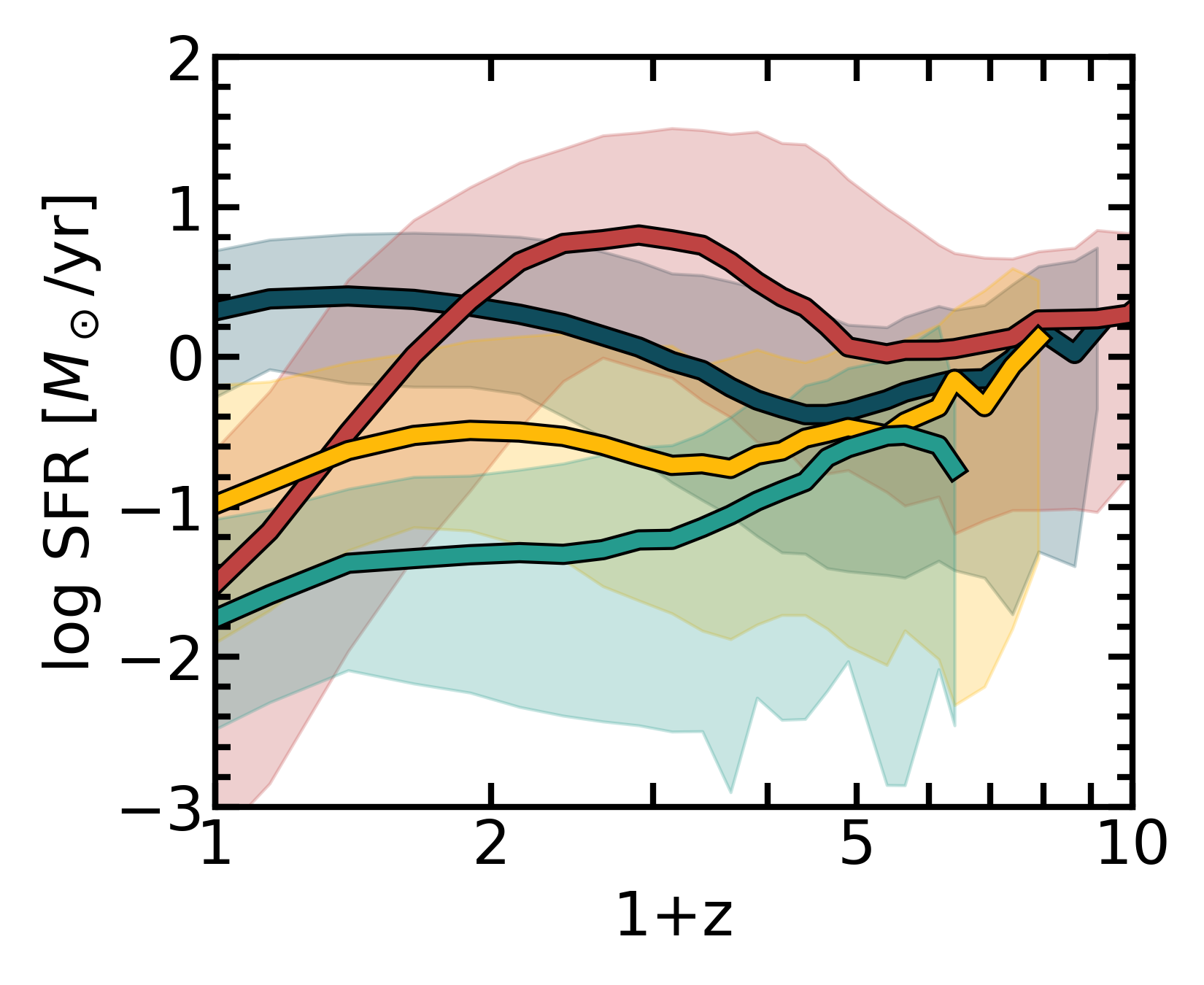}\quad
    \includegraphics[width=0.65\columnwidth]{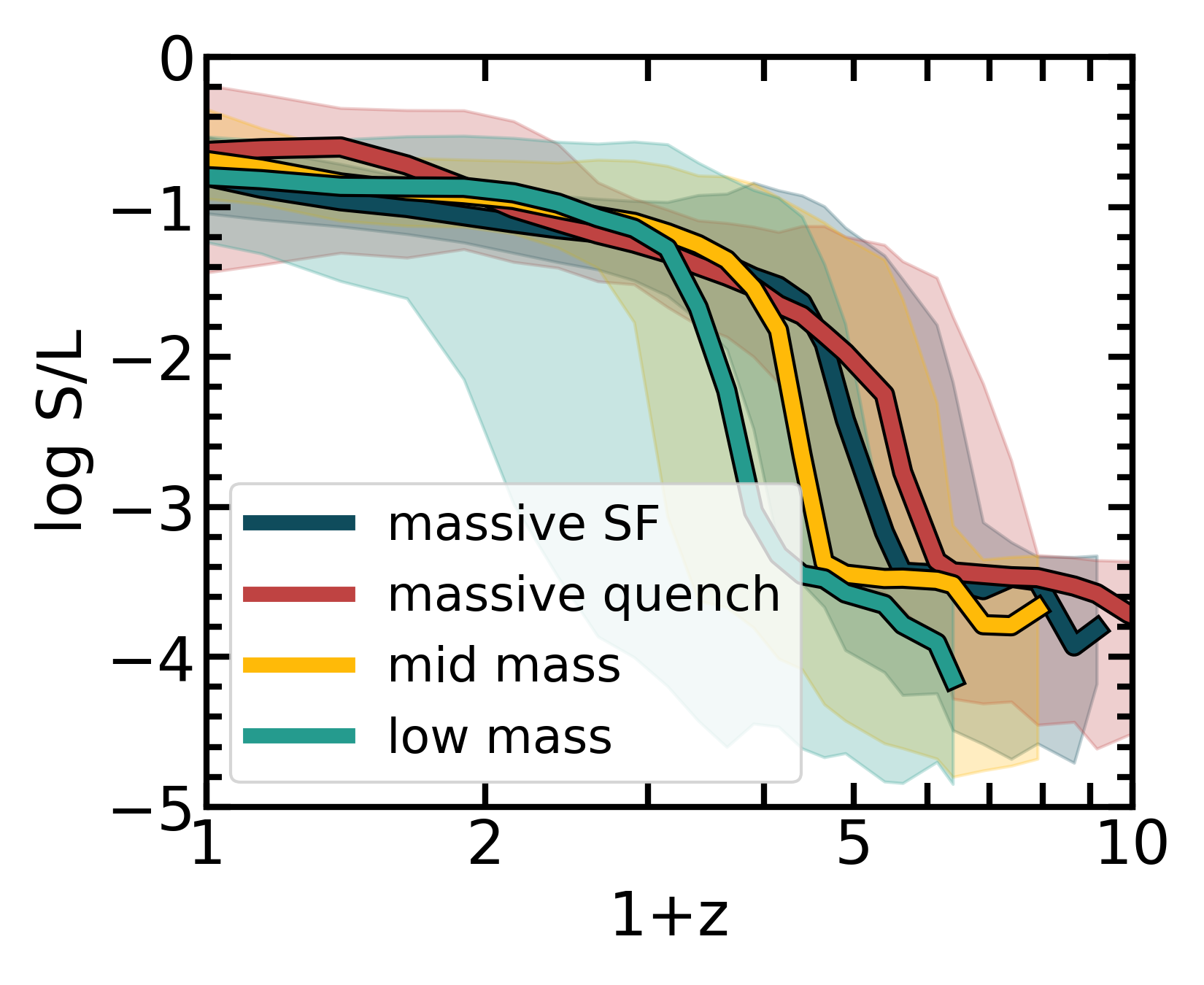}\quad
    \includegraphics[width=0.65\columnwidth]{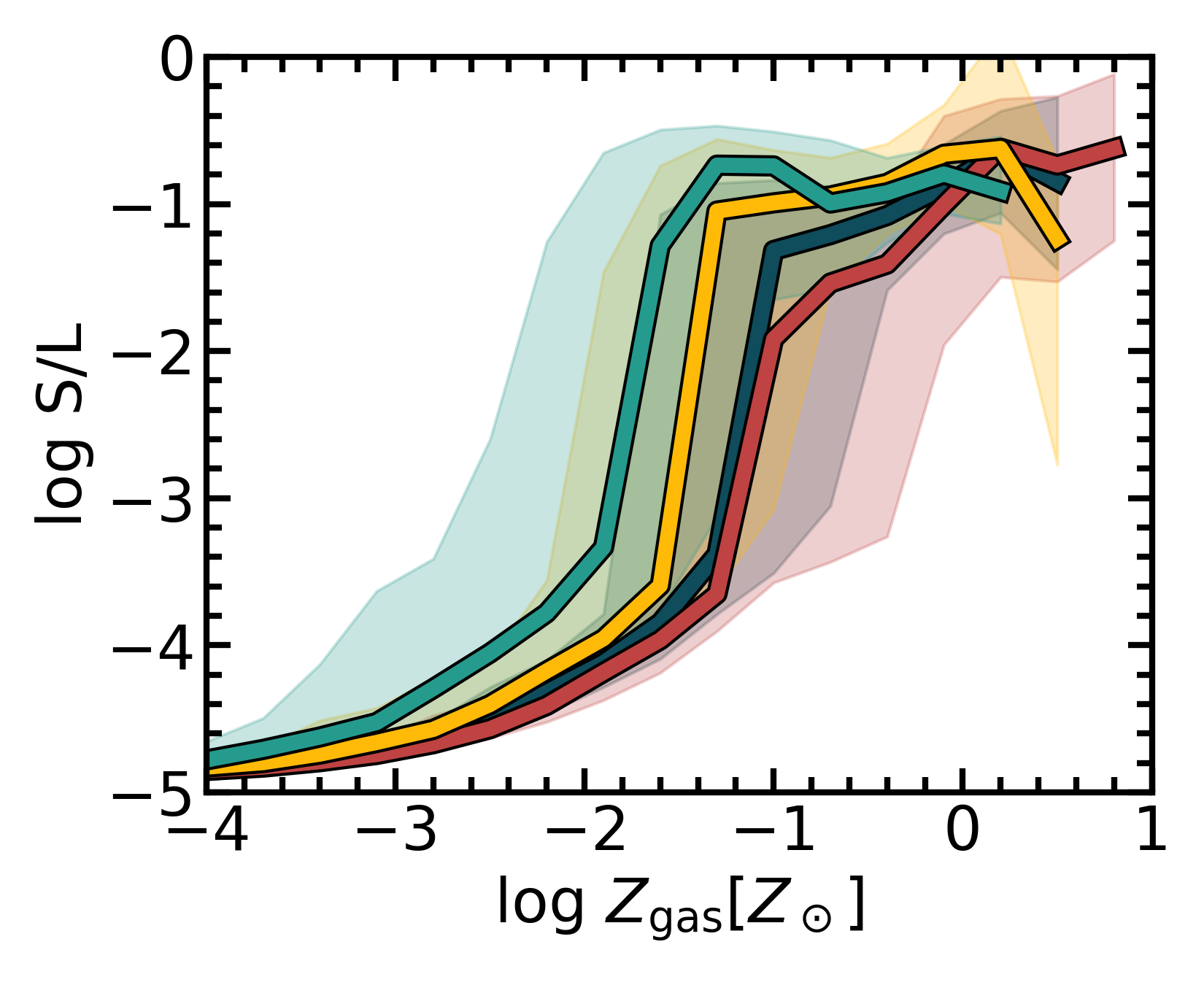}\quad

    \caption{\textbf{The small-to-large ratio evolution is similar in galaxies with different star formation histories.} Redshift evolution of the SFR (left panel), the S/L ratio (middle panel), and evolutionary tracks in the S/L–$Z_{\rm gas}$ plane (right panel) for three samples of simulated galaxies: $z=0$ massive ($M_{\rm stars} > 10^{10}\,M_\odot$) galaxies, subdivided into star-forming (${\rm sSFR} > 3 \times 10^{-11}\,\mathrm{yr}^{-1}$; blue) and quenched (${\rm sSFR} < 3 \times 10^{-12}\,\mathrm{yr}^{-1}$, red) systems, as well as intermediate-mass ($10^{9} < M_{\rm stars}/M_\odot < 10^{10}$, yellow) and low-mass ($10^{8} < M_{\rm stars}/M_\odot < 10^{9}$, green) $z=0$ galaxies. Solid lines and shaded regions denote the median values and the $16–84$th percentile ranges of the distributions.}
    \label{fig:evo}
\end{figure*}
We now study how the GSD evolves in representative samples of simulated galaxies, selected according to their $z=0$ properties. These are: \textit{(i)} massive ($M_{\rm stars} > 10^{10}\,M_\odot$) star-forming (${\rm sSFR} > 3 \times 10^{-11}\,\mathrm{yr}^{-1}$) galaxies; \textit{(ii)} massive ($M_{\rm stars} > 10^{10}\,M_\odot$) quenched (${\rm sSFR} < 3 \times 10^{-12}\,\mathrm{yr}^{-1}$) galaxies; \textit{(iii)} intermediate-mass galaxies ($10^{9} < M_{\rm stars}/M_\odot < 10^{10}$); and \textit{(iv)} low-mass galaxies ($10^{8} < M_{\rm stars}/M_\odot < 10^{9}$). The evolution of the SFR, the S/L ratio, and their tracks in the S/L$-Z_{\rm gas}$ diagram are shown in Fig.~\ref{fig:evo}.

We first note that, despite their very different star formation histories\footnote{The rising SFR of the massive star-forming sample toward low redshift reflects a selection effect: galaxies selected for high present-day sSFR preferentially assemble their mass later, resulting in more recent star formation.}, the four galaxy samples show only modest differences in their $z=0$ S/L ratios. What differs instead is the epoch at which the GSD transitions from a large-grain-dominated to a small-grain-dominated regime. This transition occurs at higher redshift in more massive systems, which reach high ISM metallicities earlier and are therefore able to efficiently activate accretion, boosting the abundance of small grains. However, these systems also exhibit higher \textit{critical metallicities}, as is evident from their evolution in the S/L$-Z_{\rm gas}$ diagram. As discussed previously, this is due to their typically shorter depletion times.

Finally, we note that, because of the slow but continuous increase in the abundance of small grains over cosmic time, highly star-forming galaxies at high redshift -- the progenitors of local quenched galaxies -- are characterized by a GSD biased toward larger grains compared to local star-forming systems.

\section{Extinction curves}
\label{sec:res:ext}

\begin{figure*}[!htb]

    \centering
    \includegraphics[width=0.9\columnwidth]{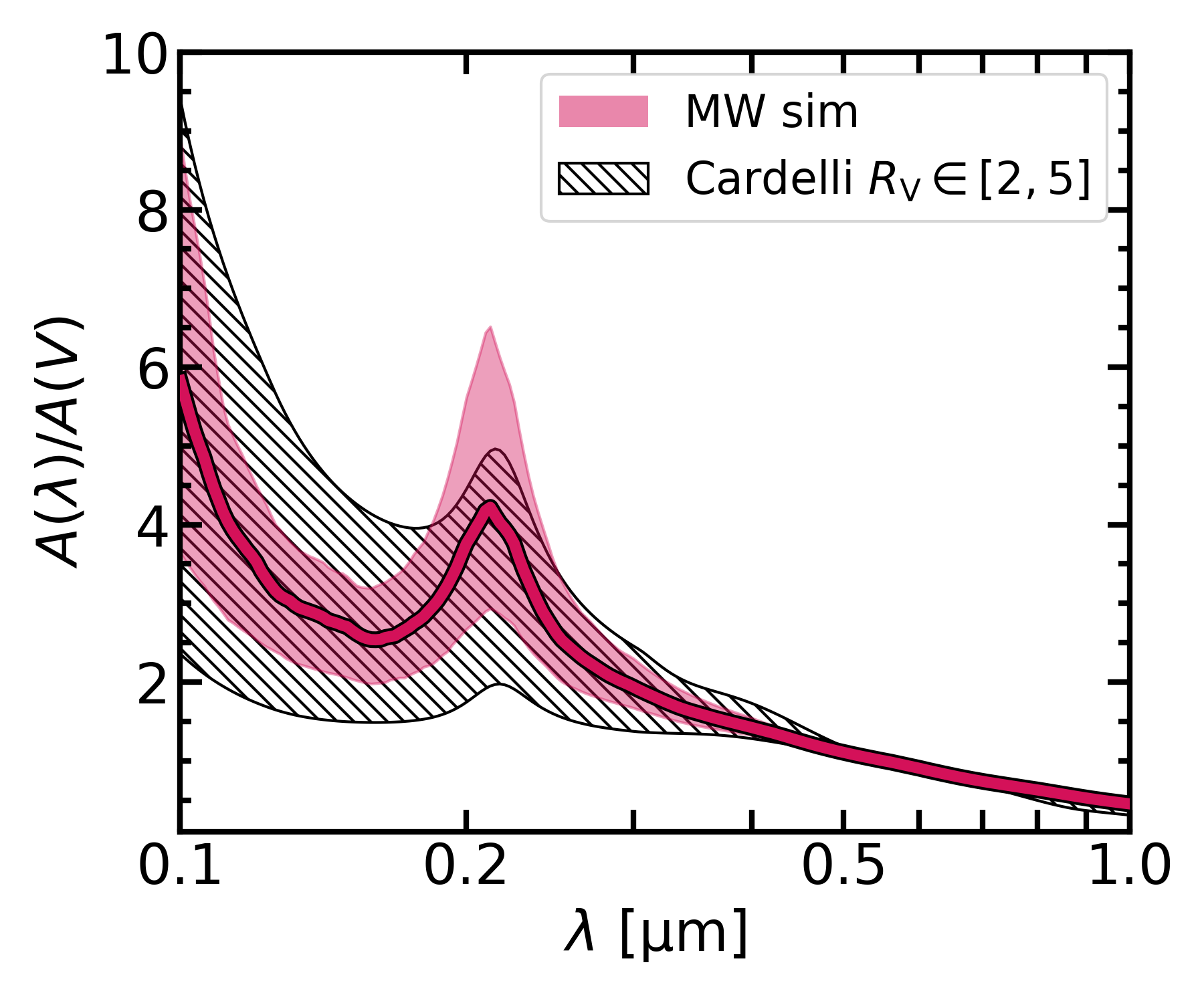}\quad
    \includegraphics[width=0.9\columnwidth]{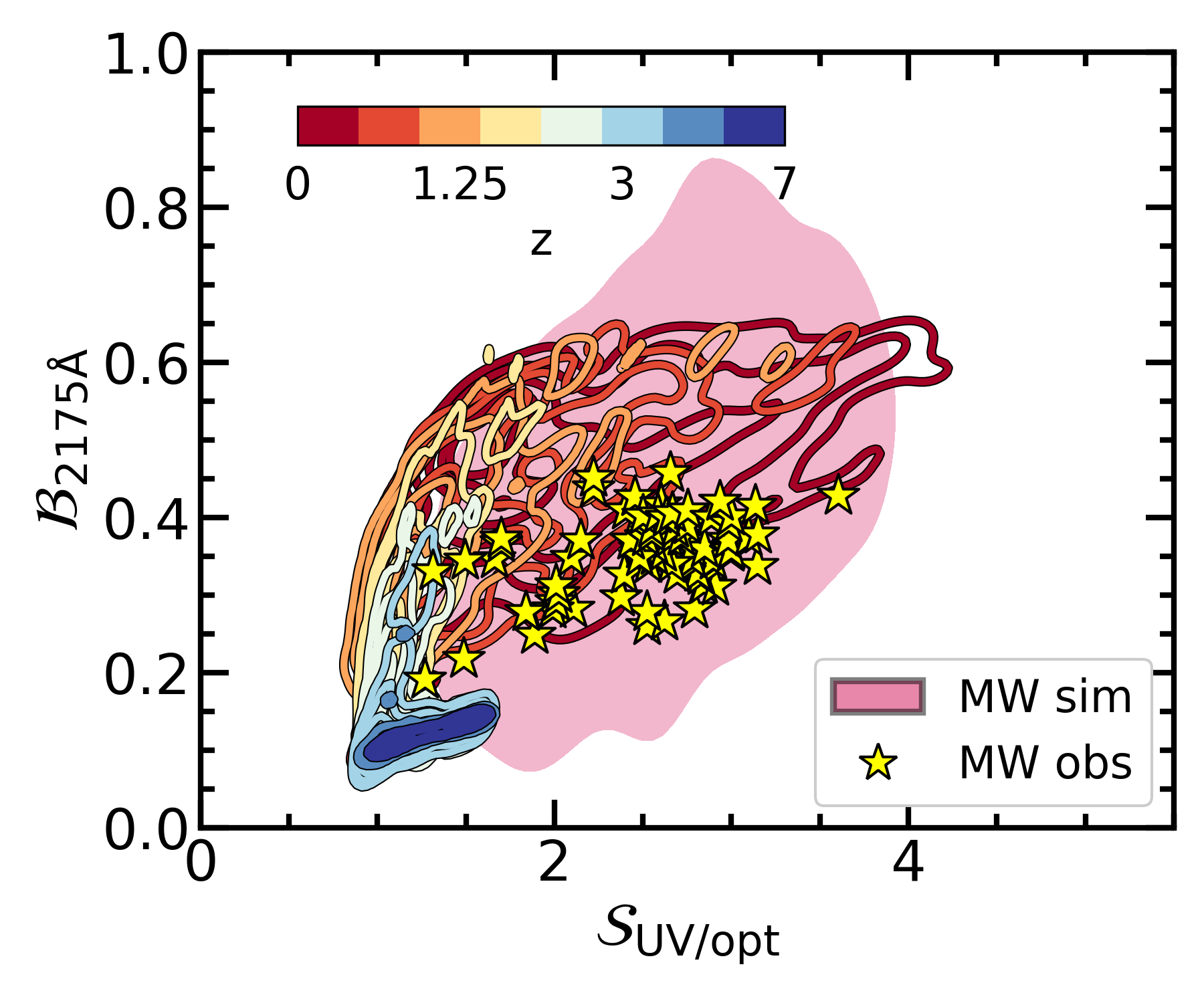}\quad

    \caption{\textbf{The $2175\,\text{\AA}$ bump and UV/optical slope increase with decreasing redshift as small grains grow, matching the MW extinction curve at $z=0$.} Left panel: extinction curves normalized to the $V$-band value for our simulated $z=0$ MW–mass galaxies (showing the median and the $16$–$84$th percentile range). The hatched region indicates the Cardelli extinction law for $R_{\rm V}$ values in the interval $[2,5]$. Right panel: UV/optical slope versus $2175\,\text{\AA}$ bump diagram for our simulated galaxies over $0<z\lesssim 7$. The area corresponding to $z=0$ MW–mass galaxies is shown in red, while star symbols indicate MW data compiled by \cite{Li2021}, which include observations from \cite{FitzMassa90, Fitz99, FitzpatrickMassa07, FitzpatrickMassa09, Nataf16}.} 
    \label{fig:ext}
\end{figure*}

In this Section, we present the model predictions for extinction curves, obtained exploiting our chemical evolution and grain-size modelling.

We compute the extinction properties of our simulated galaxies using the GSD of carbonaceous (graphite) and silicate grains, adopting the absorption and scattering coefficients from \cite{Draine84} and \cite{Laor93}, as described in Appendix D of \cite{McKinnon18} and Appendix A of \cite{Li2021}. We treat our model galaxies as if they were probed along a single sight line, thereby neglecting their internal structure, which is oversimplified in our SAM. Also, we focus only on the shape of the extinction curve -- i.e. normalized to $A_{\rm V}$ -- to avoid specifying the column density\footnote{Indeed, the column density terms cancels out when taking the ratio, as shown in Appendix A of \cite{Li2021}.} required to compute the $V$-band extinction, which is not straightforward to define in this context.\\
The results are shown in Figure \ref{fig:ext}. The extinction curve of our simulated MW-mass galaxies (left panel) agrees remarkably well with the \cite{Cardelli89} law over the range $2 \leq R_{\rm V} \leq 5$, where $R_{\rm V} = 1 / (A_B/A_V - 1)$ and $A_i$ is expressed in magnitudes in the $i$ band. This agreement is not unexpected given the previously discussed GSD and its similarity to the \citetalias{MRN} distribution. Moreover, the presence of the $2175\,\text{\AA}$ bump indicates that the abundance of (small) carbonaceous grains in our model MW-mass galaxies is sufficient to reproduce this feature at a level comparable to observations. In future work, we will also investigate the mid-infrared emission produced by PAHs, which provides complementary insights into the abundance of small carbonaceous grains. \\

A more comprehensive analysis of extinction curves and their evolution is presented in the right panel of Figure \ref{fig:ext}, where we show the $2175\,\text{\AA}$ bump ($\mathcal{B}_{2175 \,\text{\AA}}$) as a function of the UV/optical slope ($\mathcal{S}_{\rm UV/opt}$). 
These parameters provide a compact characterization of the main features of extinction curves \citep{SalimNara20} and are defined as $\mathcal{S}_{\rm UV/opt} = A_{1500 \,\text{\AA}}/A_V$ and ${\mathcal{B}}_{2175 \,\text{\AA}} = \frac{A_{2175 \,\text{\AA}}-A_{2175, \, 0}}{A_{2175, \, 0}}$, where $A_{2175, \, 0}$ represents the extinction at $2175 \,\text{\AA}$ in the absence of the bump, estimated via a simple interpolation as $A_{2175,\, 0} = 0.33 A_{1500\,\text{\AA}} + 0.67 A_{3000 \,\text{\AA}}$.

For MW-mass galaxies (red region), we find ${\mathcal{S}}_{\rm UV/opt} \approx 1-4$ and ${\mathcal{B}}_{2175\,\text{\AA}} \approx 0.2-0.8$, confirming consistency with observed MW sightlines \citep[e.g.,][]{Cardelli89, FitzpatrickMassa07}. Examining the redshift evolution over $0 \leq z \lesssim 7$ for the full simulated sample reveals a clear increase with time in both quantities. In particular, galaxies at high redshift tend to display flatter extinction curves and weaker $2175\,\text{\AA}$ bumps, reflecting the early dominance of large grains produced by stellar sources. As redshift decreases, shattering and ISM accretion become progressively more important, enhancing the abundance of small grains and consequently steepening the extinction curve. Notably, in our model small carbon grains are more abundant than small silicate grains because the threshold velocity required for shattering is lower by a factor of $\approx 2$ (Sect. \ref{sec:shacoa:vel}). As a result, the carbonaceous-driven bump naturally strengthens as shattering and accretion become more efficient.

Since in our model the bump strength and extinction steepness grow together, it is challenging to identify a physical mechanism within this framework capable of producing the steep, bump-less extinction curves characteristic of the Small Magellanic Cloud (SMC). This difficulty has been already discussed in previous studies \citep[e.g.,][]{Nozawa15, Hou16, Hirashita19}, which have suggested that amorphous carbon rather than graphite may be required.

\section{The role of dust processes}
\label{sec:res:dustproc}

\begin{table*}[t]
\centering
\setlength{\tabcolsep}{4pt} 
\begin{tabular}{cccccccc}
\hline
\hline
Dust Process & Parameter & Value & Description 
& \multicolumn{3}{c}{$\langle |\Delta_\sigma| \rangle$ from FID} 
& Figure \\
\cline{5-7}
 & & & & {\rm \quad DTG--$Z$} & {\qquad GSD} & {\qquad ext$_{\rm MW}$} & \\
\hline
\hline

\multirow{1}{6em}[0em]{\centering Stellar \\ production}
& $a_{\rm peak}^{\rm SF}$ 
& \mycol{$1 \, \mu$m \\ $0.01 \, \mu$m \\ $0.001 \, \mu$m}
& GSD produced by stars 
& \mycol{$0.09$ \\ $0.07$ \\ $0.17$} 
& \mycol{$0.42$ \\ $0.42$ \\ $0.56$} 
& \mycol{$0.13$ \\ $0.12$ \\ $0.13$} 
& \ref{fig:apeak} \\

\hline

\multirow{7}{6em}[-2em]{\centering Shattering\\and\\Coagulation} 
& sha \textsc{OFF}  
& $-$ 
& no shattering 
& \mycol{$0.58$} & \mycol{\textbf{3.5}} & \mycol{\textbf{3.2}}
& \ref{fig:oneproc:main} \\

\cline{2-8}

& $n_{\rm diff}$ 
& \mycol{0.1 cm$^{-3}$ \\ 1 cm$^{-3}$} 
& diffuse ISM density
& \mycol{0.22 \\ 0.06} & \mycol{0.69 \\ 0.25} & \mycol{0.60 \\ 0.44} 
& \ref{fig:shacoa:nT} \\

\cline{2-8}

& $v_{\rm sh, \, th}$ 
& \mycol{$\times 3$ \\ $\times 0.3$} 
& threshold velocity for shattering 
& \mycol{0.40 \\ 0.20} & \mycol{0.87 \\ 0.40} & \mycol{\textbf{1.28} \\ 0.17} 
& \ref{fig:shacoa:vrel} \\

\cline{2-8}

& coa \textsc{OFF}  
& $-$ 
& no coagulation 
& \mycol{0.24} & \mycol{0.49} & \mycol{0.80}
& \ref{fig:oneproc:main} \\

\cline{2-8}

& $n_{\rm dense}$ 
& \mycol{10$^2$ cm$^{-3}$ \\ 10$^4$ cm$^{-3}$} 
& dense ISM density 
& \mycol{0.02 \\ 0.03} & \mycol{0.06 \\ 0.07} & \mycol{0.16 \\ 0.22} 
& \ref{fig:shacoa:nT} \\

\cline{2-8}

& $v_{\rm coa, \, th}$ 
& no
& max velocity for coagulation 
& \mycol{0.05} & \mycol{0.88} & \mycol{0.25}
& \ref{fig:shacoa:vrel} \\

\cline{2-8}

& $v_{\rm gr}$ 
& $a_{\rm fix}=0.03 \, \mum$ 
& size-independent velocities 
& \mycol{0.59} & \mycol{\textbf{1.52}} & \mycol{\textbf{2.3}}
& \ref{fig:shacoa:vrel} \\

\hline

\multirow{2}{4em}{Accretion}  
& \textsc{OFF}  
& - 
& no accretion 
& \mycol{\textbf{2.2}} & \mycol{\textbf{3.22}} & \mycol{\textbf{1.2}}
& \ref{fig:oneproc:main} \\

\cline{2-8}

& $\tau_{\rm acc}$ 
& $a_{\rm fix}=0.03\, \mu{\rm m}$
& size-independent accretion 
& \mycol{0.54} & \mycol{\textbf{1.0}} & \mycol{0.37}
& \ref{fig:accSN:nosize} \\

\hline

\multirow{2}{4em}[0pt]{\centering SN \\ destruction} 
& \textsc{OFF}  
& - 
& no SN destruction 
& \mycol{0.67} & \mycol{0.89} & \mycol{\textbf{1.2}}
& \ref{fig:oneproc:main} \\

\cline{2-8}

& $\tau_{\rm SN}$
& $a_{\rm fix}=0.03\, \mu{\rm m}$ 
& size-independent destruction 
& \mycol{0.19} & \mycol{\textbf{1.2}} & \mycol{\textbf{2.0}}
& \ref{fig:accSN:nosize} \\

\hline
\end{tabular}

\caption{\small{\textbf{Suite of numerical experiments performed.}
\rm{The runs are designed to study the impact of varying parameters and parameterizations for individual dust processes. For each run, we list the dust process, the parameter, its value (when applicable), and a brief description. We also report the deviation of each run from the fiducial model at $z=0$ for the DTG$-Z$ relation, the GSD of the full sample, and the MW extinction curve, quantified using the metric defined in Eq. \ref{eq:deviation}. Values with $\langle |\Delta_\sigma| \rangle \geq 1$, indicating significant deviation, are highlighted in bold. The table also provides the figure reference where the results of each run are shown.}
}}
\label{tab:simset}
\end{table*}


\begin{figure*}[!htb]

    \centering
    \includegraphics[width=0.8\columnwidth]{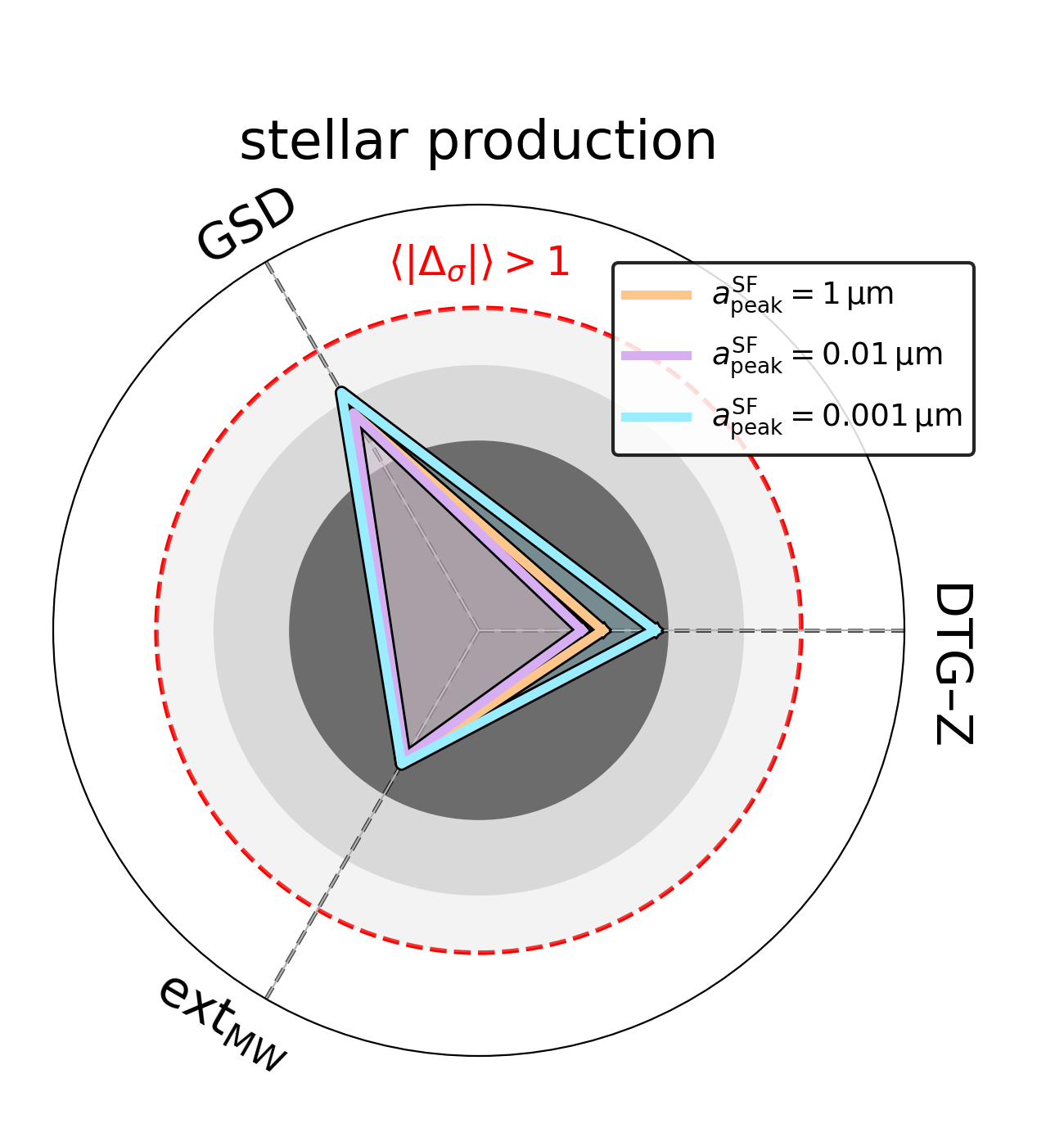}\qquad \qquad \includegraphics[width=0.8\columnwidth]{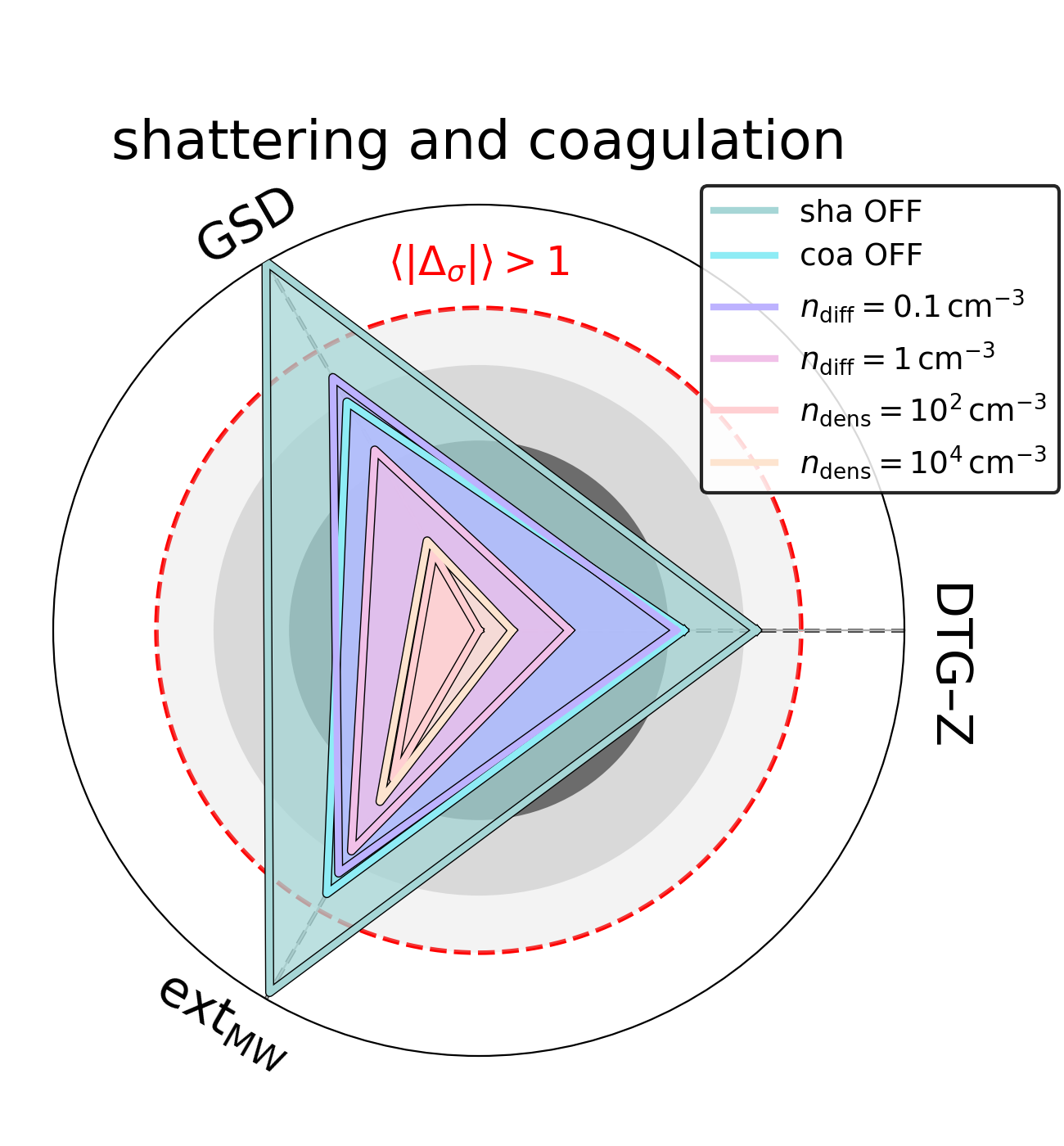}\\\vspace{5mm}
    \includegraphics[width=0.8\columnwidth]{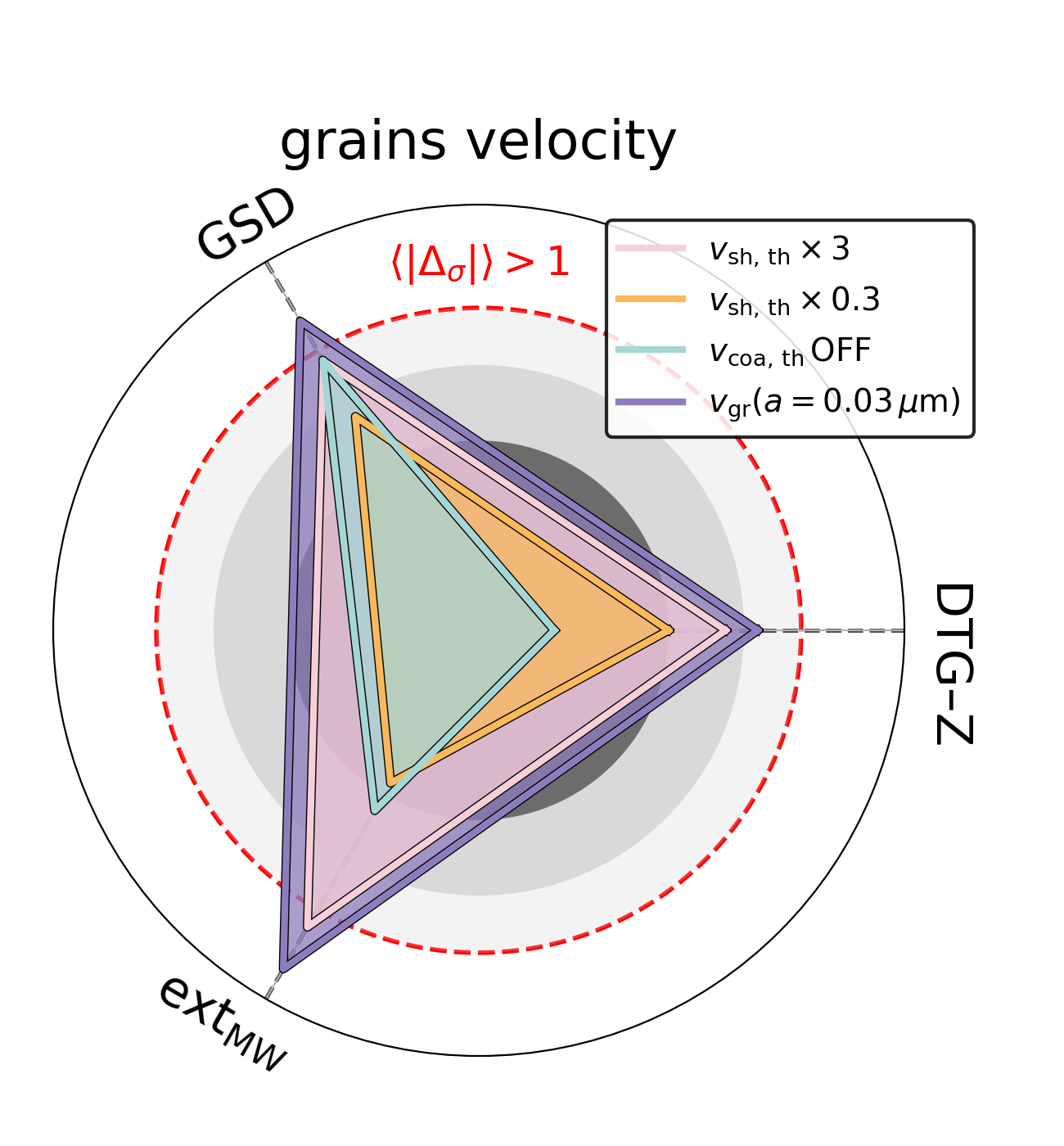}\qquad \qquad
    \includegraphics[width=0.8\columnwidth]{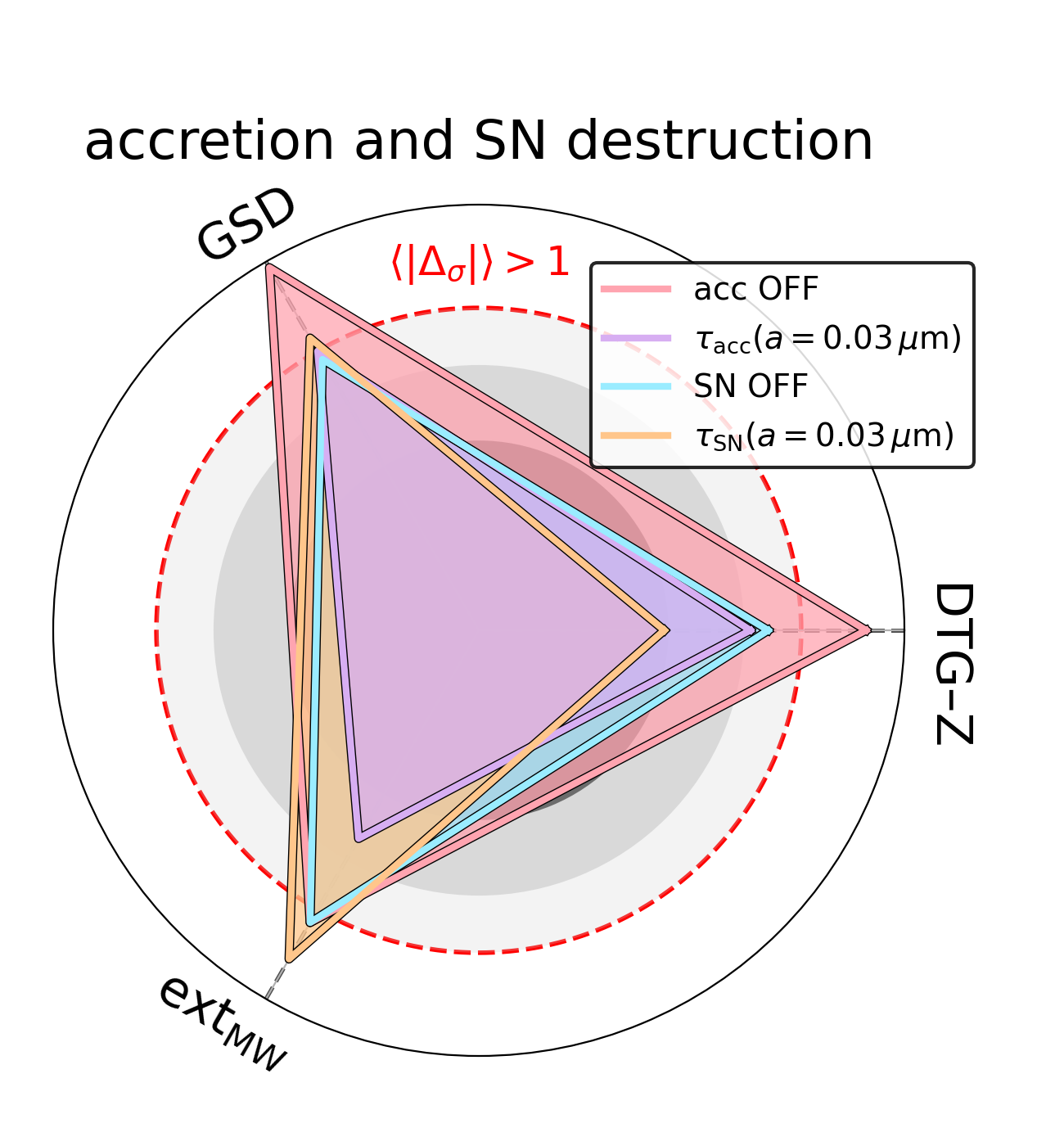}

    \caption{\textbf{Radar plots illustrating the level of agreement between each experiment and the fiducial model, where larger triangles represent more disagreement.} Each plot shows the $\langle |\Delta_\sigma| \rangle$ values (Eq. \ref{eq:deviation}) obtained by comparing the runs listed in Tab. \ref{tab:simset} with the fiducial model, for the DTG$-Z$ relation, GSD, and MW-mass extinction curves. The values along the three radial axes are shown on a logarithmic scale for clarity. The circles, shown in increasingly lighter shades of gray, correspond to $\langle |\Delta_\sigma| \rangle = 0.2$, $0.5$, and $1$, respectively. The value $1$ is used here to denote a substantial deviation and is highlighted with a dotted line.} 
    \label{fig:radar}
\end{figure*}

In this section, we qualitatively evaluate the relative importance of the physical processes that regulate some key predictions of our dust model. To this end, we perform a suite of numerical experiments -- some of which are intentionally simplified or not physically motivated -- to explicitly illustrate how commonly adopted assumptions and size-dependent parameters influence the global outcomes of the model.

While previous analytical studies have extensively investigated these processes and their parameterizations \citep[e.g.,][]{Hirashita11, HirashitaKoba13, Asano13}, here we explore their impact for the first time within a fully cosmological galaxy evolution framework. This is made possible by the relatively low computational cost of our semi-analytic approach. Our analysis is focused on three model predictions at $z=0$: the dust abundance (quantified through the DTG–$Z$ relation), the GSD, and the resulting extinction properties of MW–mass galaxies. Although differences at higher redshift naturally arise -- and are discussed where relevant -- we deliberately focus on local predictions, which are more tightly constrained by observations and best suited to evaluating the cumulative effects of these processes in an evolved galaxy population.

A summary of all simulation runs performed in this analysis is provided in Tab. \ref{tab:simset}.
For each model (subscript $\mathrm{mod}$), we quantify its deviation from the fiducial run (subscript $\mathrm{FID}$) for each of the three aforementioned relations by means of the metric
\begin{equation}
\label{eq:deviation}
\Delta_\sigma(x) =
\frac{y_{\mathrm{mod}}(x) - y_{\mathrm{FID}}(x)}
{\sqrt{\sigma_{\mathrm{mod}}^2(x) + \sigma_{\mathrm{FID}}^2(x)}},
\end{equation}
where $y(x)$ is the median relation predicted by a given simulation, and
\begin{equation}
\sigma(x) = \frac{y_{\mathrm{up}}(x) - y_{\mathrm{down}}(x)}{2}
\end{equation}
is the effective dispersion estimated from the upper and lower percentiles. In practice, the median relation and the corresponding percentiles of $y$ as a function of $x$ are the quantities shown in the figures.

As a global summary statistic, we adopt the mean absolute dispersion-normalized deviation,
$\langle |\Delta_\sigma| \rangle$.
This metric quantifies the typical separation between the median predictions of two models relative to their combined dispersions, while retaining sensitivity\footnote{Such sensitivity is not expected when using the median absolute deviation, which is intrinsically less affected by localized outliers.} to both extended and localized regions of disagreement. Values of $\langle |\Delta_\sigma| \rangle \lesssim 1$ indicate that the two models are statistically consistent within their uncertainties, whereas values $\langle |\Delta_\sigma| \rangle \gtrsim 1$ reflect increasing tension. The corresponding values of $\langle |\Delta_\sigma| \rangle$ for each simulation run are reported in Tab. \ref{tab:simset}.\\

Before discussing each set of runs in detail, we summarize their deviation from the fiducial model using radar plots in Figure \ref{fig:radar}, based on the metric described above. From this initial qualitative comparison, it is immediately clear that ISM accretion is the dominant process affecting the DTG$-Z_{\rm gas}$ relation (and thus the dust abundance). Together with shattering, it also has the largest influence on the GSD. However, both the GSD and the extinction curves of MW-mass galaxies are sensitive to grain destruction by SNe and assumptions about shattering and coagulation, such as grain velocities and gas-phase densities. The assumption on the typical size of stellar produced grains, in contrast, is always almost negligible.

\subsection{One process at time}


\begin{figure*}[!htb]

    \centering
    \includegraphics[width=0.67\columnwidth]{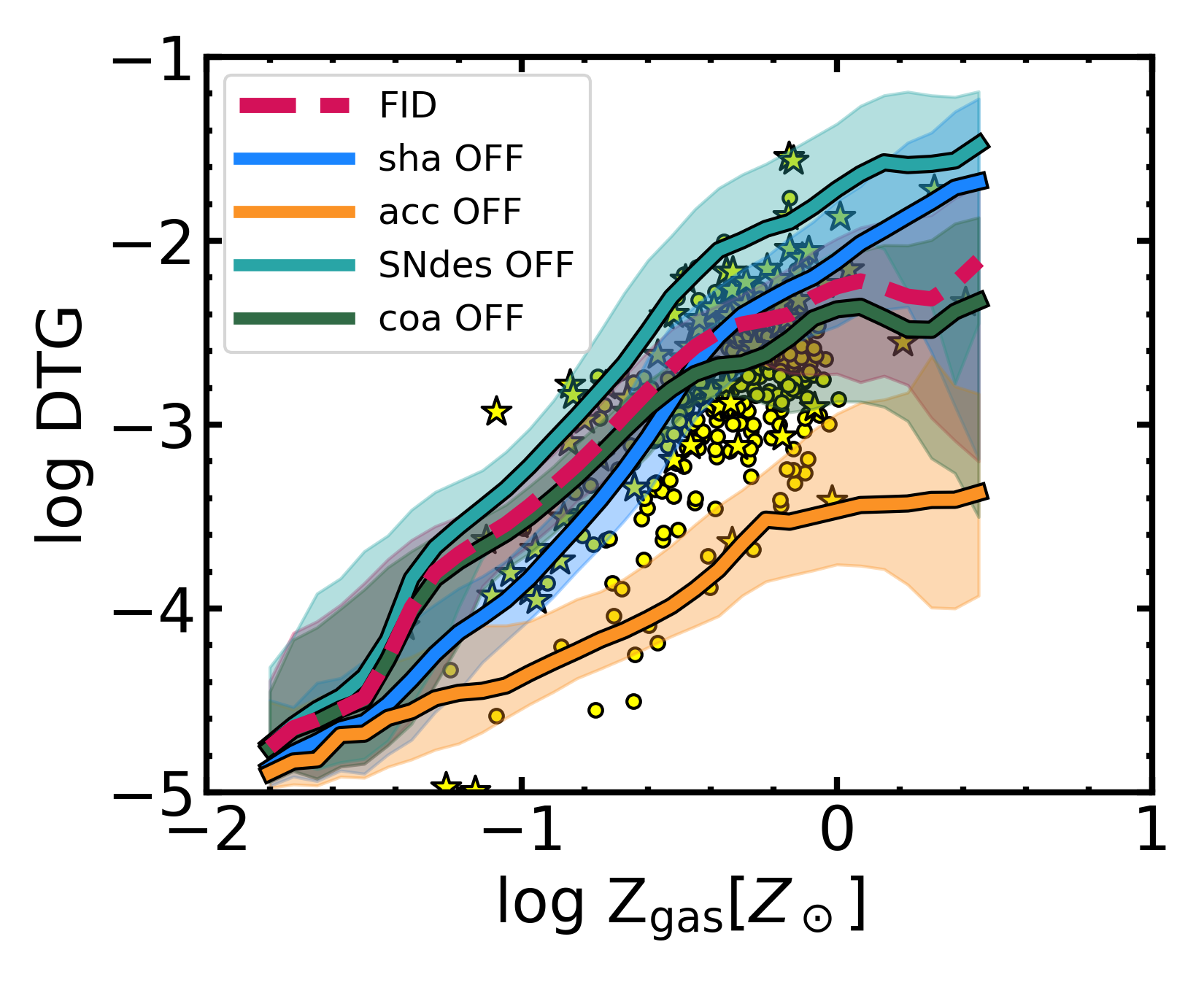}\quad
    \includegraphics[width=0.65\columnwidth]{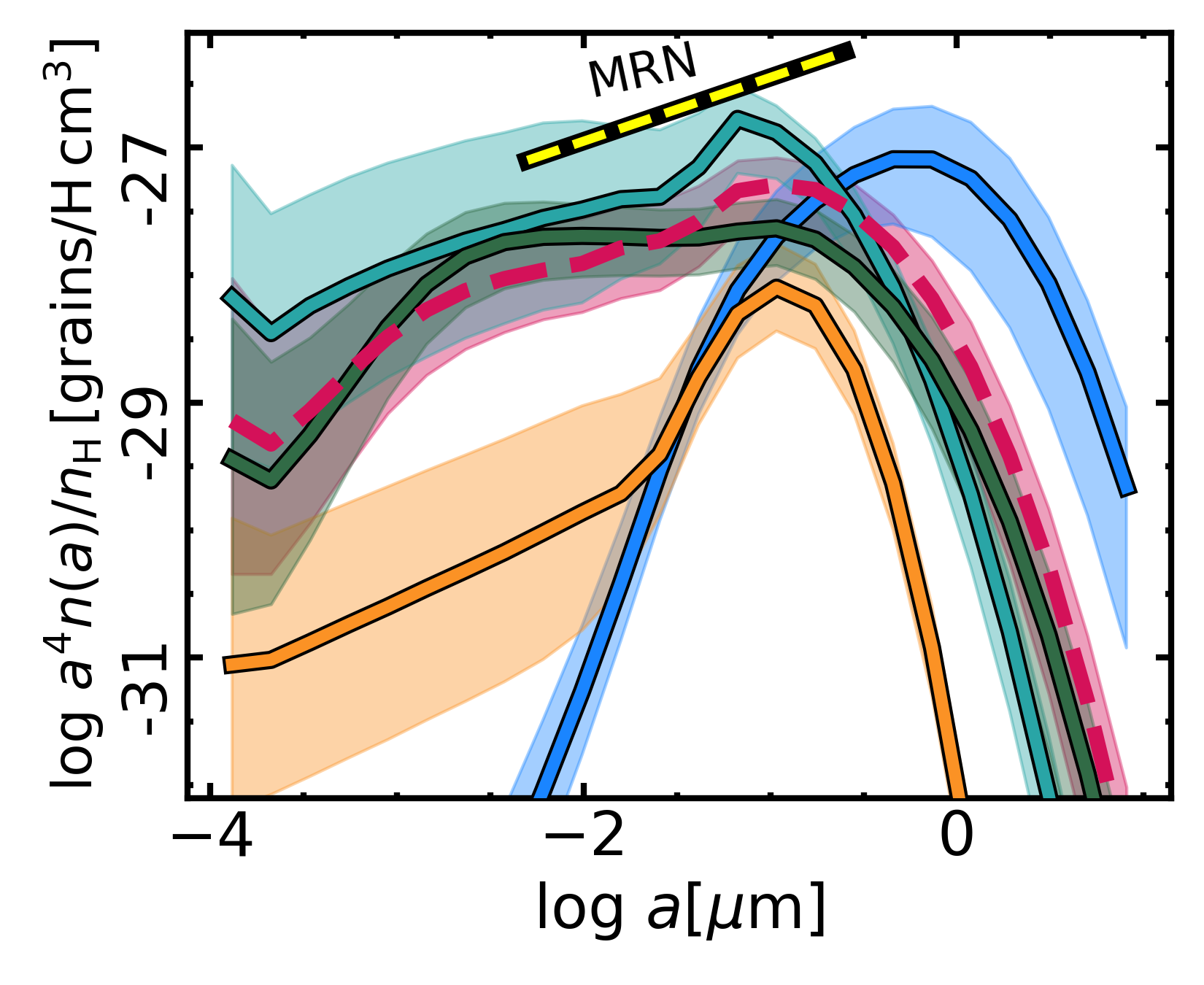}\quad
    \includegraphics[width=0.65\columnwidth]{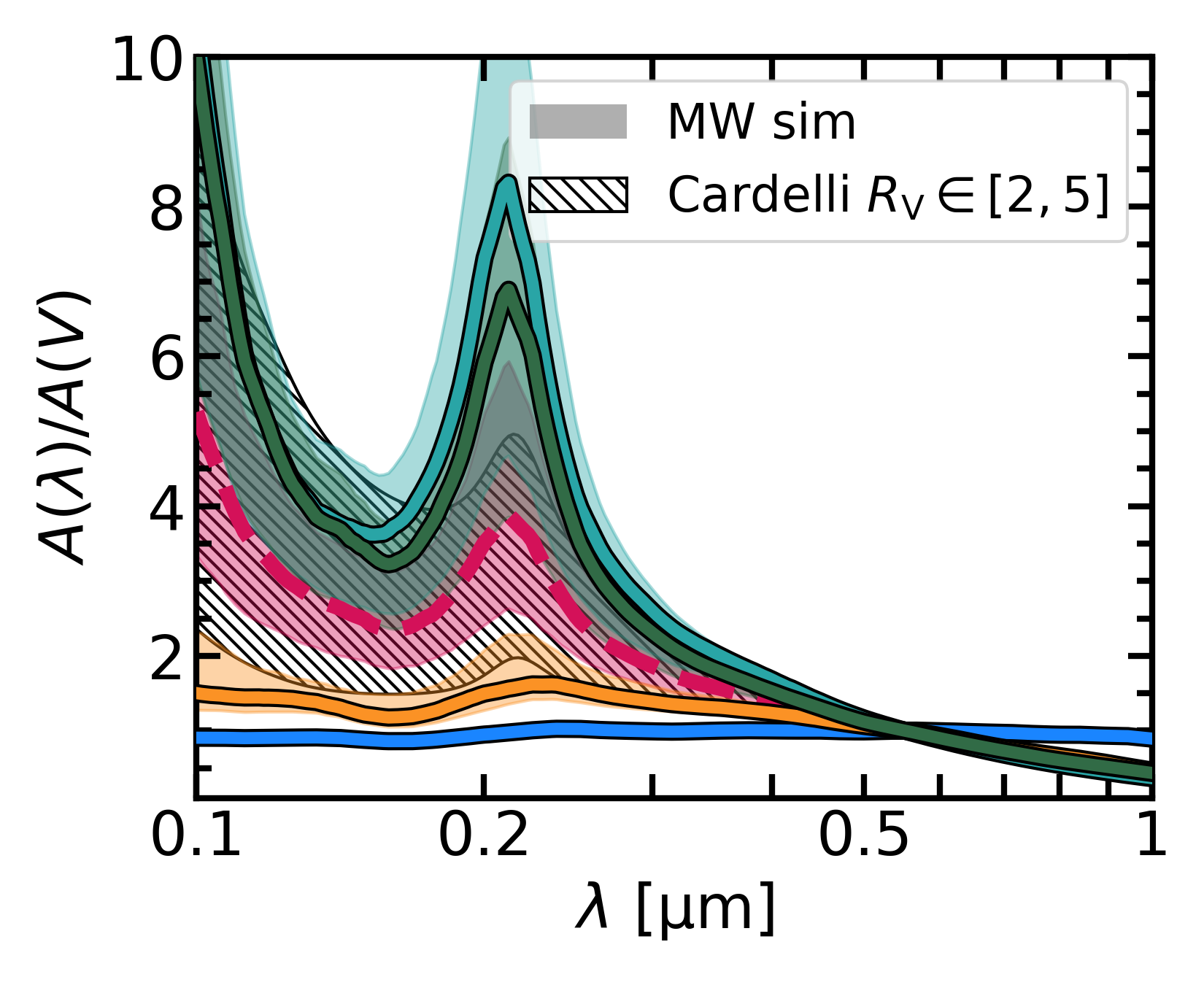}\quad

    \caption{\textbf{The impact of switching off individual dust processes is very strong, in particular for accretion.} 
    The panels show $z=0$ model predictions of the DTG–$Z_{\rm gas}$ relation (left), the GSD for the full galaxy sample (middle), and the extinction curves of MW–mass galaxies (right). In each run shown here a single dust process (shattering, coagulation, accretion, SN destruction) is turned off.  Solid lines and shaded regions indicate the median and the $16$–$84$th percentiles. Local observational are shown in the left panel, the \citetalias{MRN} slope is reported for reference in the middle panel, and the \cite{Cardelli89} extinction curve for $2 \leq R_{\rm V} \leq 5$ is shown in the right panel.} 
    \label{fig:oneproc:main}
\end{figure*}

We start by investigating the role of the different ISM processes\footnote{In this analysis we don't consider the impact of hot gas sputtering, whose role is negligible for integrated ISM quantities.} -- accretion, destruction in SN shocks, shattering, and coagulation -- by disabling them individually. Figure \ref{fig:oneproc:main} shows the results of these experiments.

Regarding the impact on the dust mass abundance, accretion is the dominant process: when it is switched off, the DTG is strongly suppressed, with a more pronounced effect in higher-metallicity galaxies. This behaviour is expected, since accretion dominates over stellar production in the dust mass assembly at $z \lesssim 6$ in many simulation-based studies \citep[see e.g., Figure 3 of ][]{Parente25rev}. Grain destruction in SN shocks has a somewhat smaller impact. In contrast, coagulation and shattering have a smaller impact on the total dust mass. In low- to intermediate-metallicity galaxies ($Z_{\rm gas} \lesssim 0.3\,Z_\odot$), shattering indirectly boosts accretion by increasing the abundance of small grains, which accrete more efficiently. At higher metallicities (and in more star forming systems), however, the enhanced small-grain population instead leads to more efficient destruction by SNe; as a consequence, turning off shattering leads to higher DTGs in this regime.

When it comes to the impact on grain sizes, shattering and accretion are the most influential processes. Shattering is essential for generating a population of small grains ($a \lesssim 0.01\,\mu{\rm m}$) through the fragmentation of large, stellar-produced grains, thereby enabling accretion to subsequently increase their mass and build up the small-grain tail of the GSD. This tail therefore arises from the combined action of shattering and accretion. It is worth noting that, in the absence of shattering, accretion alone can still increase the total dust mass, but the small-grain tail is entirely absent, potentially resulting in unrealistic extinction curves.
Coagulation and SN destruction have a minor effect on shaping the GSD: coagulation decreases the small-to-large grain ratio by a factor of $\approx 6$, while SN destruction shifts the distribution to larger sizes and suppresses mainly $a \lesssim 0.1 \, \mum$ grains abundance. However, both of these processes are important to recover MW-mass extinction curves.

\subsection{The initial stellar production}


\begin{figure*}[!htb]
    \centering
    \includegraphics[width=0.67\columnwidth]{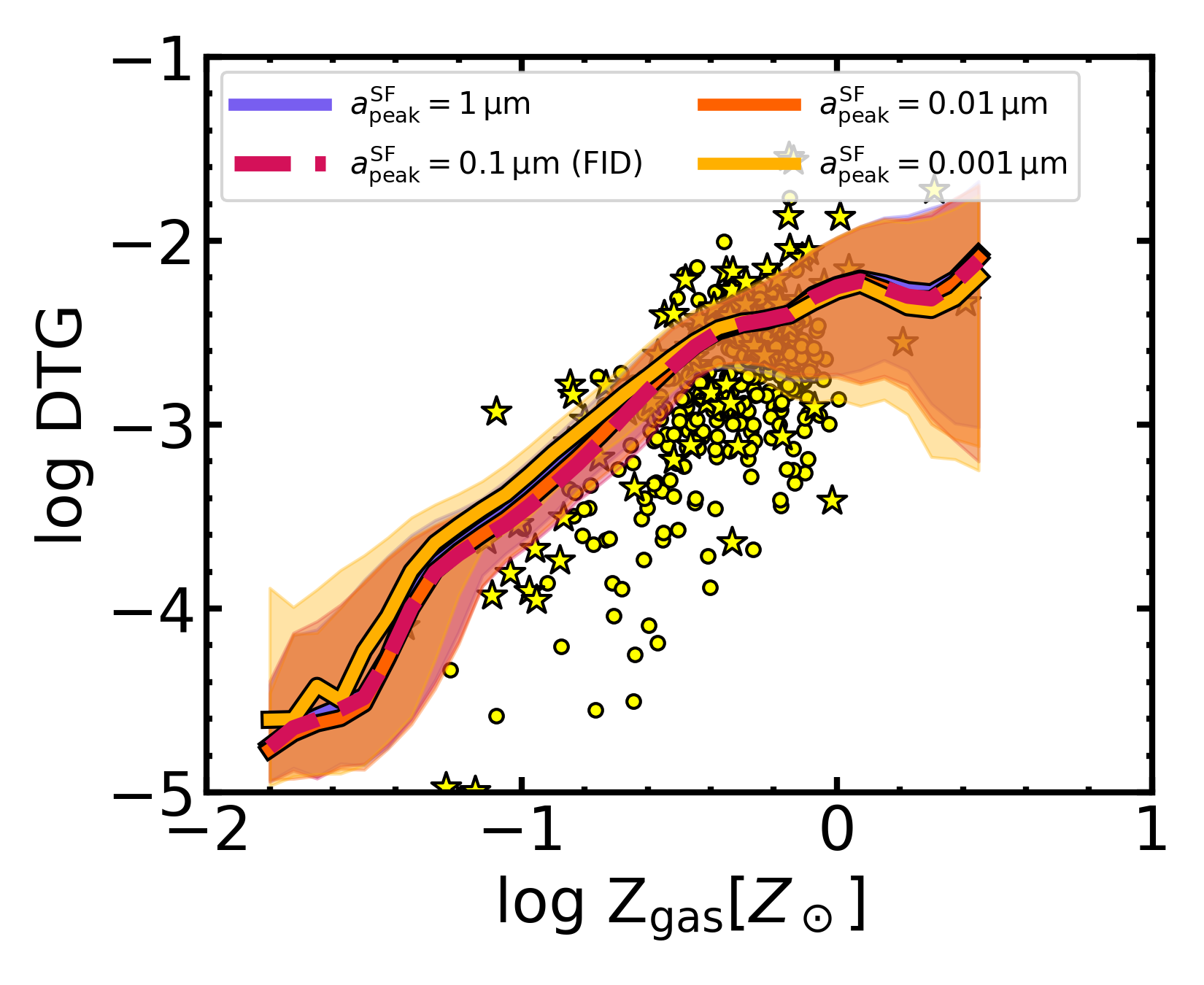}\quad
    \includegraphics[width=0.65\columnwidth]{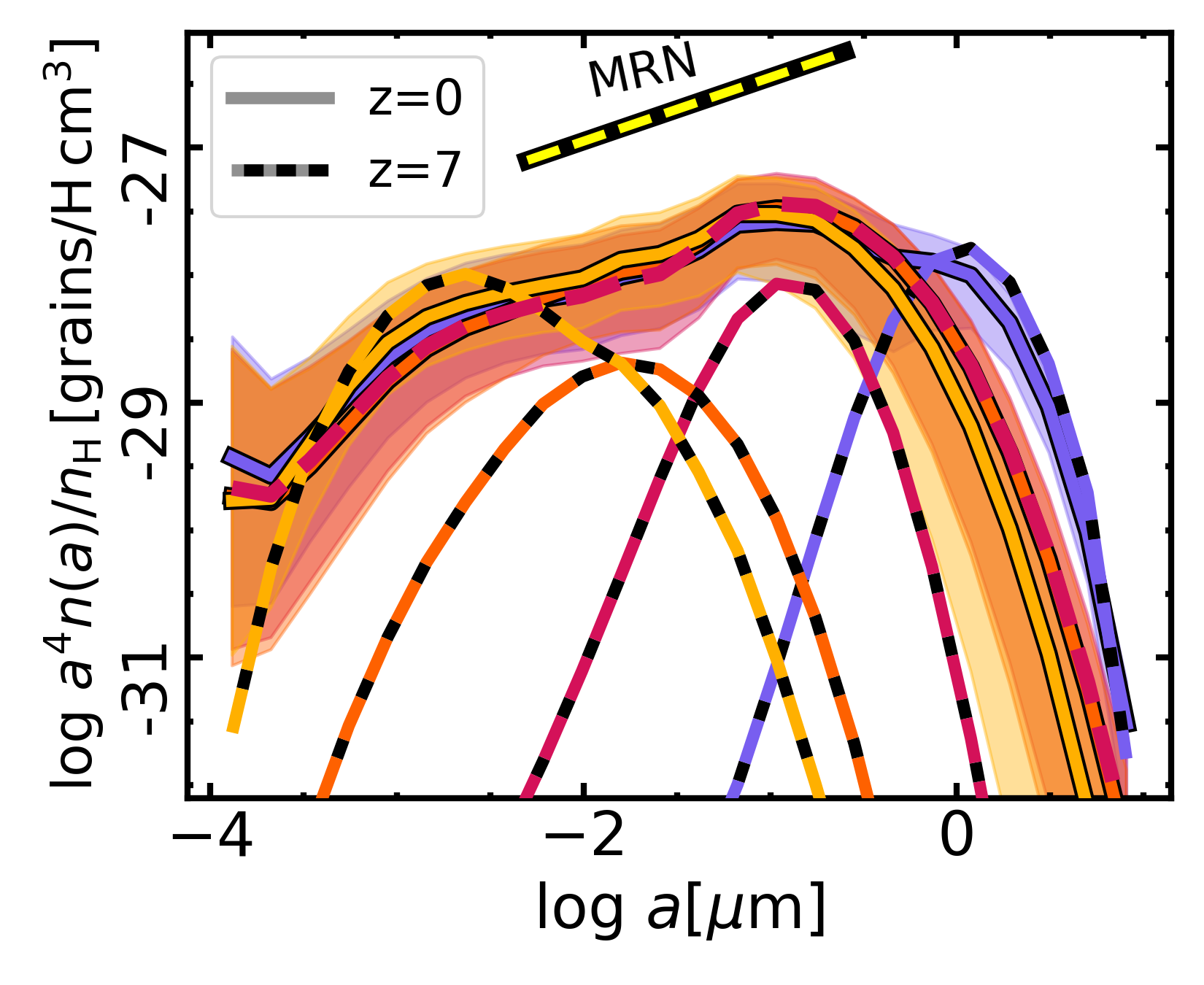}\quad
    \includegraphics[width=0.65\columnwidth]{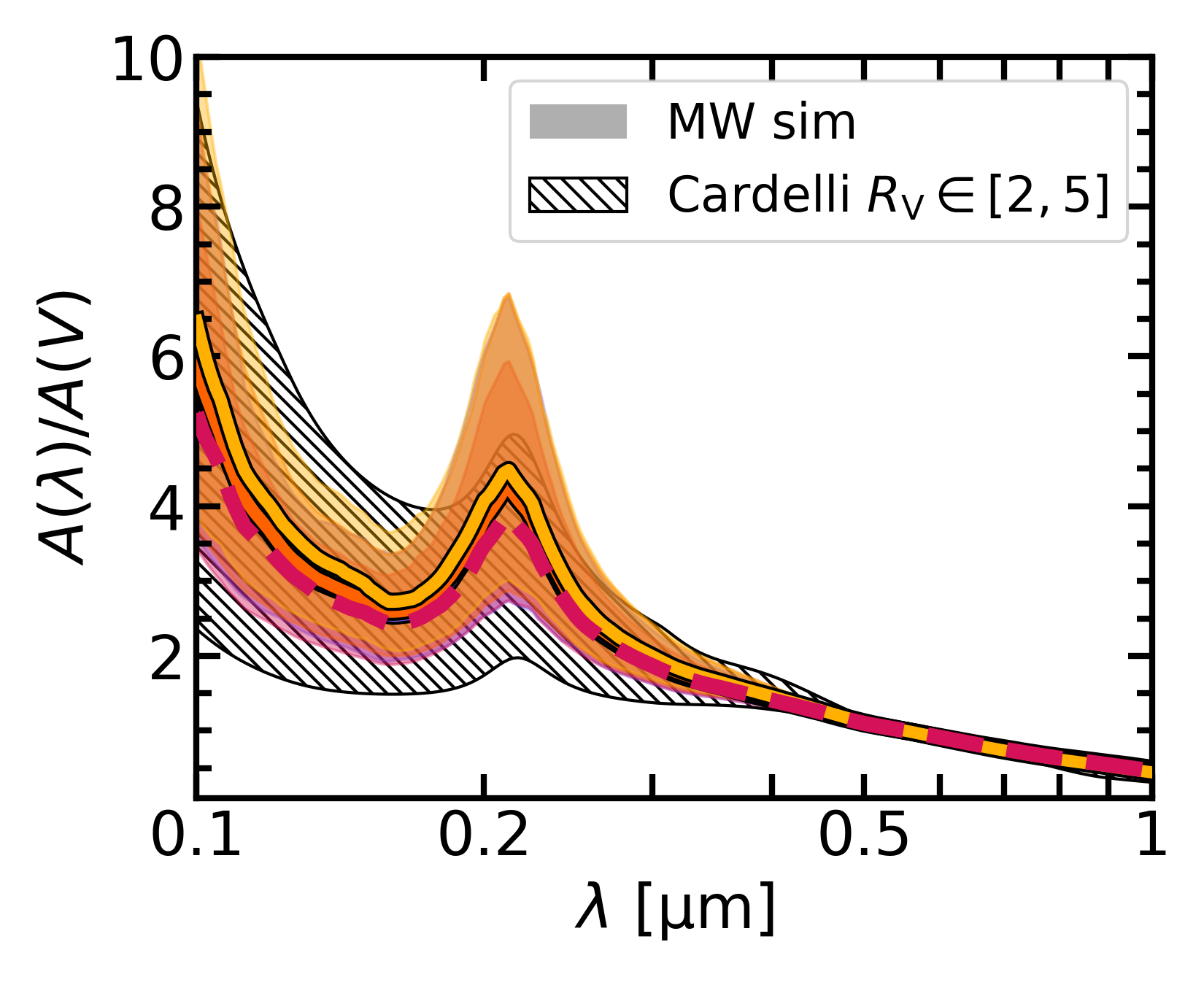}\quad

    \caption{\textbf{Assumption on stellar production of grains has a small impact on the dust of local galaxies.} The panels show $z=0$ model predictions of the DTG–$Z_{\rm gas}$ relation (left), the GSD for the full galaxy sample (middle, at both $z=0$ and $z=7$), and the extinction curves of MW–mass galaxies (right). Each run assumes a different peak $a_{\rm peak}$ of the lognormal distribution of the GSD of grains produced by stellar sources. Solid lines and shaded regions indicate the median and the $16$–$84$th percentiles. Local observational are shown in the left panel, the \citetalias{MRN} slope is reported for reference in the middle panel, and the \cite{Cardelli89} extinction curve for $2 \leq R_{\rm V} \leq 5$ is shown in the right panel.} 
    \label{fig:apeak}
\end{figure*}

We now investigate the impact of assumptions about the size of grains produced by stellar sources -- both AGB stars and SNe -- which are commonly assumed to be large ($a \gtrsim 0.1\,\mu{\rm m}$). To this end, we carry out a series of experiments in which we vary the peak $a_{\rm peak}^{\rm SF}$ of the lognormal probability distribution that defines the size distribution of stellar-produced grains in our model. Specifically, we adopt $a_{\rm peak}^{\rm SF} = 0.001,\,0.01,\,0.1,$ and $1\,\mu{\rm m}$, with the second-to-last value corresponding to our fiducial choice. The results are presented in Figure \ref{fig:apeak}.

The impact of this parameter on the total dust mass is negligible, as indicated by the DTG–$Z_{\rm gas}$ relation at $z = 0$ ($\langle |\Delta_\sigma| \rangle \lesssim 0.2$ when compared to the fiducial run). Interestingly, in evolved (i.e. metal-enriched) low-$z$ galaxies, its effect on the overall GSD is also minimal. The size of stellar-produced grains plays a role only at high redshift (with $z = 7$ shown here for reference), whereas at low redshift the resulting GSDs are similar to each other. This shows that ISM processing efficiently erases the imprint of the initial stellar grain-size distribution. As a consequence, the extinction curve properties of MW-mass galaxies are also very similar across these models.  A similar result was recovered in the cosmological zoom-in simulations by \citet{Narayanan25}.

\subsection{Shattering and Coagulation}
\label{sec:shacoa:exp}

\begin{figure*}[!htb]
    \centering
    \includegraphics[width=0.67\columnwidth]{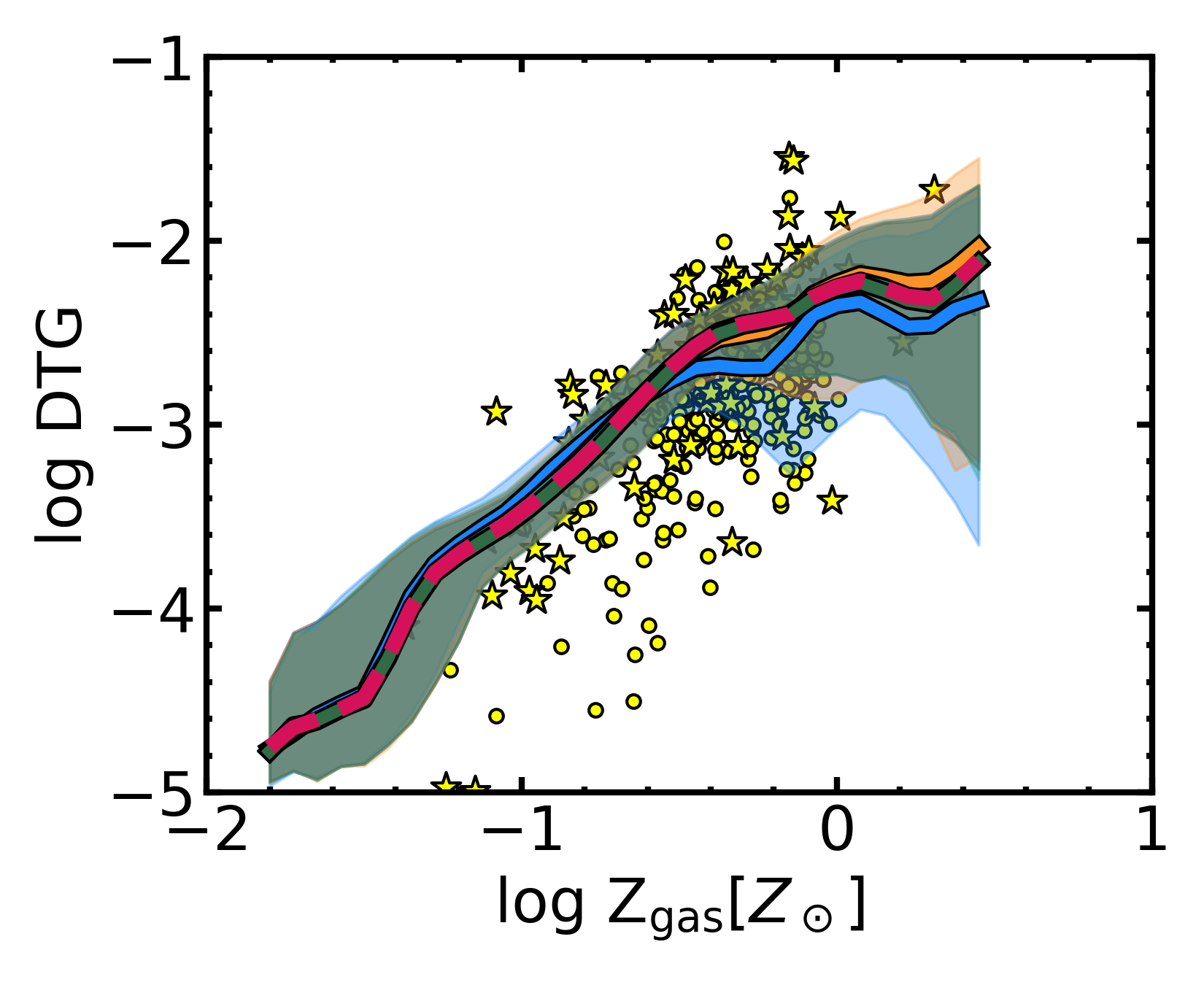}\quad
    \includegraphics[width=0.65\columnwidth]{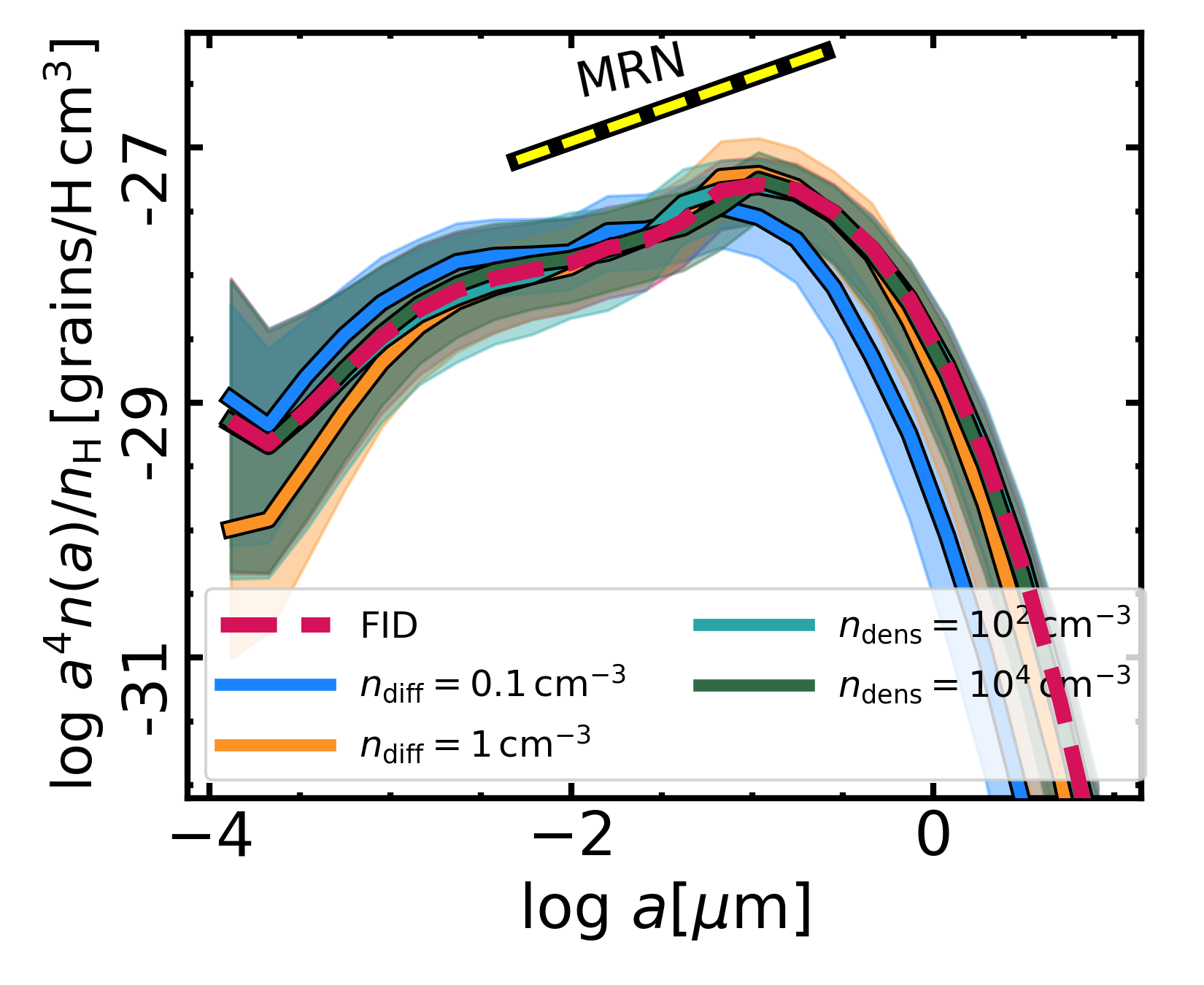}\quad
    \includegraphics[width=0.65\columnwidth]{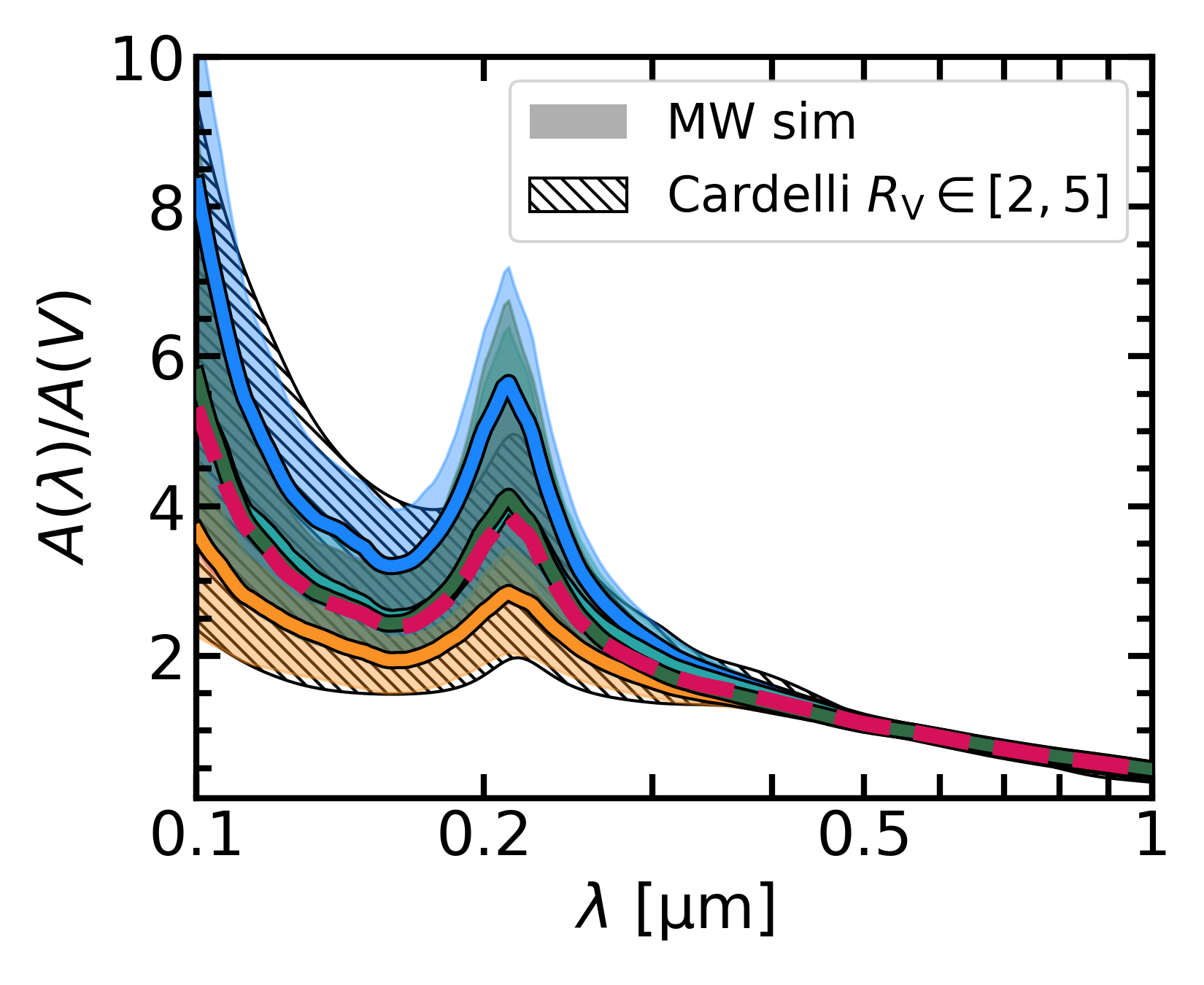}\quad

    \caption{\textbf{Assumptions on the density of the unresolved ISM have a small impact on dust properties.} The panels show $z=0$ model predictions of the DTG–$Z_{\rm gas}$ relation (left), the GSD for the full galaxy sample (middle), and the extinction curves of MW–mass galaxies (right). Each run assumes a different density for the diffuse and dense medium where shattering and coagulation occur. Solid lines and shaded regions indicate the median and the $16$–$84$th percentiles. Local observational are shown in the left panel, the \citetalias{MRN} slope is reported for reference in the middle panel, and the \cite{Cardelli89} extinction curve for $2 \leq R_{\rm V} \leq 5$ is shown in the right panel.} 
    \label{fig:shacoa:nT}
\end{figure*}

In this section we investigate the impact of assumptions related to the treatment of shattering and coagulation. Being collisional processes, these depend on the modeling of relative velocities and thermodynamic properties -- assumptions that are necessary here due to the limited resolution of the SAM.

In Figure \ref{fig:shacoa:nT}, we explore the effect of the assumed densities of the dense and diffuse ISM\footnote{While we focus here on the effects of varying density, we note that temperature and Mach number are also free parameters (at least within our model) that, when varied, can produce the same changes in grain velocities according to Eq. \ref{eq:vgrain}.}, noting that in our fiducial run the diffuse medium is modeled using the simple form in Eq. \ref{eq:rhodiff} and $n_{\rm dense}=10^3\, {\rm cm}^{-3}$. The figure shows results for $n_{\rm diff} = 0.1,\,1 \, {\rm cm^{-3}}$ and $n_{\rm dense} = 10^2,\,10^4 \, {\rm cm^{-3}}$. These variations have only a minor effect on the total dust mass predicted by the model\footnote{Here we vary only the physical properties of the dense medium as relevant for coagulation. While grain accretion (as a collisional process) also depends on these densities, it is kept fixed at the fiducial value $n_{\rm dense}=10^3 \, {\rm cm^{-3}}$ to isolate the effect of coagulation.}, as evident from the DTG$-Z_{\rm gas}$ relation.

The impact on the GSD is also limited. The effect of varying the diffuse medium density is straightforward: lower densities result in higher relative velocities, increasing shattering and thus the abundance of small grains. The effect of varying the dense medium density on the GSD shape is minimal. The $n_{\rm dens}=10^4\,(10^2)\,{\rm cm^{-3}}$ run yields a S/L ratio that is higher (lower) by a modest factor of $\approx 1.3$ compared to the fiducial run. Although this may seem counterintuitive, higher densities reduce relative velocities, which would normally favor coagulation. Since the collisional rate of coagulation is proportional to velocity, its efficiency actually decreases, resulting in less coagulation and higher S/L ratios.
These variations are reflected in the extinction curves of MW-mass galaxies, which remain broadly similar and consistent with observational constraints.


\begin{figure*}[!htb]

    \centering
    \includegraphics[width=0.67\columnwidth]{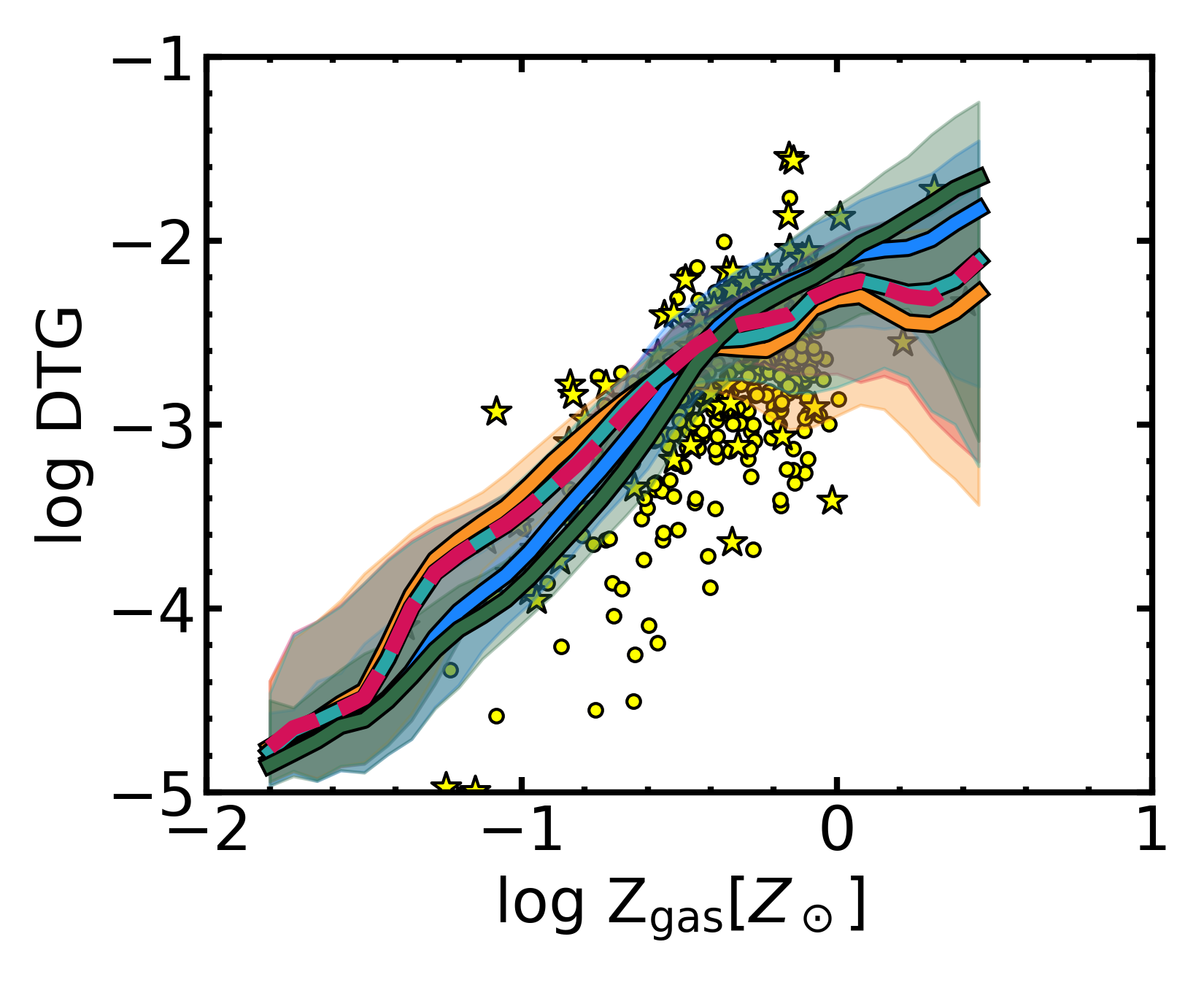}\quad
    \includegraphics[width=0.65\columnwidth]{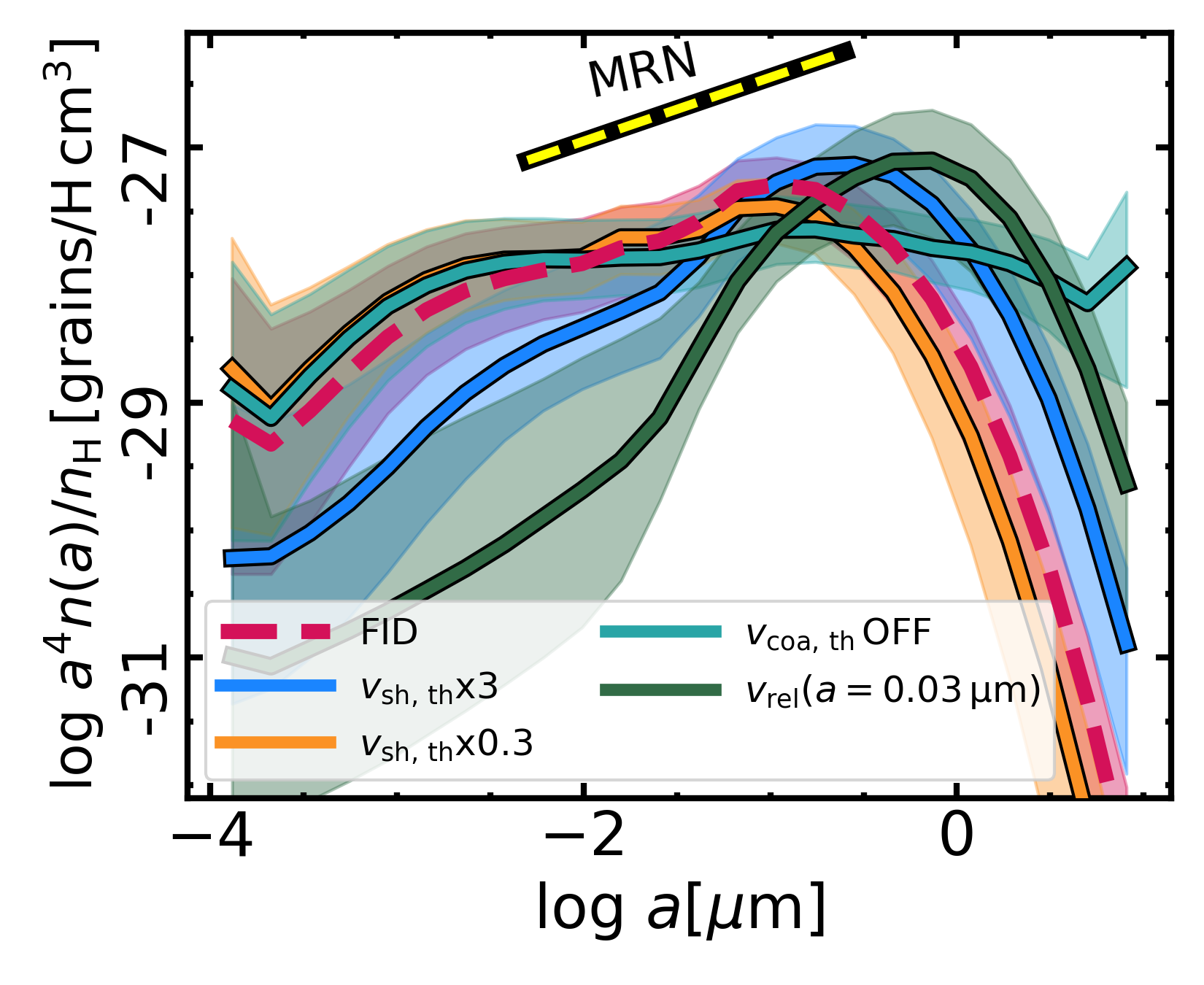}\quad
    \includegraphics[width=0.65\columnwidth]{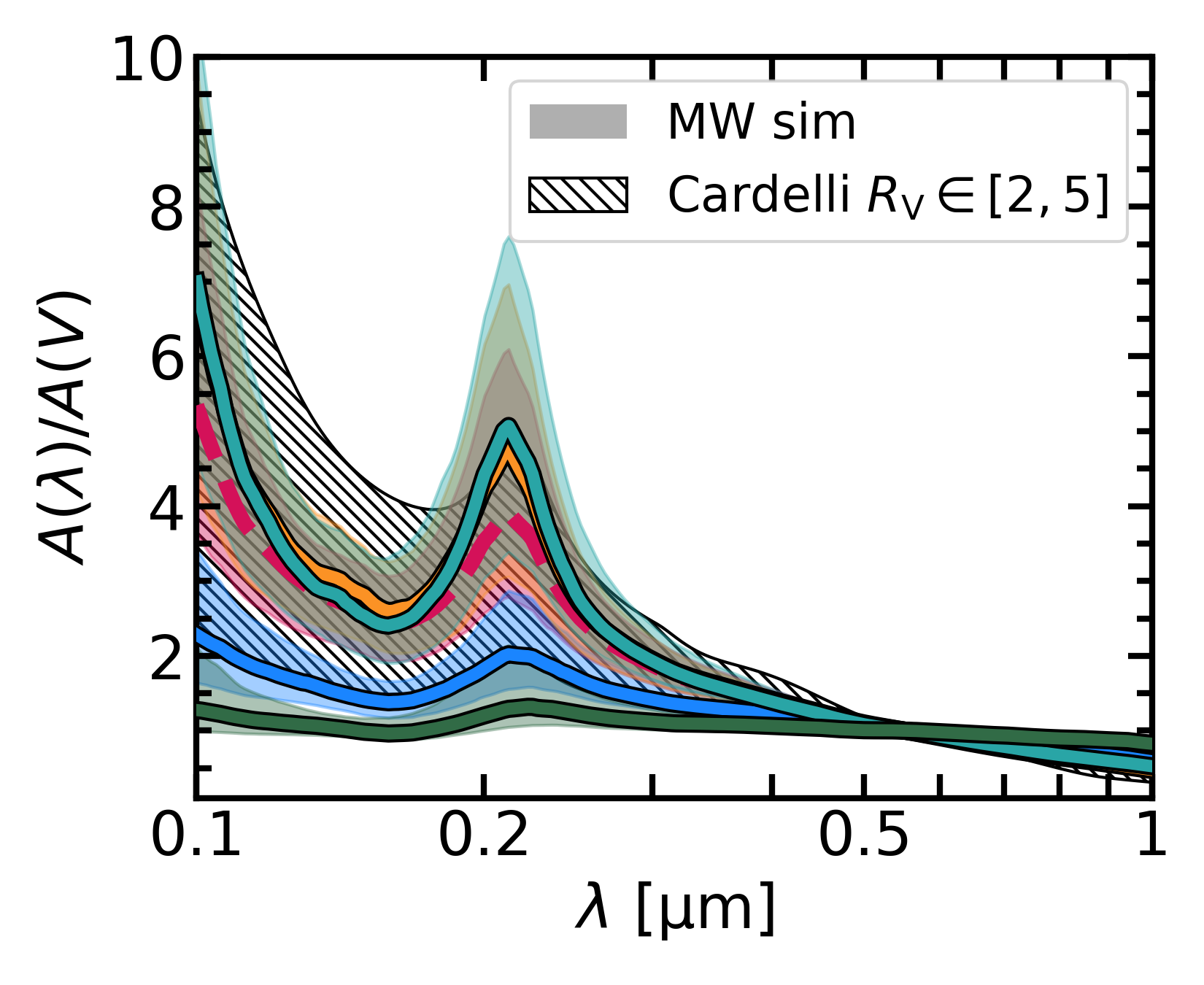}\quad

    \caption{\textbf{Grain velocities and thresholds for shattering and coagulation influence the GSD and the extinction curve.} The panels show $z=0$ model predictions of the DTG–$Z_{\rm gas}$ relation (left), the GSD for the full galaxy sample (middle), and the extinction curves of MW–mass galaxies (right). The velocity threshold for shattering and  coagulation is modified, and we also include a run in which the grain velocity is assumed to be size–independent and fixed to its value at $a = 0.03\,\mu{\rm m}$. Solid lines and shaded regions indicate the median and the $16$–$84$th percentiles. Local observational are shown in the left panel, the \citetalias{MRN} slope is reported for reference in the middle panel, and the \cite{Cardelli89} extinction curve for $2 \leq R_{\rm V} \leq 5$ is shown in the right panel.} 

    \label{fig:shacoa:vrel}
\end{figure*}


\begin{figure*}[t!]

    \centering
    \includegraphics[width=0.66\columnwidth]{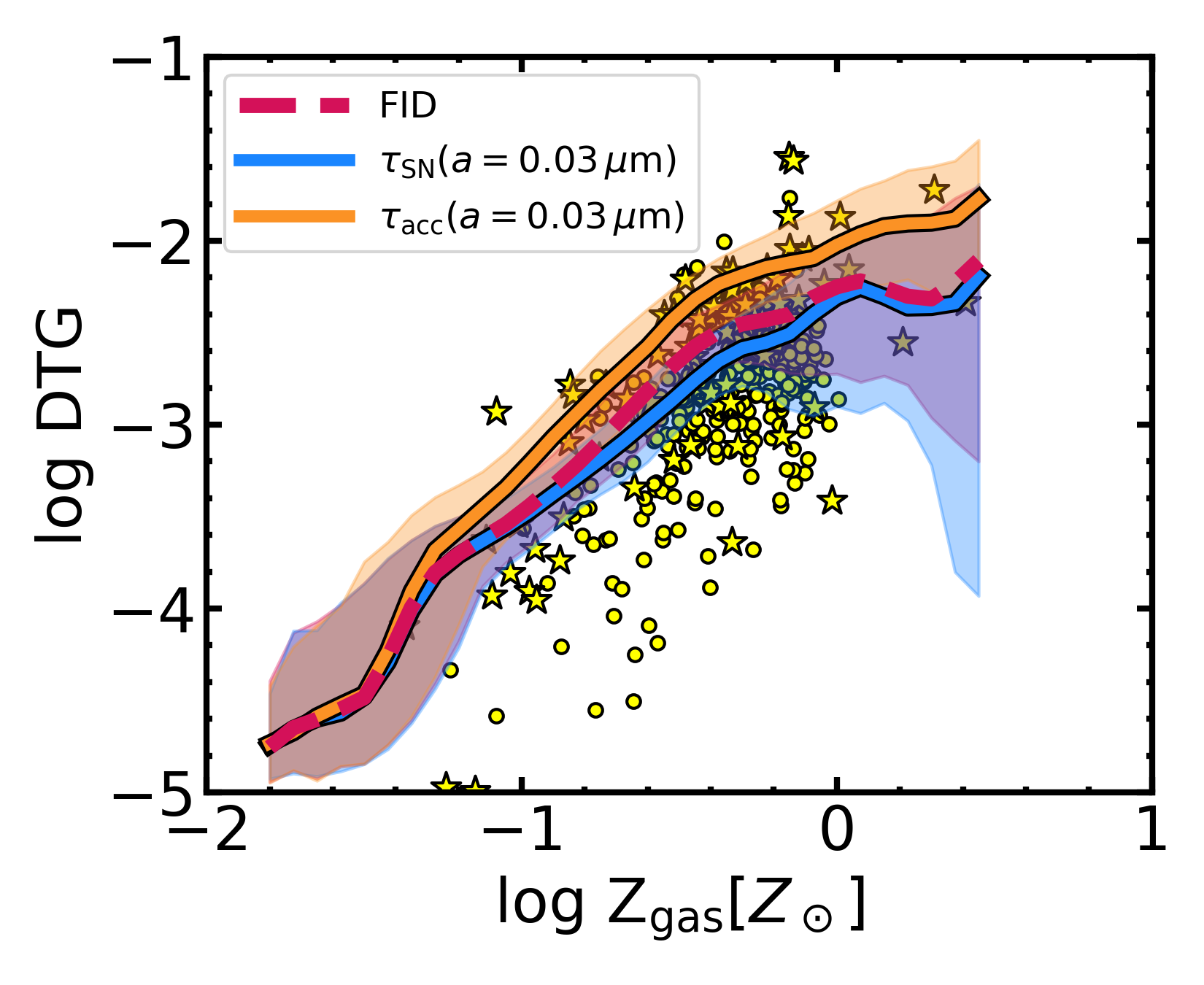}\quad
    \includegraphics[width=0.65\columnwidth]{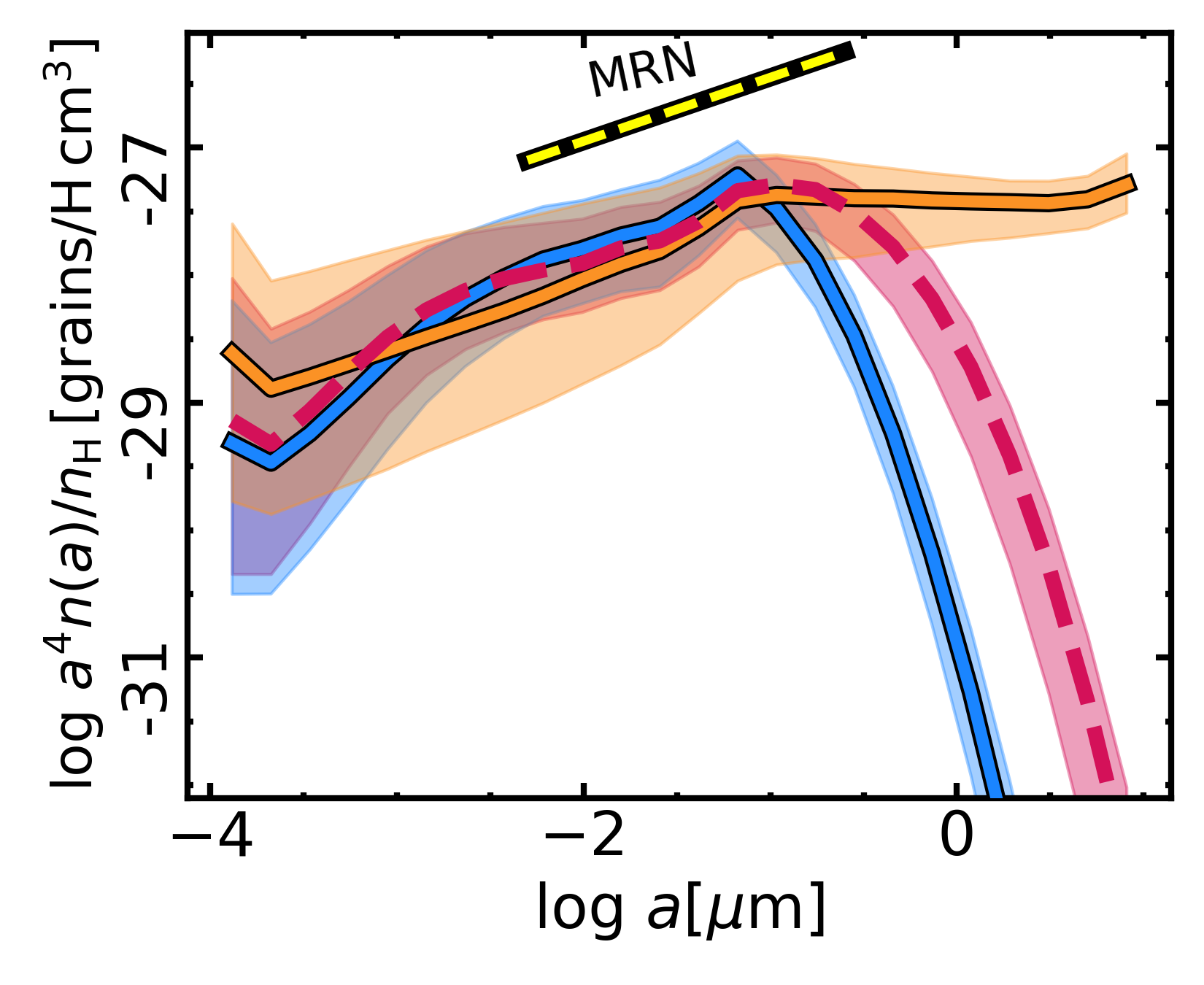}
    \includegraphics[width=0.65\columnwidth]{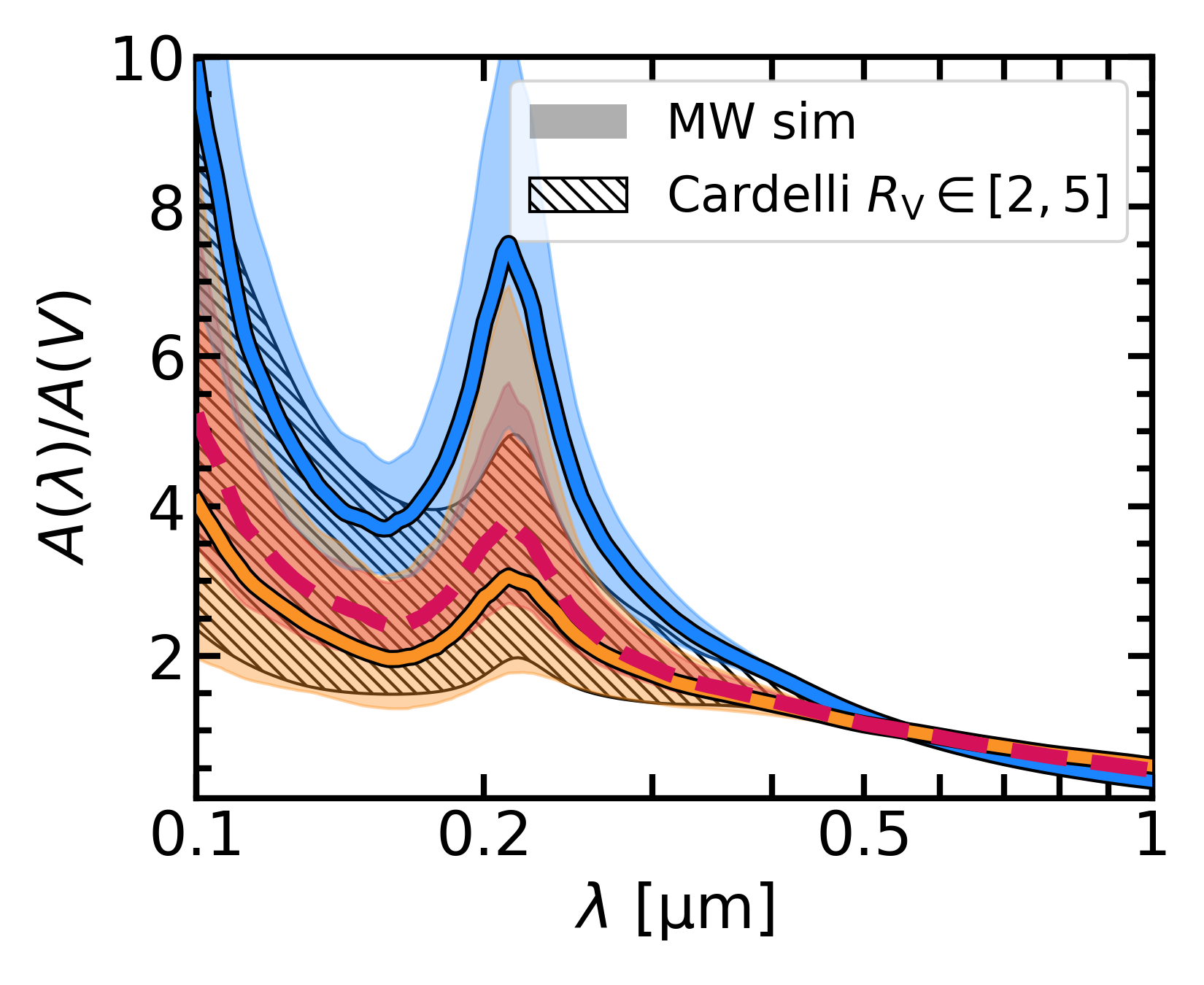}\quad

    \caption{\textbf{The size-dependence of accretion and SN destruction efficiency is important to reproduce the GSD and extinction curves.} The panels show $z=0$ model predictions of the DTG–$Z_{\rm gas}$ relation (left), the GSD for the full galaxy sample (middle, at both $z=0$ and $z=7$), and the extinction curves of MW–mass galaxies (right). In these experiments the accretion and SN destruction timescales are computed at the fixed value of $a=0.03 \, \mum$. Solid lines and shaded regions indicate the median and the $16$–$84$th percentiles. Local observational are shown in the left panel, the \citetalias{MRN} slope is reported for reference in the middle panel, and the \cite{Cardelli89} extinction curve for $2 \leq R_{\rm V} \leq 5$ is shown in the right panel.} 
    \label{fig:accSN:nosize}
\end{figure*}

Another key assumption in the modeling of shattering and coagulation concerns their velocity dependence. These processes are activated above or below specific velocity thresholds, which are material-dependent (Sect. \ref{sec:shacoa:vel}). In addition, the relative velocity of grains depends on their size, with larger grains moving faster because they are more easily coupled to large-scale motions.

In Figure \ref{fig:shacoa:vrel} we illustrate the impact of these velocity-related assumptions by varying the shattering threshold velocity, removing the velocity threshold for coagulation, and eliminating the size dependence of grain velocities by adopting the velocity of $0.03\,\mu{\rm m}$ grains for all sizes. In these experiments, the total dust mass remains largely unchanged, with only a slight decrease at mid-to-low metallicities when shattering efficiency is reduced (either by increasing the threshold velocity or by fixing the velocity for all grain sizes). These differences become even smaller at higher redshift, because shattering only takes place in the diffuse ISM, whose fraction tends to grow toward $z=0$. On the other hand, at higher metallicities the dust abundance is slightly enhanced, due to the presence of larger grains that are less efficiently destroyed by SN shocks.

The impact of these experiments on the GSD and extinction curves is more relevant. Ignoring the size dependence of grain velocities has the largest impact: because the velocities of large grains are reduced, shattering is significantly suppressed, preventing the formation of a small-grain tail, hence too flat extinction curves. Allowing all grains to coagulate\footnote{This corresponds to the sticky coagulation model of \cite{HirashitaLi13}, motivated by the fact that grains coated with water ice can stick together at higher velocities.} enables the distribution to extend to grains $\gtrsim 1\,\mu{\rm m}$. Nevertheless, this run still yields an extinction curve consistent with both the fiducial model and observations, and is in fact slightly steeper, with a more pronounced bump. This apparently counterintuitive outcome arises because, although the GSD extends to larger grains -- whose contribution to the extinction curve at $\lambda \lesssim 0.5 ,\mum$ is minimal -- the shattering of these grains enhances the population of small grains, hence amplifying these features in the extinction curve. 

\subsection{Accretion and SN destruction}

Finally, we investigate the role of the size dependence of ISM metal accretion and SN shock destruction -- both of which preferentially affect small grains (Eqs. \ref{eq:SNeps} and \ref{eq:acc}) -- on the model predictions. As a numerical experiment, we run two simulations in which these processes are made size-independent by fixing the grain radius to $a = 0.03\,\mu{\rm m}$. The net effect is that both accretion and SN destruction become less efficient for small grains and more efficient for large grains. The results are shown in Figure \ref{fig:accSN:nosize}.

The total dust mass is only weakly sensitive to this assumption. At high metallicity, the model with size-independent accretion yields slightly higher DTG ratios. The differences become more pronounced at high redshift (not shown), where accretion becomes efficient at earlier times, as it no longer relies on shattering to produce small grains.

In contrast, the GSD is strongly impacted. Removing the size dependence of SN destruction shifts the peak of the distribution toward smaller grains, as destruction becomes more efficient at relatively large sizes, resulting in steeper extinction curves. Making accretion size-independent also promotes the growth of grains beyond the stellar-dominated regime, i.e. $a \gtrsim 0.1\,\mu{\rm m}$. Interestingly, despite these changes, an \citetalias{MRN}-like shape is still recovered, allowing the model to reproduce a realistic MW extinction curve. Although the smallest grains are not accreted or destroyed with the same efficiency, they retain the shape of shattered fragments (Eq. \ref{eq:fragments}), which indeed closely resembles the \citetalias{MRN}.

\section{Summary}
\label{sec:summary}
In this work we have introduced a new framework for modeling the evolution of dust grain size distributions (GSDs) within the \textsc{L-Galaxies} cosmological semi-analytic model (SAM) of galaxy evolution. By embedding a multi-bin treatment of grain sizes into a computationally efficient SAM, we bridge the gap between detailed dust physics -- previously explored mainly in idealized or zoom-in simulations -- and population-wide studies of galaxies across cosmic time. This allows, for the first time, an efficient investigation of the GSD and grain-size–dependent processes in a statistically representative galaxy sample.

Our model follows the evolution of dust mass and grain size through stellar production and interstellar processing. Grains produced by AGB stars and core-collapse SNe provide an initial reservoir dominated by large sizes, while subsequent evolution in the ISM is governed by shattering, coagulation, accretion of gas-phase metals, and destruction in SN shocks and hot gas. These processes are treated in a size-dependent manner and coupled self-consistently to the evolving physical conditions of galaxies predicted by the SAM.
Our main results follow.

\begin{itemize}
    \item \textbf{Dust abundance and size distribution\\}The model reproduces observed dust abundances and predicts a GSD that evolves toward an \citetalias{MRN}-like slope in low-redshift, MW-mass galaxies. Accretion of metals onto grains in the ISM powers the transition from a GSD dominated by large grains to one dominated by small grains, with local galaxies featuring a small-to-large grain mass ratios consistent with observational estimates. This transition is closely tied to the overall star formation history and galaxy evolution, with more massive galaxies at $z=0$ experiencing it at earlier cosmic times and larger metallicities.
    \item \textbf{Extinction curves\\}Extinction curves of MW–mass galaxies at $z =0$ are consistent with observed UV–optical slopes and show a prominent $2175 \, \text{\AA}$ bump. The redshift evolution of extinction properties follows a physically intuitive trend: early galaxies, dominated by large grains produced by stellar sources, feature flatter extinction curves and weaker bumps, while continued ISM processing progressively steepens the curves and strengthens the carbon-driven feature at later times.
    \item \textbf{The impact of dust processes\\}Accretion in the ISM is the dominant contributor to the total dust budget, and its inclusion is essential for reproducing the observed dust abundance and dust-metallicity relation in local galaxies. Shattering, while contributing little to the total dust mass, plays a critical role in shaping the GSD by enabling the formation of a small-grain population. Coagulation and SN destruction act as secondary processes, redistributing mass across grain sizes or preferentially suppressing the smallest grains, with a minor impact on global dust abundances.
    \item \textbf{The impact of modeling choices\\}Many commonly adopted modeling choices have only a minor impact on low-redshift dust properties. Variations in the initial stellar dust size distribution, the efficiency of coagulation, or the densities of diffuse and dense ISM phases do not substantially affect dust abundance, GSD or extinction curves in evolved galaxies. In contrast, assumptions that remove or strongly modify the size dependence of key processes -- particularly grain velocities in collisional interactions -- can significantly affect the GSD, highlighting the importance of using physically motivated grain size scalings even in low-resolution models such as our semi-analytic framework.
    
\end{itemize}

Overall, our results indicate that dust mass predictions in galaxy formation models are relatively robust, provided that grain growth by accretion is included. Predictions for GSDs and extinction properties are more sensitive, but still converge toward observationally-supported solutions under a wide (non extreme) range of assumptions. Importantly, incorporating the full GSD has only a limited effect on the predicted dust abundances. Its main role is instead in accurately modelling the extinction curve, and thus in determining how galaxies are observed.

While this work primarily focuses on presenting the numerical model and constraining it against key observables, the model will be extensively exploited to investigate grain-size–dependent dust physics. This includes, but is not limited to, the impact of different grain populations on attenuation curves and the formation and emission of PAHs. Owing to its computational efficiency, future work will perform these studies in a fully cosmological context up to the earliest epochs, providing unique predictions to be tested against forthcoming observations, e.g., from JWST. \vspace{7mm}

\section*{Acknowledgments}

We thank the referee for the detailed and constructive report which improved the clarity of the work. We thank Caleb R. Choban and Gian Luigi Granato for stimulating discussions and for their careful reading of the manuscript. We also thank Monica Relaño for providing observational data. This work was funded by NASA ATP programs 80NSSC22K0716 (PI: PT) and 80NSSC24K1223 (PI: DN).

\bibliography{GSD_param}{}
\bibliographystyle{aasjournal} 

\appendix

\section{Numerical convergence}
\label{app:convergence}

\begin{figure}[htb!]
    \centering
    \includegraphics[width=0.4\columnwidth]{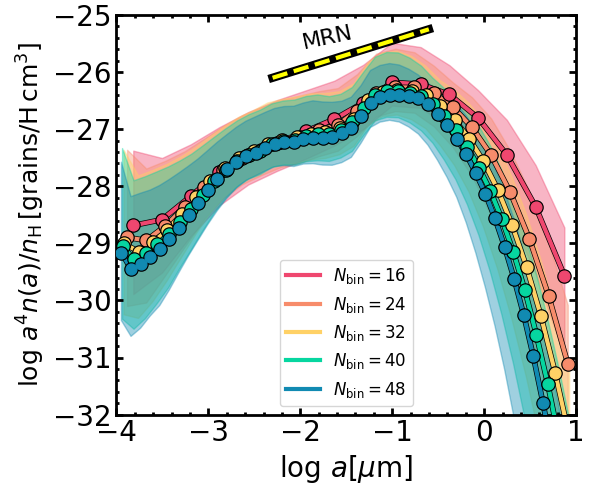}\quad
    \includegraphics[width=0.4\columnwidth]{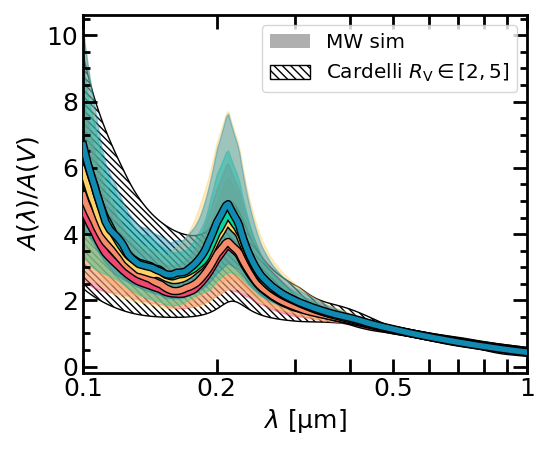}\quad
    \caption{{\bf Demonstration of convergence as a function of number of size bins.} GSD (left panel) and MW-mass extinction curves (right panel) predicted by our model at $z=0$ for the full sample of galaxies when adopting different number of grain size bins, namely $N=16, \, 24, \, 32, \, 40, \, {\rm and} \, 48$.}
    \label{fig:convergence:GSD}
\end{figure}

In Figure \ref{fig:convergence:GSD}, we show several realizations of the model obtained by varying the number of bins used to discretize the grain-size range $[10^{-4},\,10]\,\mu\mathrm{m}$. The results show excellent convergence in terms of GSD for grain sizes $a \lesssim 0.1\,\mu\mathrm{m}$, while convergence at larger sizes is slower, though still good. This convergence is further confirmed by the extinction curve properties of MW-mass galaxies, which also show good agreement for $N_{\rm bin} \geq 32$.


\section{S/L and DTM relation}
\label{app:SL-DTM}


\begin{figure}[]

    \centering
    \includegraphics[width=0.4\columnwidth]{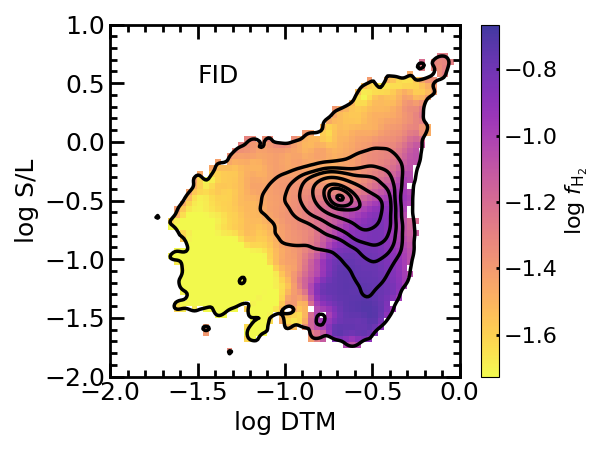}\quad
    \includegraphics[width=0.4\columnwidth]{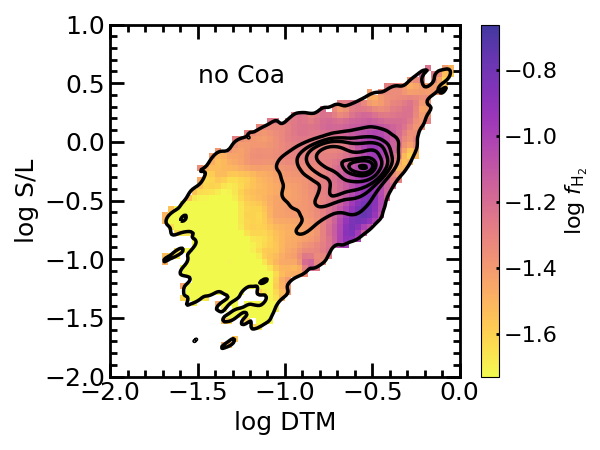}\quad

    \caption{\textbf{The S/L ratio strongly correlates with DTM, since accretion boosts both quantities.} Both panels display the relation between the S/L and DTM ratios for simulated $z=0$ galaxies. Contours indicate the galaxy distribution, while colors represent the molecular fraction. Results are shown for the fiducial model (left panel) and for a model in which coagulation is turned off (right panel).} 
    \label{fig:app:SLDTM}
\end{figure}

Figure \ref{fig:app:SLDTM} shows the relation between the S/L ratio and the DTM ratio for simulated $z=0$ galaxies, color-coded by their molecular fraction. As discussed in the main text, the two quantities are correlated, since both are enhanced by ISM accretion, which converts gas-phase metals into dust grains and is more efficient for small grains. Here, we further find that coagulation -- active in dense environments and therefore in galaxies with high molecular fractions -- plays a significant role in driving the scatter of this relation. Indeed, when coagulation is switched off, both the scatter and its dependence on molecular fraction are substantially reduced.\\

\end{document}